\pdfoutput=1

\documentclass[12pt,a4paper]{article}

\usepackage{ifthen} % for conditional statements
\newboolean{pdflatex}
\setboolean{pdflatex}{true} % False for eps figures 

\newboolean{articletitles}
\setboolean{articletitles}{true} % False removes titles in references

\newboolean{uprightparticles}
\setboolean{uprightparticles}{false} %True for upright particle symbols

\newboolean{inbibliography}
\setboolean{inbibliography}{false} %True once you enter the bibliography

\usepackage{microtype}
\usepackage{lineno}  % for line numbering during review
\usepackage{xspace} % To avoid problems with missing or double spaces after
\usepackage{caption} %these three command get the figure and table captions automatically small

\usepackage{graphicx}  % to include figures (can also use other packages)
\usepackage{color}
\usepackage{colortbl}
\graphicspath{{./figs/}} % Make Latex search fig subdir for figures

\usepackage{amsmath} % Adds a large collection of math symbols
\usepackage{amssymb}
\usepackage{amsfonts}
\usepackage{upgreek} % Adds in support for greek letters in roman typeset

\newcommand*\patchAmsMathEnvironmentForLineno[1]{%
\expandafter\let\csname old#1\expandafter\endcsname\csname #1\endcsname
\expandafter\let\csname oldend#1\expandafter\endcsname\csname
end#1\endcsname
 \renewenvironment{#1}%
   {\linenomath\csname old#1\endcsname}%
   {\csname oldend#1\endcsname\endlinenomath}%
}
\newcommand*\patchBothAmsMathEnvironmentsForLineno[1]{%
  \patchAmsMathEnvironmentForLineno{#1}%
  \patchAmsMathEnvironmentForLineno{#1*}%
}
\AtBeginDocument{%
\patchBothAmsMathEnvironmentsForLineno{equation}%
\patchBothAmsMathEnvironmentsForLineno{align}%
\patchBothAmsMathEnvironmentsForLineno{flalign}%
\patchBothAmsMathEnvironmentsForLineno{alignat}%
\patchBothAmsMathEnvironmentsForLineno{gather}%
\patchBothAmsMathEnvironmentsForLineno{multline}%
\patchBothAmsMathEnvironmentsForLineno{eqnarray}%
}

\usepackage{hyperref}    % Hyperlinks in references
\usepackage[all]{hypcap} % Internal hyperlinks to floats.

\usepackage{xspace} 
\usepackage{upgreek}

\def\lhcb {\mbox{LHCb}\xspace}

\def\MagUp {\mbox{\em Mag\kern -0.05em Up}\xspace}

\ifthenelse{\boolean{uprightparticles}}%
{

 \def\Pmu         {\ensuremath{\upmu}\xspace}

 \def\Ppi         {\ensuremath{\uppi}\xspace}

 \def\Ppsi        {\ensuremath{\uppsi}\xspace}

 \def\PDelta      {\ensuremath{\Delta}\xspace}                 
 \def\PXi      {\ensuremath{\Xi}\xspace}                 
 \def\PLambda      {\ensuremath{\Lambda}\xspace}                 
 \def\PSigma      {\ensuremath{\Sigma}\xspace}                 
 \def\POmega      {\ensuremath{\Omega}\xspace}                 
 \def\PUpsilon      {\ensuremath{\Upsilon}\xspace}                 
 
 \def\PB      {\ensuremath{\mathrm{B}}\xspace}                 
                  
 \def\PD      {\ensuremath{\mathrm{D}}\xspace}

 \def\PJ      {\ensuremath{\mathrm{J}}\xspace}                 
 \def\PK      {\ensuremath{\mathrm{K}}\xspace}

 \def\Pb      {\ensuremath{\mathrm{b}}\xspace}                 
 \def\Pc      {\ensuremath{\mathrm{c}}\xspace}

 \def\Pi      {\ensuremath{\mathrm{i}}\xspace}

 \def\Ps      {\ensuremath{\mathrm{s}}\xspace}

}
{

 \def\Pmu         {\ensuremath{\mu}\xspace}

 \def\Ppi         {\ensuremath{\pi}\xspace}

 \def\Ppsi        {\ensuremath{\psi}\xspace}                 
                  
 \mathchardef\PDelta="7101
 \mathchardef\PXi="7104
 \mathchardef\PLambda="7103
 \mathchardef\PSigma="7106
 \mathchardef\POmega="710A
 \mathchardef\PUpsilon="7107
                  
 \def\PB      {\ensuremath{B}\xspace}                 
                  
 \def\PD      {\ensuremath{D}\xspace}

 \def\PJ      {\ensuremath{J}\xspace}                 
 \def\PK      {\ensuremath{K}\xspace}

 \def\Pb      {\ensuremath{b}\xspace}                 
 \def\Pc      {\ensuremath{c}\xspace}

 \def\Pi      {\ensuremath{i}\xspace}

 \def\Ps      {\ensuremath{s}\xspace}

}

\makeatletter
\ifcase \@ptsize \relax% 10pt
  \newcommand{\miniscule}{\@setfontsize\miniscule{4}{5}}% \tiny: 5/6
\or% 11pt
  \newcommand{\miniscule}{\@setfontsize\miniscule{5}{6}}% \tiny: 6/7
\or% 12pt
  \newcommand{\miniscule}{\@setfontsize\miniscule{5}{6}}% \tiny: 6/7
\fi
\makeatother

\DeclareRobustCommand{\optbar}[1]{\shortstack{{\miniscule (\rule[.5ex]{1.25em}{.18mm})}
  \\ [-.7ex] $#1$}}

\def\mup        {{\ensuremath{\Pmu^+}}\xspace}
\def\mun        {{\ensuremath{\Pmu^-}}\xspace} % muon negative (\mum is taken)
\def\mumu       {{\ensuremath{\Pmu^+\Pmu^-}}\xspace}

\def\ellm       {{\ensuremath{\ell^-}}\xspace}
\def\ellp       {{\ensuremath{\ell^+}}\xspace}
\def\squark    {{\ensuremath{\Ps}}\xspace}

\def\cquark    {{\ensuremath{\Pc}}\xspace}

\def\bquark    {{\ensuremath{\Pb}}\xspace}

\def\pion   {{\ensuremath{\Ppi}}\xspace}

\def\pim    {{\ensuremath{\pion^-}}\xspace}

\def\kaon    {{\ensuremath{\PK}}\xspace}
  \def\Kbar    {{\kern 0.2em\overline{\kern -0.2em \PK}{}}\xspace}

\def\KorKbar    {\kern 0.18em\optbar{\kern -0.18em K}{}\xspace}

\def\Kp      {{\ensuremath{\kaon^+}}\xspace}

\def\Kstarz  {{\ensuremath{\kaon^{*0}}}\xspace}
\def\Kstarzb {{\ensuremath{\Kbar{}^{*0}}}\xspace}

  \def\Dbar    {{\kern 0.2em\overline{\kern -0.2em \PD}{}}\xspace}

\def\DorDbar    {\kern 0.18em\optbar{\kern -0.18em D}{}\xspace}

\def\B       {{\ensuremath{\PB}}\xspace}
\def\Bbar    {{\ensuremath{\kern 0.18em\overline{\kern -0.18em \PB}{}}}\xspace}

\def\BorBbar    {\kern 0.18em\optbar{\kern -0.18em B}{}\xspace}
\def\Bz      {{\ensuremath{\B^0}}\xspace}
\def\Bzb     {{\ensuremath{\Bbar{}^0}}\xspace}

\def\Bd      {{\ensuremath{\B^0}}\xspace}

\def\Bsb     {{\ensuremath{\Bbar{}^0_\squark}}\xspace}

\def\jpsi     {{\ensuremath{{\PJ\mskip -3mu/\mskip -2mu\Ppsi\mskip 2mu}}}\xspace}
\def\psitwos  {{\ensuremath{\Ppsi{(2S)}}}\xspace}

  \def\Y#1S{\ensuremath{\PUpsilon{(#1S)}}\xspace}% no space before {...}!

\def\Lbar        {{\ensuremath{\kern 0.1em\overline{\kern -0.1em\PLambda}}}\xspace}
\def\LorLbar    {\kern 0.18em\optbar{\kern -0.18em \PLambda}{}\xspace}

\newcommand{\decay}[2]{\ensuremath{#1\!\to #2}\xspace}         % {\Pa}{\Pb \Pc}

\def\to                 {\ensuremath{\rightarrow}\xspace}

\def\qsq       {{\ensuremath{q^2}}\xspace}

\def\CP                {{\ensuremath{C\!P}}\xspace}

\def\BdToKstmm    {\decay{\Bd}{\Kstarz\mup\mun}}

\def\BdToJPsiKst  {\decay{\Bd}{\jpsi\Kstarz}}

\def\AT#1     {\ensuremath{A_{\mathrm{T}}^{#1}}\xspace}           % 2

\def\ctl       {\ensuremath{\cos{\theta_\ell}}\xspace}
\def\ctk       {\ensuremath{\cos{\theta_K}}\xspace}

\def\C#1      {\ensuremath{\mathcal{C}_{#1}}\xspace}                       % 9
\def\Cp#1     {\ensuremath{\mathcal{C}_{#1}^{'}}\xspace}                    % 7
\def\Ceff#1   {\ensuremath{\mathcal{C}_{#1}^{\mathrm{(eff)}}}\xspace}        % 9  
\def\Cpeff#1  {\ensuremath{\mathcal{C}_{#1}^{'\mathrm{(eff)}}}\xspace}       % 7
\def\Ope#1    {\ensuremath{\mathcal{O}_{#1}}\xspace}                       % 2
\def\Opep#1   {\ensuremath{\mathcal{O}_{#1}^{'}}\xspace}                    % 7

\newcommand{\tev}{\ifthenelse{\boolean{inbibliography}}{\ensuremath{~T\kern -0.05em eV}\xspace}{\ensuremath{\mathrm{\,Te\kern -0.1em V}}}\xspace}
\newcommand{\gev}{\ensuremath{\mathrm{\,Ge\kern -0.1em V}}\xspace}
\newcommand{\mev}{\ensuremath{\mathrm{\,Me\kern -0.1em V}}\xspace}
\newcommand{\kev}{\ensuremath{\mathrm{\,ke\kern -0.1em V}}\xspace}
\newcommand{\ev}{\ensuremath{\mathrm{\,e\kern -0.1em V}}\xspace}
\newcommand{\gevc}{\ensuremath{{\mathrm{\,Ge\kern -0.1em V\!/}c}}\xspace}
\newcommand{\mevc}{\ensuremath{{\mathrm{\,Me\kern -0.1em V\!/}c}}\xspace}
\newcommand{\gevcc}{\ensuremath{{\mathrm{\,Ge\kern -0.1em V\!/}c^2}}\xspace}
\newcommand{\gevgevcccc}{\ensuremath{{\mathrm{\,Ge\kern -0.1em V^2\!/}c^4}}\xspace}
\newcommand{\mevcc}{\ensuremath{{\mathrm{\,Me\kern -0.1em V\!/}c^2}}\xspace}

\def\mum  {\ensuremath{{\,\upmu\mathrm{m}}}\xspace}

\def\invfb   {\ensuremath{\mbox{\,fb}^{-1}}\xspace}

\def\deriv {\ensuremath{\mathrm{d}}}

\def\gsim{{~\raise.15em\hbox{$>$}\kern-.85em
          \lower.35em\hbox{$\sim$}~}\xspace}
\def\lsim{{~\raise.15em\hbox{$<$}\kern-.85em
          \lower.35em\hbox{$\sim$}~}\xspace}

\def\ptot       {\mbox{$p$}\xspace}
\def\pt         {\mbox{$p_{\mathrm{ T}}$}\xspace}

\def\evtgen     {\mbox{\textsc{EvtGen}}\xspace}

\def\geant      {\mbox{\textsc{Geant4}}\xspace}

\def\photos     {\mbox{\textsc{Photos}}\xspace}

\def\pythia     {\mbox{\textsc{Pythia}}\xspace}

\def\tell1  {TELL1\xspace}
\def\ukl1   {UKL1\xspace}

\def\invgevc{\ensuremath{\gev^{-1}c}\xspace}

\def\swave {S-wave\xspace}

\def\KstarENT{\ensuremath{\kaon^{*}(892)^{0}}\xspace}

\def\KstarFT{\ensuremath{\kaon_{0}^{*}(1430)^{0}}\xspace}

\newcommand{\mkpi}   {\ensuremath{m_{K\pi}}\xspace}
\newcommand{\mkpimm} {\ensuremath{m_{K\pi\mu\mu}}\xspace}

\def\btosll               {\ensuremath{\decay{\bquark}{\squark\,\ellp\ellm}}\xspace}

\def\BdToKstENTmm         {\ensuremath{{\decay{\Bd}{\KstarENT\mup\mun}}}\xspace}

\def\BdToKpimm            {\ensuremath{{\decay{\Bd}{\Kp\pim\mumu}}}\xspace}

\def\BsbToKstmm            {\ensuremath{{\decay{\Bsb}{\Kstarz\mumu}}}\xspace}

\def\BdToJPsiKstENT {\ensuremath{{\decay{\Bd}{\jpsi\KstarENT}}}\xspace}

\def\BsbToJPsiKst {\ensuremath{{\decay{\Bsb}{\jpsi\Kstarz}}}\xspace}
\def\BdToJPsiKstarz          {\ensuremath{{\decay{\Bd}{\jpsi\Kstarz}}}\xspace}

\def\BdToPsiTwoSKstarz        {\ensuremath{{\decay{\Bd}{\psitwos\Kstarz}}}\xspace}

\def\BdToJPsiMuMuKst  {\decay{\Bd}{\jpsi(\to\mu^+\mu^-)\Kstarz}}
\newcommand{\FS}[2]{\ensuremath{\left.F_{\rm S}\right|_{#1}^{#2}}\xspace}
\def\phih{\ensuremath{\phi'}\xspace}
\def\diff{\ensuremath{\textrm{d}}\xspace}
\def\dqsq{\ensuremath{\diff\qsq}\xspace}
\def\dmkpi{\ensuremath{\diff\mkpi}\xspace}

\newcommand{\thl}{\ensuremath{\theta_{\ell}}\xspace}
\newcommand{\thk}{\ensuremath{\theta_{K}}\xspace}

\newcommand{\ctksq}{\ensuremath{\cos^2\thk}\xspace}
\newcommand{\stlsq}{\ensuremath{\sin^2\thl}\xspace}
\newcommand{\stksq}{\ensuremath{\sin^2\thk}\xspace}
\newcommand{\cttl}{\ensuremath{\cos{2\thl}}\xspace}

\usepackage{cite} % Allows for ranges in citations
\usepackage{mciteplus}

\makeatletter
\DeclareRobustCommand*{\bfseries}{%
  \not@math@alphabet\bfseries\mathbf
  \fontseries\bfdefault\selectfont
  \boldmath
}
\makeatother

\usepackage{longtable} % only for template; not usually to be used in PAPERs

\begin{document}

\renewcommand{\thefootnote}{\fnsymbol{footnote}}
\setcounter{footnote}{1}

\begin{titlepage}
  \pagenumbering{roman}

% Header ---------------------------------------------------
  \vspace*{-1.5cm}
  \centerline{\large EUROPEAN ORGANIZATION FOR NUCLEAR RESEARCH (CERN)}
  \vspace*{1.5cm}
  \noindent
  \begin{tabular*}{\linewidth}{lc@{\extracolsep{\fill}}r@{\extracolsep{0pt}}}
    \ifthenelse{\boolean{pdflatex}}% Logo format choice
    {\vspace*{-2.7cm}\mbox{\!\!\!\includegraphics[width=.14\textwidth]{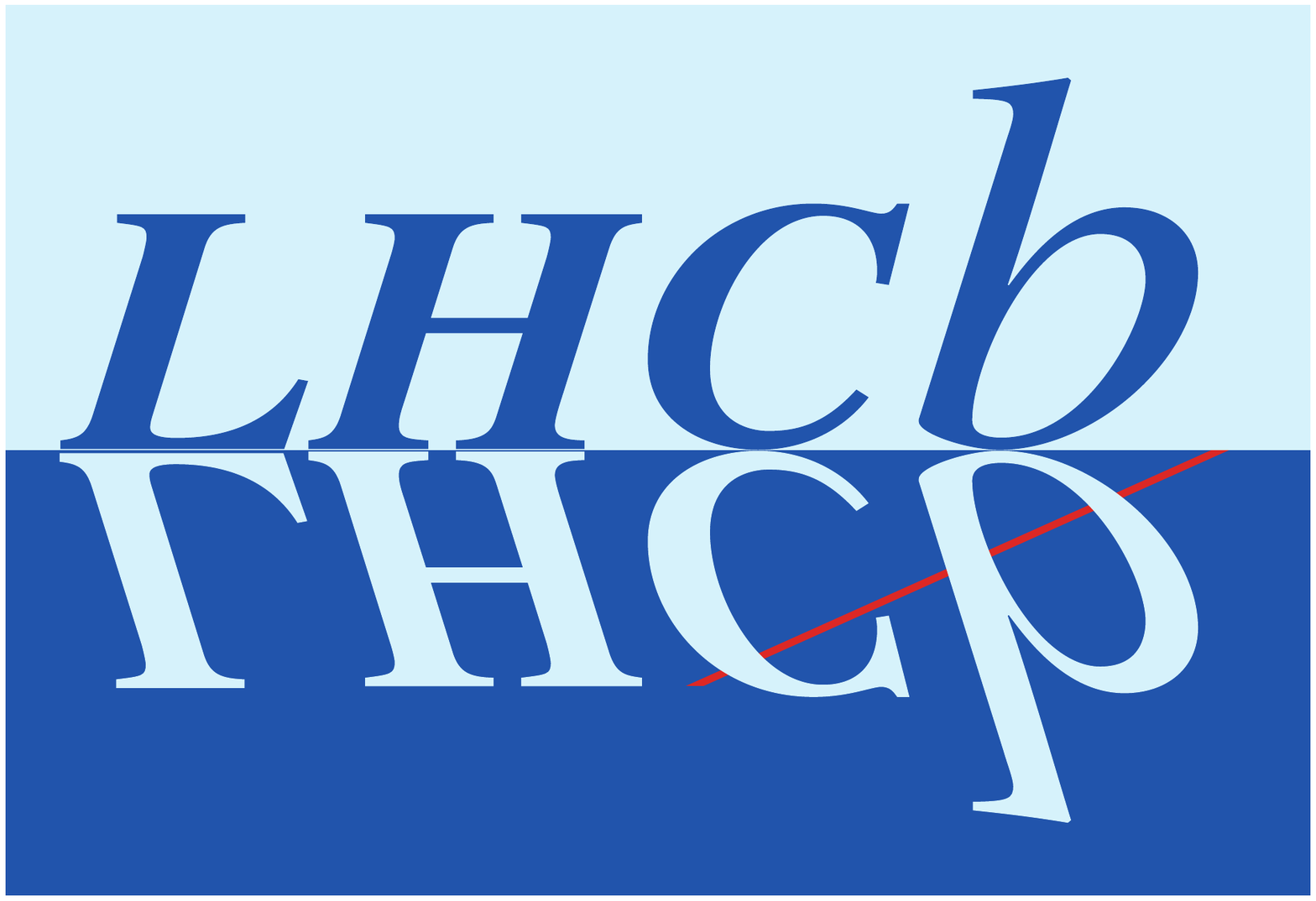}} & &}%
    {\vspace*{-1.2cm}\mbox{\!\!\!\includegraphics[width=.12\textwidth]{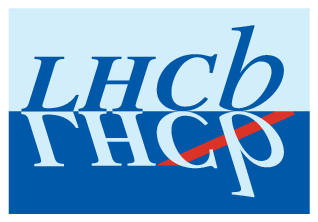}} & &}%
    \\
    & & CERN-EP-2016-141 \\  % ID 
    & & LHCb-PAPER-2016-012 \\  % ID 
    & & 15 June 2016\\ % Date - Can also hardwire e.g.: 23 March 2010
    & & \\
% not in paper \hline
  \end{tabular*}

  %\vspace*{4.0cm}
  \vspace*{0.9cm}

% Title --------------------------------------------------
  {\normalfont\bfseries\boldmath\LARGE
    \begin{center}
      Measurements of the S-wave fraction in $B^{0}\rightarrow K^{+}\pi^{-}\mu^{+}\mu^{-}$ decays and
      the $B^{0}\rightarrow K^{\ast}(892)^{0}\mu^{+}\mu^{-}$ differential branching fraction
    \end{center}
  }

  \vspace*{0.90cm}

  \begin{center}
    The LHCb collaboration\footnote{Authors are listed at the end of this paper.}
  \end{center}

  \vspace{\fill}

% Abstract -----------------------------------------------
  \begin{abstract}
    \noindent
    A measurement of the differential branching fraction of the decay
    ${B^{0}\rightarrow K^{\ast}(892)^{0}\mu^{+}\mu^{-}}$ is presented together with a determination of the
    S-wave fraction of the $K^+\pi^-$ system in the decay $B^{0}\rightarrow K^{+}\pi^{-}\mu^{+}\mu^{-}$.
    The analysis is based on $pp$-collision
    data corresponding to an integrated luminosity of 3\,fb$^{-1}$ collected
    with the LHCb experiment. 
    The measurements are made in bins of the invariant mass squared of
    the dimuon system, $q^2$. Precise theoretical predictions for the
    differential branching fraction of  $B^{0}\rightarrow K^{\ast}(892)^{0}\mu^{+}\mu^{-}$ decays are
    available for the  $q^2$ region \mbox{$1.1<q^2<6.0\,{\rm GeV}^2/c^4$}.
    In this $q^2$ region, for the $K^+\pi^-$ invariant mass
    range \mbox{$796 < m_{K\pi} < 996\,{\rm MeV}/c^2$}, the S-wave fraction of the
    $K^+\pi^-$ system 
    in $B^{0}\rightarrow K^{+}\pi^{-}\mu^{+}\mu^{-}$ decays is found to be
    \begin{equation*}
      F_{\rm S} = 0.101\pm0.017({\rm stat})\pm0.009 ({\rm syst}),
    \end{equation*}
    and the differential branching fraction of $B^{0}\rightarrow K^{\ast}(892)^{0}\mu^{+}\mu^{-}$ decays is
    determined to be 
    \begin{equation*}
      {\rm d}\mathcal{B}/{\rm d} q^2 = (0.342_{\,-0.017}^{\,+0.017}({\rm
      stat})\pm{0.009}({\rm syst})\pm0.023({\rm norm}))\times 10^{-7}c^{4}/{\rm GeV}^{2}.
    \end{equation*} 
    The differential branching fraction measurements presented are the
    most precise to date and are found to be in agreement with Standard Model predictions.
  \end{abstract}

  \vspace*{0.3cm}

  \begin{center}
    Published in JHEP {\bf 11} (2016) 047
  \end{center}

  \vspace{\fill}

  {\footnotesize 
    \centerline{\copyright~CERN on behalf of the LHCb collaboration, licence \href{http://creativecommons.org/licenses/by/4.0/}{CC-BY-4.0}.}}
    \vspace*{2mm}

  \end{titlepage}

%%%%%%%%%%%%%%%%%%%%%%%%%%%%%%%%
%%%%%  EOD OF TITLE PAGE  %%%%%%
%%%%%%%%%%%%%%%%%%%%%%%%%%%%%%%%

%  empty page follows the title page ----
  \newpage
  \setcounter{page}{2}
  \mbox{~}

  \cleardoublepage

\renewcommand{\thefootnote}{\arabic{footnote}}
\setcounter{footnote}{0}

\pagestyle{plain} % restore page numbers for the main text
\setcounter{page}{1}
\pagenumbering{arabic}

\section{Introduction}
\label{sec:Introduction}

The decay $\decay{\Bd}{\Kstarz\mumu}$  proceeds via a \btosll flavour-changing neutral-current transition. In the Standard Model (SM), this transition is forbidden at tree level and must therefore occur via a loop-level process.  Extensions to the SM predict new particles that can contribute to the \btosll process and affect the rate and angular distribution of the decay. Recently, global analyses of measurements involving \btosll processes have reported significant deviations from SM predictions~\cite{Hurth:2016fbr,QuimLatest,Descotes-Genon:2013wba,Altmannshofer:2013foa,Beaujean:2013soa,Hurth:2013ssa,Jager:2012uw,Descotes-Genon:2014uoa,Lyon:2014hpa,Altmannshofer:2014cfa,Crivellin:2015mga,Gauld:2013qja,AltAndStraubLatest,Mahmoudi:2014mja,Datta:2013kja}. These deviations could be explained either by new particles~\cite{Descotes-Genon:2013wba,Altmannshofer:2013foa,Altmannshofer:2014cfa,Crivellin:2015mga,Mahmoudi:2014mja,Datta:2013kja,Buttazzo:2016kid} or by unexpectedly large hadronic effects~\cite{Ciuchini:2015qxb,AltAndStraubLatest,Lyon:2014hpa}.

In this paper, the symbol $\Kstarz$ denotes any neutral strange meson in an excited state that decays to a $K^+$ and a $\pi^-$.\footnote{Inclusion of charge conjugate processes is implied throughout this paper unless otherwise noted.}
For invariant masses of the $K^+\pi^-$ system in the range considered in this analysis, the $\Kstarz$ decay products are predominantly found in a P- or S-wave state. The fractional size of the scalar (S-wave) component of the $\Kp\pim$ {\mbox{system ($F_{\rm S}$)}} depends on the squared invariant mass of the dimuon system ($\qsq$). This dependence is expected to be similar to that of the longitudinal polarisation fraction ($F_{\rm L}$) of the $\KstarENT$ meson~\cite{Doring:2013wka,WeiWang,Becirevic:2012dp}. 

The S-wave fraction is predicted to be maximal in the $\qsq$ range \mbox{$1.0<\qsq<6.0\gevgevcccc$}~\cite{Doring:2013wka,WeiWang,Becirevic:2012dp}. A previous analysis by the LHCb collaboration set an upper limit of $F_{\rm S}<0.07$
at 68\% confidence level for invariant masses of the $K^+\pi^-$ system in the range \mbox{$792<\mkpi<992\mevcc$}~\cite{LHCb-PAPER-2013-019}. The measurement was performed by exploiting the phase shift of the $\KstarENT$ Breit--Wigner function around the corresponding pole mass.

In all previous determinations of the differential branching fraction of \mbox{$\decay{\Bd}{\KstarENT\mumu}$} decays~\cite{LHCb-PAPER-2013-019,LHCb-PAPER-2014-006,CMSKstmm,Wei:2009zv,Lees:2012tva}, the $\KstarENT$ was selected by requiring a window of size 80--380\mevcc around the known \KstarENT mass, but no correction was made for the scalar fraction. This fraction was assumed to be small and was treated as a systematic uncertainty. The measurements of the differential branching fraction of \mbox{$\decay{\Bd}{\KstarENT\mumu}$} decays are included in global analyses of \btosll processes. As these analyses make use of theory predictions which are made purely for the resonant P-wave part of the $K^+\pi^-$ system, an accurate assessment of the S-wave component in $\BdToKstmm$ decays is critical.

In this paper, the first measurement of $F_{\rm S}$ in $\decay{\Bd}{\Kstarz\mumu}$ decays is presented. The measurement is performed through a fit to the kaon helicity angle~\cite{LHCb-PAPER-2013-019,Altmannshofer:2008dz}, $\thk$,
and the \mkpi spectrum, in the range $644<\mkpi<1200\mevcc$. Motivated by previous estimates of the S-wave fraction~\cite{LHCb-PAPER-2013-019,Doring:2013wka,WeiWang,Becirevic:2012dp}, 
$F_{\rm S}$ is also determined in a narrower window of  \mbox{$796<\mkpi<996\mevcc$}. The values of $F_{\rm S}$  are reported in eight bins of $\qsq$ of approximately $2\gevgevcccc$ width, and in two larger bins $1.1<\qsq<6.0\gevgevcccc$ and $15.0<\qsq<19.0\gevgevcccc$. The choice of $\qsq$ bins is identical to that of Ref.~\cite{LHCb-PAPER-2015-051}. 

The measurements of $F_{\rm S}$ allow the determination of the differential branching fraction of the \BdToKstENTmm decay. 
The differential branching fraction is determined by normalising the $\BdToKstENTmm$ yield in each $\qsq$ bin to the total event yield of the $\BdToJPsiKstarz$ control channel, where the $J/\psi\to\mu^+\mu^-$ decay mode is used. The measurements are made using a $pp$-collision data sample recorded by the LHCb experiment in Run 1, corresponding to an integrated luminosity of 3\invfb. These data were collected at centre-of-mass energies of $7$ and $8\tev$ during 2011 and 2012 respectively. The differential branching fraction measurement is complementary to the angular analysis presented in Ref.~\cite{LHCb-PAPER-2015-051}, and supersedes that of Ref.~\cite{LHCb-PAPER-2013-019}. The latter analysis was performed on a 1\invfb subset of the Run 1 data sample.

This paper is organised as follows. Section~\ref{sec:model} describes the angular and \mkpi distributions of \BdToKstmm decays with the $K^+\pi^-$ system in a P- or S-wave state. Section~\ref{sec:detector} describes the \lhcb detector and the procedure used to generate simulated data. The reconstruction and selection of \BdToKpimm candidates are described in Sec.~\ref{sec:selection}. Section~\ref{sec:massdistr} describes the parameterisation of the mass distributions and Sec.~\ref{sec:fsmeasurement} describes the determination of $F_{\rm S}$, including the method used to correct for the detection and selection biases. The measurement of the differential branching fraction of  $\mbox{\decay{\Bd}{\KstarENT\mumu}}$ decays is presented in Sec.~\ref{sec:bfmeas}. The systematic uncertainties affecting the measurements are discussed in Sec.~\ref{sec:systs}. Finally, the conclusions are presented in Sec.~\ref{sec:conclusions}.

\section{The angular distribution and $F_{\rm S}$}
\label{sec:model}

The final state of the \BdToKstmm decay is completely described by
$\qsq$, and the three decay angles, $\vec\Omega \equiv
(\ctk,\,\ctl,\,\phi)$~\cite{LHCb-PAPER-2013-019}. The angle between the $\mu^+$
($\mu^-$) and the direction opposite to that of the $\Bz$ ($\Bzb$)
meson in the rest frame of the dimuon system is denoted
by $\theta_{\ell}$. The angle between the direction of the $K^+$ ($K^-$)
and the $\Bz$ ($\Bzb$) meson in the rest frame of the $\Kstarz$
($\Kstarzb$) is denoted by $\theta_{K}$. The angle between the plane
defined by the dimuon pair and the plane defined by the kaon and pion
in the $\Bz$ ($\Bzb$) rest frame is denoted by $\phi$. 

In the limit that the dimuon mass is large compared to the mass of the
muons (\mbox{$\qsq\gg4m_{\mu}^{2}$}), this choice of the angular basis allows
the differential decay rates of \mbox{\decay{\Bz}{\Kstarz\mumu} and
\decay{\Bzb}{\Kstarzb\mumu}} decays to be written as
\begin{equation}
\begin{split}
\!\!\!\!\!\!\!\frac{\deriv^{5}(\Gamma+\overline{\Gamma})}{\deriv\mkpi\deriv\qsq\,\deriv\vec\Omega}
= \frac{9}{32\pi}\left[ \frac{}{} \right.&\!\!\!\!\left.\frac{}{} {(I_{1}^{s}+\bar{I}_{1}^{s})}\sin^{2} \theta_{K}(1+3\cos 2\theta_{\ell}) + {(I_{1}^{c}+\bar{I}_{1}^{c})}\cos^{2}\theta_{K}(1-\cos 2\theta_{\ell}) ~+\right. \\ 
&\!\!\!\!\left. \frac{}{} {(I_{3}+\bar{I}_{3})}  \sin^{2}\theta_{K}\sin^{2} \theta_{\ell} \cos 2\phi+{{(I_{4}+\bar{I}_{4}) \sin 2\theta_{K} \sin 2\theta_{\ell} \cos\phi}} ~+\right.\\
&\!\!\!\!\frac{}{} \left. {{{(I_{5}+\bar{I}_{5})} \sin 2\theta_{K}\sin\theta_{\ell}\cos\phi}} + (I_{6s}+\bar{I}_{6s})\sin^{2}\theta_{K}  \cos\theta_{\ell} ~+\right. \\ 
&\!\!\!\!\frac{}{} \left. {{{(I_{7}+\bar{I}_{7})}  \sin 2\theta_{K} \sin\theta_{\ell} \sin\phi}} +{{{(I_{8}+\bar{I}_{8})} \sin 2\theta_{K} \sin 2\theta_{\ell}\sin\phi}} ~+\right. \\ 
&\!\!\!\!\frac{}{} \left. (I_{9}+\bar{I}_{9}) \sin^{2}\theta_{K}\sin^{2}\theta_{\ell} \sin 2\phi+(I_{10}+\bar{I}_{10})(1-\cos2\theta_{\ell})~+\right. \\
&\!\!\!\!\frac{}{} \left. (I_{11}+\bar{I}_{11})\cos\theta_{K}(1-\cos2\theta_{\ell})~+\right. \\ 
&\!\!\!\!\frac{}{} \left. (I_{14}+\bar{I}_{14})\sin\theta_{K}\sin2\theta_{\ell}\cos\phi+(I_{15}+\bar{I}_{15})\sin\theta_{K}\sin\theta_{\ell}\cos\phi~+\right.\\ 
&\!\!\!\!\frac{}{}  \left. (I_{16}+\bar{I}_{16})\sin\theta_{K}\sin\theta_{\ell}\sin\phi+(I_{17}+\bar{I}_{17})\sin\theta_{K}\sin2\theta_{\ell}\sin\phi\frac{}{}\right],\\
\end{split}
\label{eq:fullangular}
\end{equation}
\noindent where $\Gamma$ and $\overline{\Gamma}$ denote the decay rates of the $\Bz$ and
$\Bzb$ respectively. The 15 coefficients $I_j$ ($\bar{I}_j$) are
bilinear combinations of the $\Kstarz$ ($\Kstarzb$) decay amplitudes
and vary with $\qsq$ and $\mkpi$. The numbering of the coefficients
follows the convention used in
Ref.~\cite{LHCb-PAPER-2015-051}. Coefficients $I_j$ with $j\leq9$ involve P-wave
amplitudes only, coefficient $I_{10}$ involves S-wave amplitudes only
and coefficients with $11\leq j \leq 17$ describe the interference between P- and S-wave amplitudes~\cite{Hofer:2015kka}.

The polarity of the LHCb dipole magnet, discussed in
Sec.~\ref{sec:detector}, is reversed periodically. Coupled with the
fact that $\Bz$ and $\Bzb$ decays are studied simultaneously, this
results in a symmetric detection efficiency in $\phi$. Therefore, the
angular distribution is simplified by performing a
transformation of the $\phi$ angle such that
\begin{equation}
\phih=
\begin{cases} \phi + \pi & \text{~if~} \phi < 0
\\
\phi & \text{~otherwise},
\end{cases}
\label{eq:folding}
\end{equation}
which results in the cancellation of terms in Eq.~\ref{eq:fullangular} that have a
$\sin\phi$ or $\cos\phi$ dependence.  

The remaining $I_j$ and $\bar{I}_j$ coefficients can be written in terms of the decay
amplitudes given in Ref.~\cite{LHCb-PAPER-2015-051}. Defining
$\vec\Omega' \equiv (\ctk,\,\ctl,\,\phih)$, the resulting differential
decay rate has the form
\begin{equation}\label{eqn:dGGGGG}
  \begin{split}
    \frac{\deriv^{5}(\Gamma+\overline{\Gamma})}{\deriv\mkpi\deriv\qsq\,\deriv\vec\Omega'}  =&\, 
    \frac1{4\pi}G_{\rm S}\left|f_{\rm LASS}(\mkpi)\right|^2  (1-\cttl)~+\\
    &\frac3{4\pi}G_{\rm P}^{0}\left|f_{\rm BW}(\mkpi)\right|^2\ctksq(1-\cttl)~+\\
    &\frac{\sqrt3}{2\pi}{\rm Re}\left[\left(G_{\rm SP}^{\rm Re} + iG_{\rm SP}^{\rm Im}\right) f_{\rm LASS}(\mkpi)f^\ast_{\rm BW}(\mkpi)\right]\ctk(1-\cttl)+~\\
    &\frac9{16\pi}G_{\rm P}^{\perp\parallel}\left|f_{\rm BW}(\mkpi)\right|^2\stksq\left(1+\frac13\cttl\right)+~\\
    &\frac3{8\pi}S_{3}(G_{\rm P}^{0}+G_{\rm P}^{\perp\parallel})\left|f_{\rm BW}(\mkpi)\right|^2 \stksq\stlsq\cos{2\phih}+~\\
    &\frac3{2\pi}A_{\rm FB}(G_{\rm P}^{0}+G_{\rm P}^{\perp\parallel})\left|f_{\rm BW}(\mkpi)\right|^2 \stksq\ctl+~\\
    &\frac3{4\pi}S_9 (G_{\rm P}^{0}+G_{\rm P}^{\perp\parallel})\left|f_{\rm BW}(\mkpi)\right|^2 \stksq\stlsq\sin{2\phih},
  \end{split} \end{equation} where $f_{\rm BW}(\mkpi) $ denotes the
$\mkpi$ dependence of the resonant P-wave component, which is modelled
using a relativistic Breit--Wigner function. The S-wave component is
modelled using the LASS parameterisation~\cite{lass},
$f_{\rm LASS}(\mkpi)$. The exact definitions of the P- and S-wave
line shapes are given in Appendix~\ref{app:mkpi}. 
The real-valued coefficients $G_{\rm S}^{\phantom R}$, $G_{\rm SP}^{\rm Re}$,
$G_{\rm SP}^{\rm Im}$, $G_{\rm P}^{0}$ and $G_{\rm
  P}^{\perp\parallel}$ are bilinear combinations of the \qsq-dependent
parts of the $\Kstarz$  ($\Kstarzb$) helicity amplitudes
$A_{i}^{L,R}(\qsq)$ ($\overline{A}_{i}^{L,R}(\qsq)$) and are given by
\begin{align}\label{eqn:thegs}
    G_{\rm S}^{\phantom R}=&\,|A_{S}^{L}(\qsq)|^2+|A_{S}^{R}(\qsq)|^2 +|\overline{A}_{S}^{L}(\qsq)|^2+|\overline{A}_{S}^{R}(\qsq)|^2,\nonumber\\[5pt]
   G_{\rm SP}^{\rm Re} + iG_{\rm SP}^{\rm Im}=&\,A_{S}^{L}A_{0}^{L\ast}+A_{S}^{R}A_{0}^{R\ast}+\overline{A}_{S}^{L}\overline{A}_{0}^{L\ast}+\overline{A}_{S}^{R}\overline{A}_{0}^{R\ast},\nonumber\\[5pt]
    G_{\rm P}^{0}=&\,|A_{0}^{L}(\qsq)|^2+|A_{0}^{R}(\qsq)|^2 +|\overline{A}_{0}^{L}(\qsq)|^2+|\overline{A}_{0}^{R}(\qsq)|^2,\\[5pt]
  G_{\rm P}^{\perp\parallel}=&\displaystyle\sum\limits_{i=\perp,\parallel}^{} |A_{i}^{L}(\qsq)|^2+|A_{i}^{R}(\qsq)|^2 +|\overline{A}_{i}^{L}(\qsq)|^2+|\overline{A}_{i}^{R}(\qsq)|^2,\nonumber
\end{align}
where $L$ and $R$ denote the (left- and right-handed) chiralities of
the dimuon system. These coefficients are determined through the extended
maximum likelihood fit described in Sec.~\ref{sec:fsfit}. The
coefficients $S_{3}$, $A_{\rm FB}$ and $S_9$ are \CP-averaged
observables that are defined in Ref.~\cite{LHCb-PAPER-2015-051}.  The
integral of Eq.~\ref{eqn:dGGGGG} with respect to $\ctl$ and
$\phih$ is independent of these observables.  However, detection
effects that are either asymmetric or non-uniform in $\ctl$ and $\phih$ introduce a
residual dependence on these observables.  In this analysis, $S_{3}$, $A_{\rm FB}$ and $S_9$ are set
to their measured values~\cite{LHCb-PAPER-2015-051}. The
systematic uncertainty associated with this choice is negligible.

Using the definitions of Eq.~\ref{eqn:thegs}, the S-wave fraction
$F_{\rm S}$ in the range \mbox{$a<\mkpi<b$} can be determined from the
coefficients $G_{\rm S}^{\phantom R}$ and $G_{\rm P}^{0, \perp\parallel}$, through
\begin{equation}\label{eqn:FSdefinition}
  F_{\rm S}|_{a}^{b} = \frac{ G_{\rm S}^{\phantom R}\int_a^b{\dmkpi\left|f_{\rm LASS}(\mkpi)\right|^2}}
  {G_{\rm S}\int_a^b \dmkpi\left|f_{\rm LASS}(\mkpi)\right|^2+\left(G_{\rm P}^{0}+G_{\rm P}^{\perp\parallel}\right)\int_a^b \dmkpi\left|f_{\rm BW}(\mkpi)\right|^2}.
\end{equation}

\section{Detector and simulation}
\label{sec:detector}

The \lhcb detector~\cite{Alves:2008zz,LHCb-DP-2014-001} is a single-arm forward
spectrometer covering the \mbox{pseudorapidity} range $2<\eta <5$,
designed for the study of particles containing \bquark or \cquark quarks.
The detector includes a high-precision tracking system divided
into three sub-systems: a silicon-strip vertex detector 
surrounding the $pp$ interaction region, a large-area silicon-strip detector that is located
upstream of a dipole magnet with a bending power of about
$4{\mathrm{\,Tm}}$, and three stations of silicon-strip detectors and straw
drift tubes situated downstream of the magnet.
%Firstly, a silicon-strip vertex detector 
%surrounds the $pp$ interaction region, %~\cite{LHCb-DP-2014-001},
%and is followed by a large-area silicon-strip detector that is located
%upstream of a dipole magnet with a bending power of about
%$4{\mathrm{\,Tm}}$. Finally, three stations of silicon-strip detectors and straw
%drift tubes %~\cite{LHCb-DP-2013-003} 
%are situated downstream of the magnet.
The tracking system provides a measurement of the momentum, \ptot, of charged particles with
a relative uncertainty that varies from 0.5\% at low momentum to 1.0\% at 200\gevc.
The minimum distance of a track to a \mbox{primary vertex (PV)}, the impact parameter, is measured with a resolution of $(15+29/\pt)\mum$,
where \pt is the component of the momentum transverse to the beam, in\,\gevc.
Different types of charged hadrons are distinguished using information
from two ring-imaging Cherenkov {\mbox{(RICH) detectors}}. %~\cite{LHCb-DP-2012-003}. 
Photons, electrons and hadrons are identified by a calorimeter system consisting of
scintillating-pad and preshower detectors, an electromagnetic
calorimeter and a hadronic calorimeter. Muons are identified by a
system composed of alternating layers of iron and multiwire
proportional chambers. % ~\cite{LHCb-DP-2012-002}.
The online event selection is performed by a trigger~\cite{LHCb-DP-2012-004}, 
which consists of a hardware stage, based on information from the calorimeter and muon
systems, followed by a software stage, which applies a full event
reconstruction.

A large sample of simulated events is used to determine the
effect of the detector geometry, trigger, and the selection criteria
on the angular distribution of the signal, and to determine the ratio
of efficiencies between the signal and the $\BdToJPsiKstarz$ normalisation mode. 
In the simulation, $pp$ collisions are generated using
\pythia~\cite{Sjostrand:2006za,*Sjostrand:2007gs} with a specific
\lhcb configuration~\cite{LHCb-PROC-2010-056}.  The decay of the \Bd
meson is described by \evtgen~\cite{Lange:2001uf}, which generates
final-state radiation using
\photos~\cite{Golonka:2005pn}.  As described in
Ref.~\cite{LHCb-PROC-2011-006}, the \geant
toolkit~\cite{Allison:2006ve, *Agostinelli:2002hh} is used to
implement the interaction of the generated particles with the detector
and the detector response. Data-driven corrections
are applied to the simulation following the procedure of
Ref.~\cite{LHCb-PAPER-2015-051}. These corrections account for the small
level of mismodelling of the detector occupancy, the $\Bz$ momentum and
vertex quality, and the particle identification (PID)
performance.

\section{Selection of signal candidates}
\label{sec:selection}

The $\decay{\Bd}{\Kstarz\mumu}$ signal candidates are first required
to pass the hardware trigger, which selects events containing at least
one muon with transverse momentum $\pt>1.48\gevc$ in the 7\tev data or
$\pt>1.76\gevc$ in the 8\tev data.  In the subsequent software
trigger, at least one of the final-state particles is
required to have $\pt>1.7\gevc$ in the 7\tev data or
$\pt>1.6\gevc$ in the 8\tev data, unless the particle is identified
as a muon in which case $\pt>1.0\gevc$ is required. The final-state
particles that satisfy these transverse momentum criteria are also
required to have an impact parameter larger than $100\mum$ with
respect to all PVs in the event.
Finally, the tracks of two or more of the final-state
particles are required to form a vertex that is significantly
displaced from the PVs.

Signal candidates are formed from a pair of oppositely charged tracks that are identified as muons, combined with a $\Kstarz$ meson candidate. The $\Kstarz$ candidate is formed from two oppositely charged tracks that are identified as a kaon and a pion. These signal candidates are required to pass a set of loose preselection requirements, which are identical to those described in Ref.~\cite{LHCb-PAPER-2015-051}, with the exception that the $\Kstarz$ candidate is required to have an invariant mass in the wider $644<\mkpi<1200\mevcc$ range.  The preselection requirements exploit the decay topology of \BdToKstmm transitions and restrict the data sample to candidates with good quality vertex and track fits. Candidates are required to have a reconstructed $\Bz$ invariant mass ($\mkpimm$) in the range $5170<\mkpimm<5780\mevcc$. 

The backgrounds formed by combining particles from different $b$- and $c$-hadron decays are referred to as combinatorial. Such backgrounds are suppressed with the use of a Boosted Decision Tree~(BDT)~\cite{Breiman,AdaBoost}. The BDT used for the present analysis is identical to that described in Ref.~\cite{LHCb-PAPER-2015-051} and the same working point is used. The BDT selection has a signal efficiency of 90\% while removing 95\% of the combinatorial background surviving the preselection. The efficiency of the BDT is uniform with respect to \mkpimm in the above mass range.

Specific background processes can mimic the signal if their final states are misidentified or misreconstructed.  The requirements of Ref.~\cite{LHCb-PAPER-2015-051} are reassessed and found to reduce the sum of all backgrounds from such decay processes to a level of less than 2\% of the expected signal yield.  The only requirement that is modified in the present analysis is that responsible for removing genuine \BdToKstmm decays, where the track of the genuine pion is reconstructed with the kaon hypothesis and vice versa.  These misidentified signal candidates occur more often in the wider $\mkpi$ window used for the present analysis, and are reduced by tightening the requirements made on the kaon and pion PID information provided by the RICH detectors. After the application of all the selection criteria, this specific background process is reduced to less than 1\% of the level of the signal.

\section{The \texorpdfstring{$K^+\pi^-\mu^+\mu^-$}{K+pi-mu+mu-} and
  \texorpdfstring{$K^+\pi^-$}{K+pi-} mass distributions}
\label{sec:massdistr}

\begin{figure}
 \centering
 \includegraphics[width=0.495\textwidth]{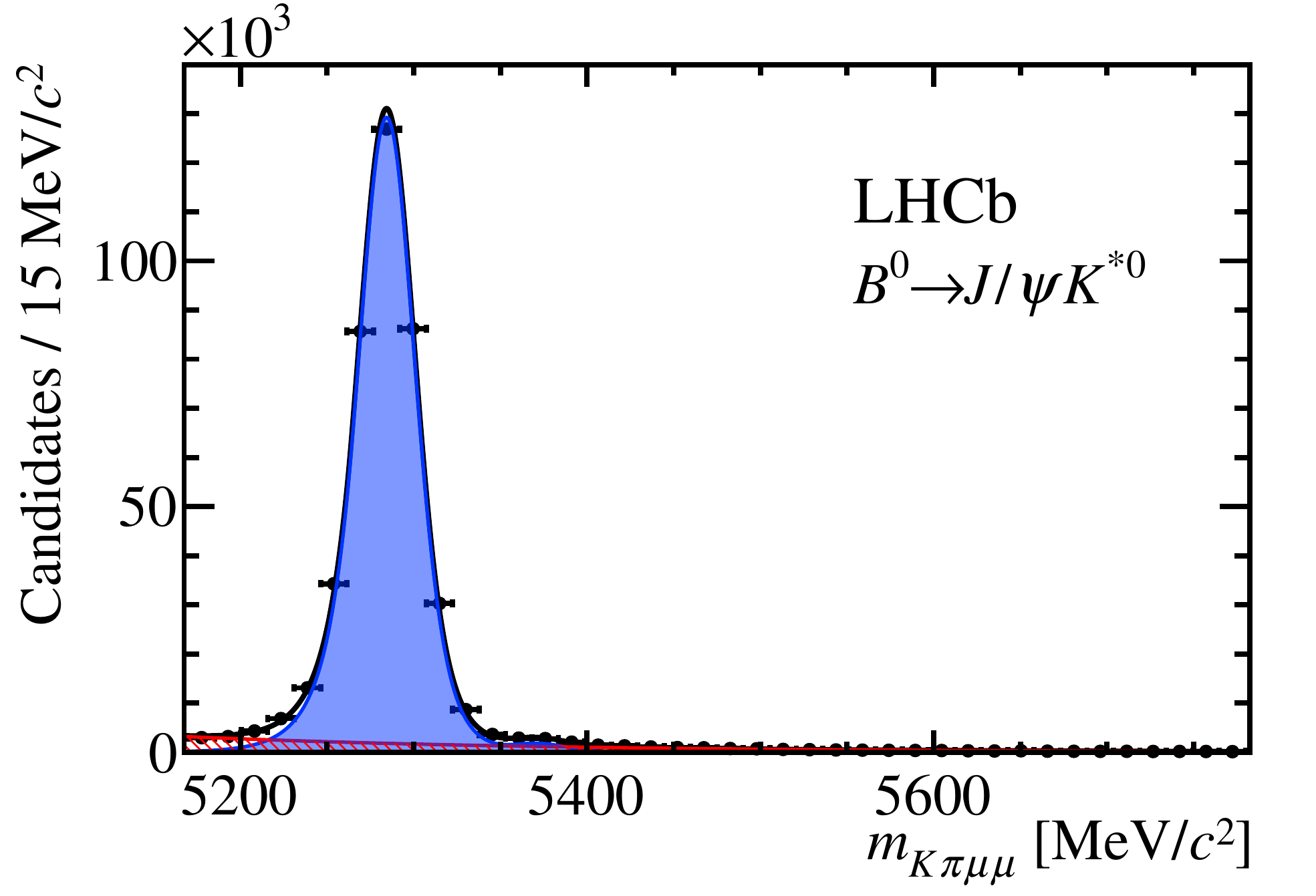}
 \includegraphics[width=0.495\textwidth]{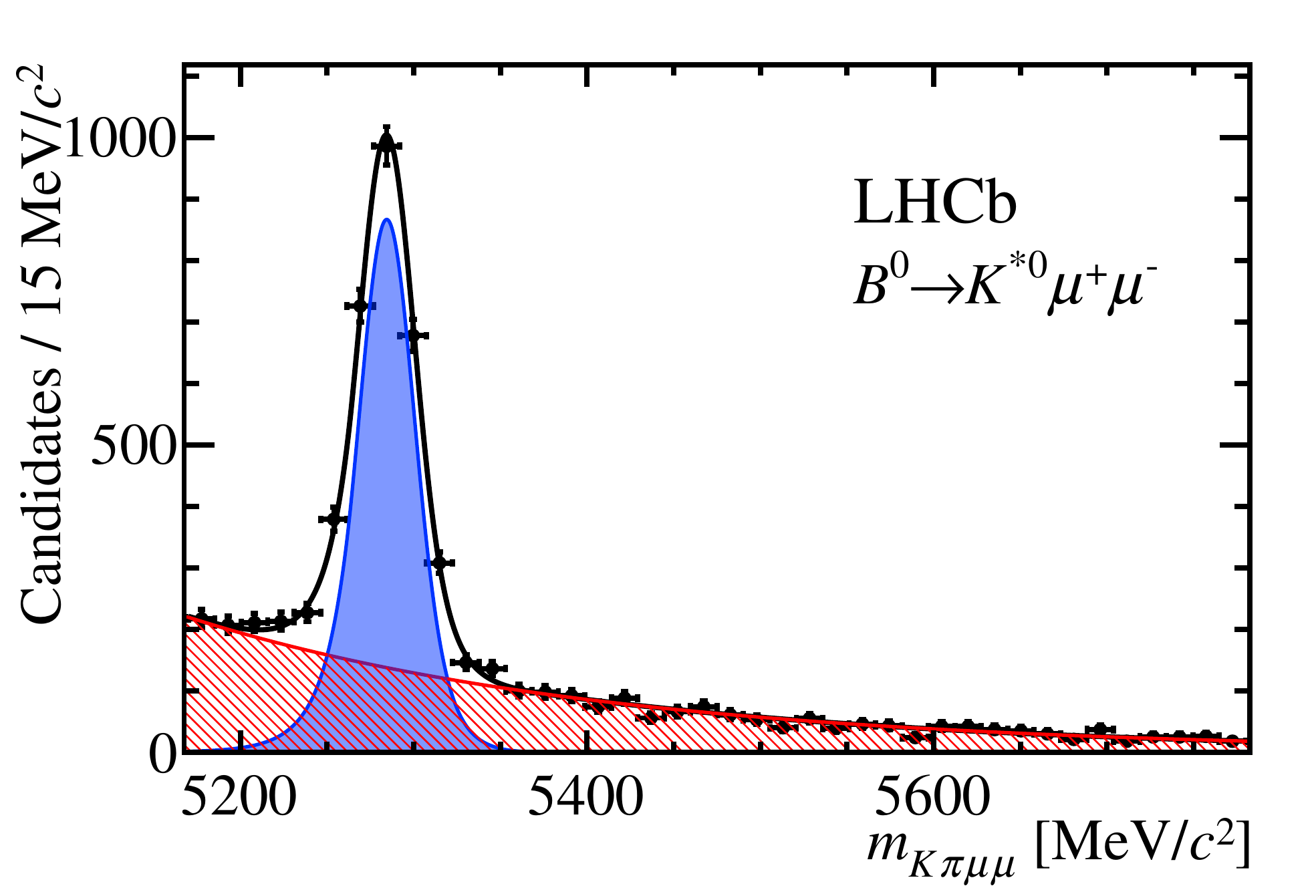}
 \caption{
   Invariant mass $\mkpimm$ of 
   (left) the $\BdToJPsiKst$ decay and (right) the signal decay $\BdToKstmm$ integrated
   over the $\qsq$ regions described in the text. The individual signal (blue shaded area) and
   background (red hatched area) components are shown. The solid line denotes
   the total fitted distribution. \label{fig:mkpimmfit}
 }
\end{figure}

The $K^+\pi^-\mu^+\mu^-$ invariant mass is used to discriminate
between signal and background. The distribution of the signal
candidates is modelled using the sum of two Gaussian functions with a
common mean, each with a power law tail on the lower side. The
parameters describing this model are determined from fits to
$\BdToJPsiKst$ data in a $\qsq$ range $9.22<\qsq<9.96\gevgevcccc$ and 
with an $\mkpi$ range of $644<\mkpi<1200\mevcc$, shown in
Fig.~\ref{fig:mkpimmfit}. 
These parameters are fixed for the subsequent fits to the $\BdToKstmm$
candidates in the same $\mkpi$ range. In samples of simulated
$\BdToKstmm$ decays, the $\mkpimm$ resolution is observed 
to differ from that in $\BdToJPsiKst$ decays by 2 to 8\% depending on $\qsq$.
A correction factor is therefore derived from
the simulation and is applied to the widths of the Gaussian functions
in the different $\qsq$ bins.
%A correction factor derived from
%simulated events is applied to the widths of the Gaussian functions in order to account for the
%2--8\% variation of the mass resolution with $\qsq$. 
In the fits to $\BdToJPsiKst$ decays, an
additional component is included to account for the $\BsbToJPsiKst$
process. The size of this additional component is taken to be
0.8\% of the $\BdToJPsiKst$ signal~\cite{LHCb-PAPER-2012-014}. 
The fit to the $\BdToJPsiKst$ mode gives $389\,577\pm649$ decays.
In the fits to $\BdToKstmm$ decays, the $\BsbToKstmm$ contribution is
neglected. The systematic uncertainty related to ignoring
this background process is negligible. For both $\BdToJPsiKst$ and $\BdToKstmm$
decays, the combinatorial background in the $K^+\pi^-\mu^+\mu^-$
invariant mass spectrum is described by an exponential function. 
The $\BdToKstmm$ yield integrated over the $\qsq$ ranges
$0.1<\qsq<8.0\gevgevcccc$, $11.0<\qsq<12.5\gevgevcccc$ and
$15.0<\qsq<19.0\gevgevcccc$ is determined to be $2593\pm60$.
The $\qsq$ regions $8.0<\qsq<11.0\gevgevcccc$ and $12.5<\qsq<15.0\gevgevcccc$
are dominated by the contributions from $\BdToJPsiKst$ and $\BdToPsiTwoSKstarz$ decays respectively
and are therefore excluded in the fits to the signal $\BdToKstmm$ decays.

As discussed in Sec.~\ref{sec:model}, the $K^+\pi^-$ invariant mass
distribution of the signal candidates is modelled with two
distributions. A relativistic Breit--Wigner function is used for the
P-wave component and the LASS parameterisation for the S-wave
component. The parameters of these functions are fixed to the values
determined in $\BdToJPsiKst$ decays using the model described in 
Ref.~\cite{LHCb-PAPER-2014-014}. A systematic uncertainty is assigned
for this choice.

The $K^+\pi^-$ invariant mass distribution of the combinatorial
background is modelled using an empirical threshold function of the form
\begin{equation}
f_{\rm bkg}(\mkpi)=(\mkpi-m_{\rm thr})^{1/\alpha},
\end{equation}
where $m_{\rm thr}=634\mevcc$ is given by the sum of the pion and kaon masses~\cite{PDG2014}, and
$\alpha$ is a parameter determined from fits to the data. This model has
been validated on data from the upper $\mkpimm$ sideband, defined as
$5350<\mkpimm<5780\mevcc$, where no resonant structure in the $\mkpi$
spectrum is observed.

\section{Determination of the S-wave fraction} 
\label{sec:fsmeasurement}

\subsection{Efficiency correction}
\label{sec:efficiency}
The trigger, selection, and detector geometry bias the
distributions of the decay angles $\ctk$, $\ctl$, $\phih$, as well as
the $\qsq$ and $\mkpi$ distributions.  The dominant sources of bias
are the geometrical acceptance of the detector and the requirements on
the track momentum, the impact parameter, and the PID of the hadrons.

\begin{figure}[t]
  \centering
  \includegraphics[width=0.495\textwidth]{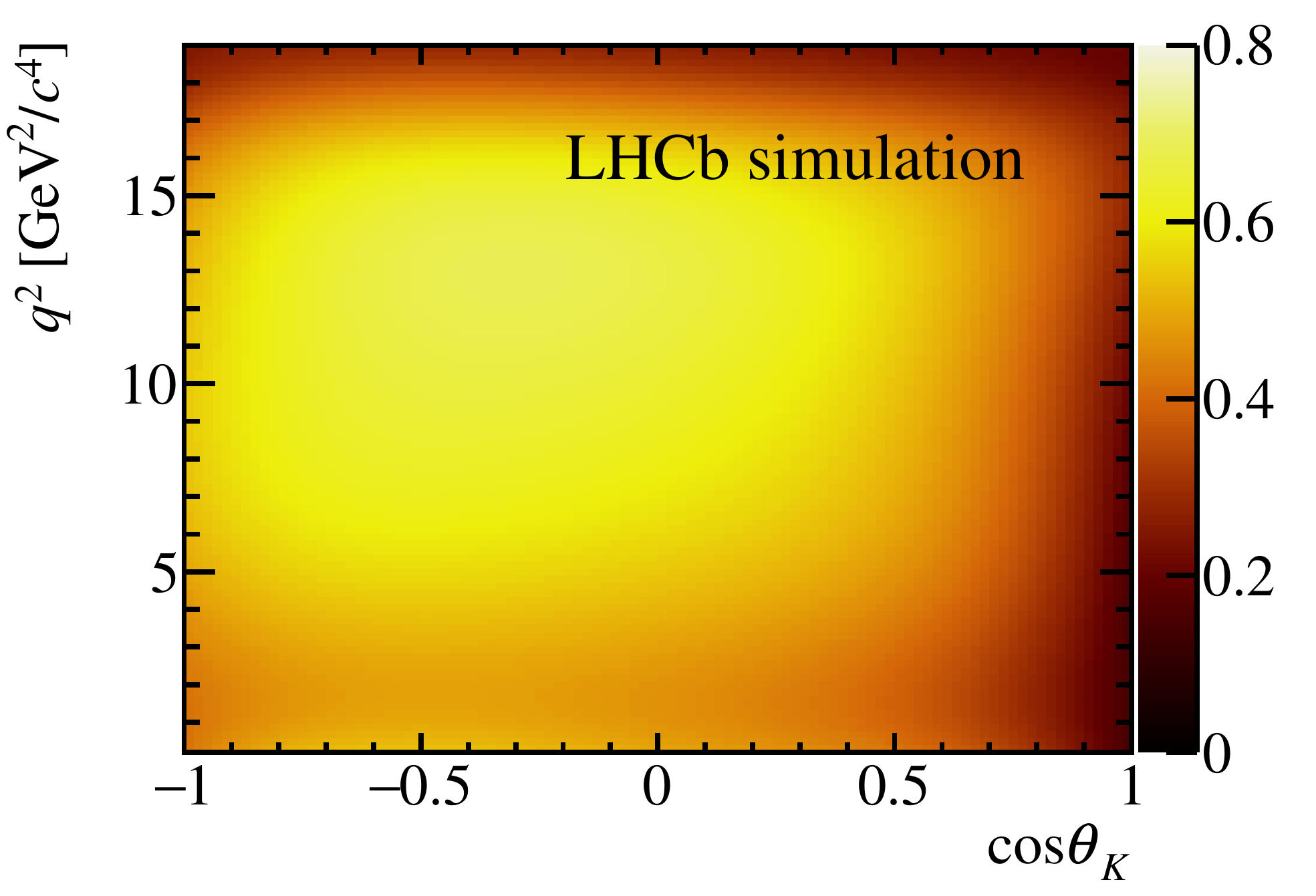}
  \includegraphics[width=0.495\textwidth]{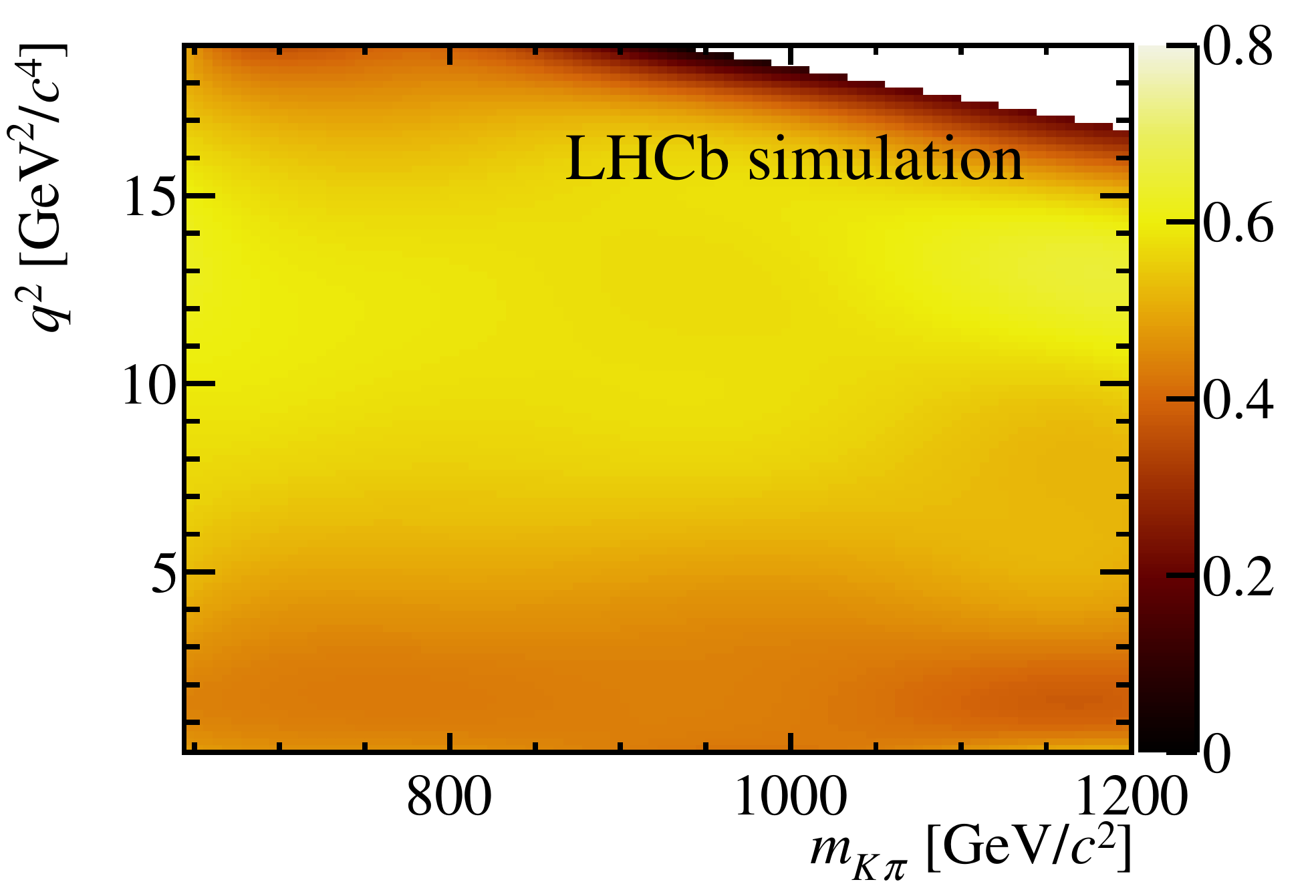}
  \caption{
    Two-dimensional projections of the efficiency (left) in the
    $\ctk$--$\qsq$ plane and (right) in the $\mkpi$--$\qsq$ plane,
    determined from a principal moments analysis of simulated four-body
    $\BdToKpimm$ phase-space decays. The colour scale denotes the
    efficiency in arbitrary units. The lack of entries in the top
    right corner of the $\mkpi$--$\qsq$ distribution is due to the
    limited phase space available in the decay of the $\Bz$ meson.
    \label{fig:acc_angles_mkpi_qsq}
  }
\end{figure}

The method for obtaining the efficiency
correction, described in Ref~\cite{LHCb-PAPER-2015-051}, 
is extended to also include the \mkpi dimension.
The detection efficiency is expressed in terms of orthonormal
Legendre polynomials of order $n$, $P_n(x)$, as
\begin{equation}\label{eqn:angular_eff}
  \epsilon(\qsq,\mkpi,\vec\Omega') = \sum\limits_{g,h,i,j,k}^{}c_{ghijk} P_g(\mkpi) P_h(\ctl) P_i(\ctk) P_j(\phih) P_k({\qsq}).
\end{equation}
As the polynomials are orthonormal over the domain
$x\in[-1,1]$, the observables $\mkpi$,
$\phih$, and $\qsq$ are linearly transformed to lie within this domain when
evaluating the efficiency. The sum in
Eq.~\ref{eqn:angular_eff} runs up to 5$^{\rm th}$ order for
$\ctk$ and $\phih$, and up to 8$^{\rm th}$, 7$^{\rm th}$
and 6$^{\rm th}$ order for $\ctl$, $\qsq$ and $\mkpi$ respectively. The coefficients $c_{ghijk}$
are determined using a principal moment analysis of simulated
four-body \BdToKpimm phase-space decays. Two-dimensional projections
of the detection efficiency as a function of $\ctk$--$\qsq$ and
$\mkpi$--$\qsq$ are shown in Fig.~\ref{fig:acc_angles_mkpi_qsq}.

\subsection{Fit to  the mass and angular distributions}
\label{sec:fsfit}
An extended maximum likelihood fit to $\mkpimm$, $\mkpi$ and $\ctk$ is performed
in each bin of $\qsq$ in order to determine the coefficients $G_{\rm
  S}^{\phantom R}$, $G_{\rm SP}^{\rm Re}$,
$G_{\rm SP}^{\rm Im}$ and $G_{\rm P}^{\perp\parallel}$ averaged over
the $\qsq$ bin. Given these coefficients, the S-wave fraction
$F_{\rm S}$ is extracted using Eq.~\ref{eqn:FSdefinition}. The angular
distribution of the signal is described  by 
Eq.~\ref{eqn:dGGGGG} multiplied by the efficiency model
evaluated at the centre of the $\qsq$ bin ($q^{2}_{\rm bc}$). Integrating over $\ctl$
and $\phih$ simplifies the fit, while retaining the sensitivity to the
parameters related to $F_{\rm S}$. The resulting angular and $\mkpi$ distribution of the
signal, $P_{\rm sig}$, within a bin $q^{2}_{\rm min}<\qsq<q^{2}_{\rm max}$, is given by
\begin{equation}
\label{eqn:sig_pdf}
P_{\rm sig}(\mkpi,\ctk)=\int_{q^{2}_{\rm min}}^{q^{2}_{\rm
    max}}\int_{0}^{\pi}\int_{-1}^{1}\deriv\!\ctl {\rm d}\phih\dqsq\left[\frac{\deriv^{5}(\Gamma+\overline{\Gamma})}{\deriv\mkpi\deriv\qsq\,\deriv\vec\Omega'}\times
  \epsilon(q^{2}_{\rm bc},\mkpi,\vec\Omega')\right],
\end{equation}
The overall scale of $P_{\rm sig}$ is set by fixing the parameter $G_{\rm P}^{0}$ to an arbitrary value.
The \mkpimm distribution of the signal is assumed to factorise with $P_{\rm sig}(\mkpi,\ctk)$.
This assumption is validated using simulated events.

\begin{figure}[t]
  \centering
  \includegraphics[width=0.495\textwidth]{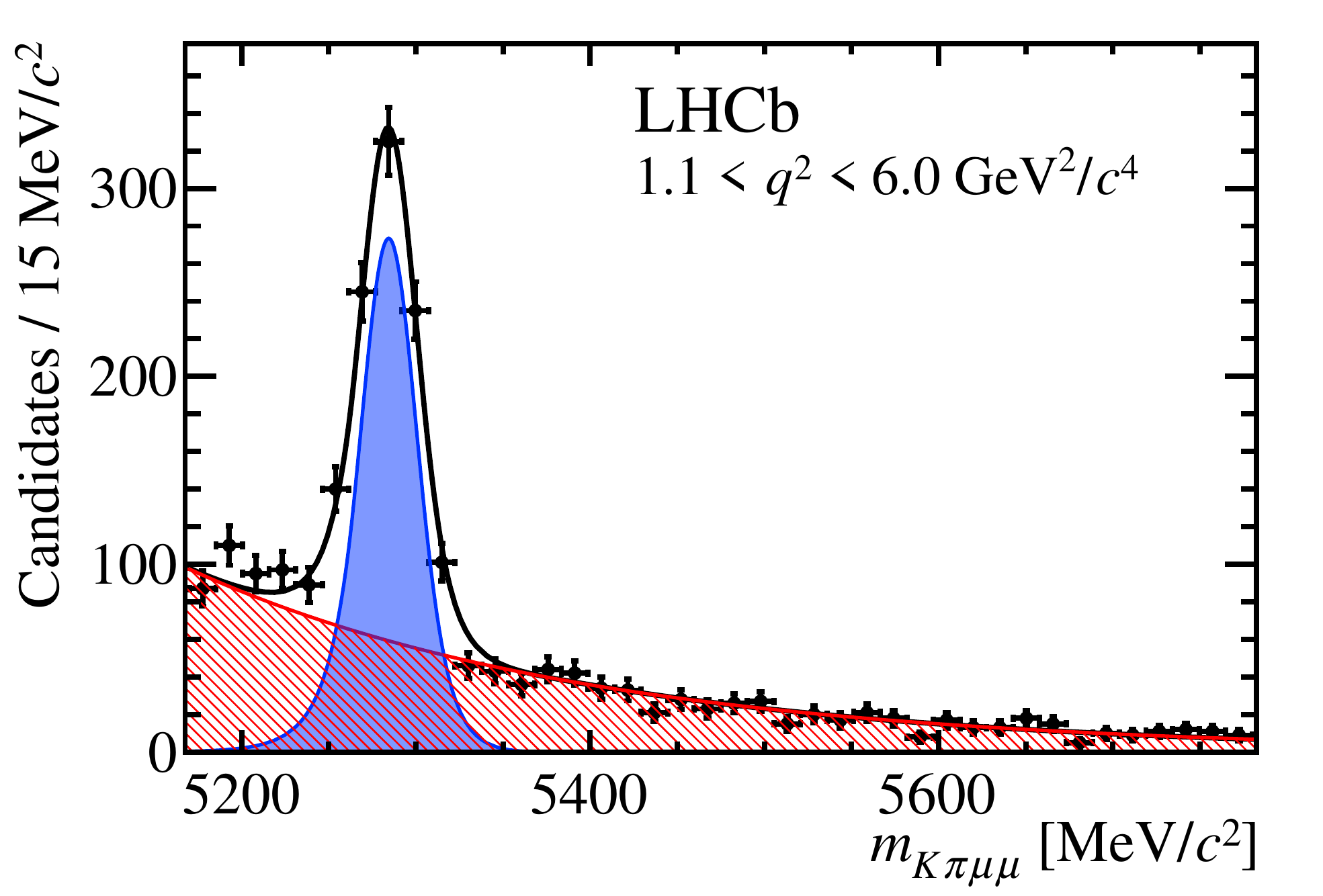}
  \includegraphics[width=0.495\textwidth]{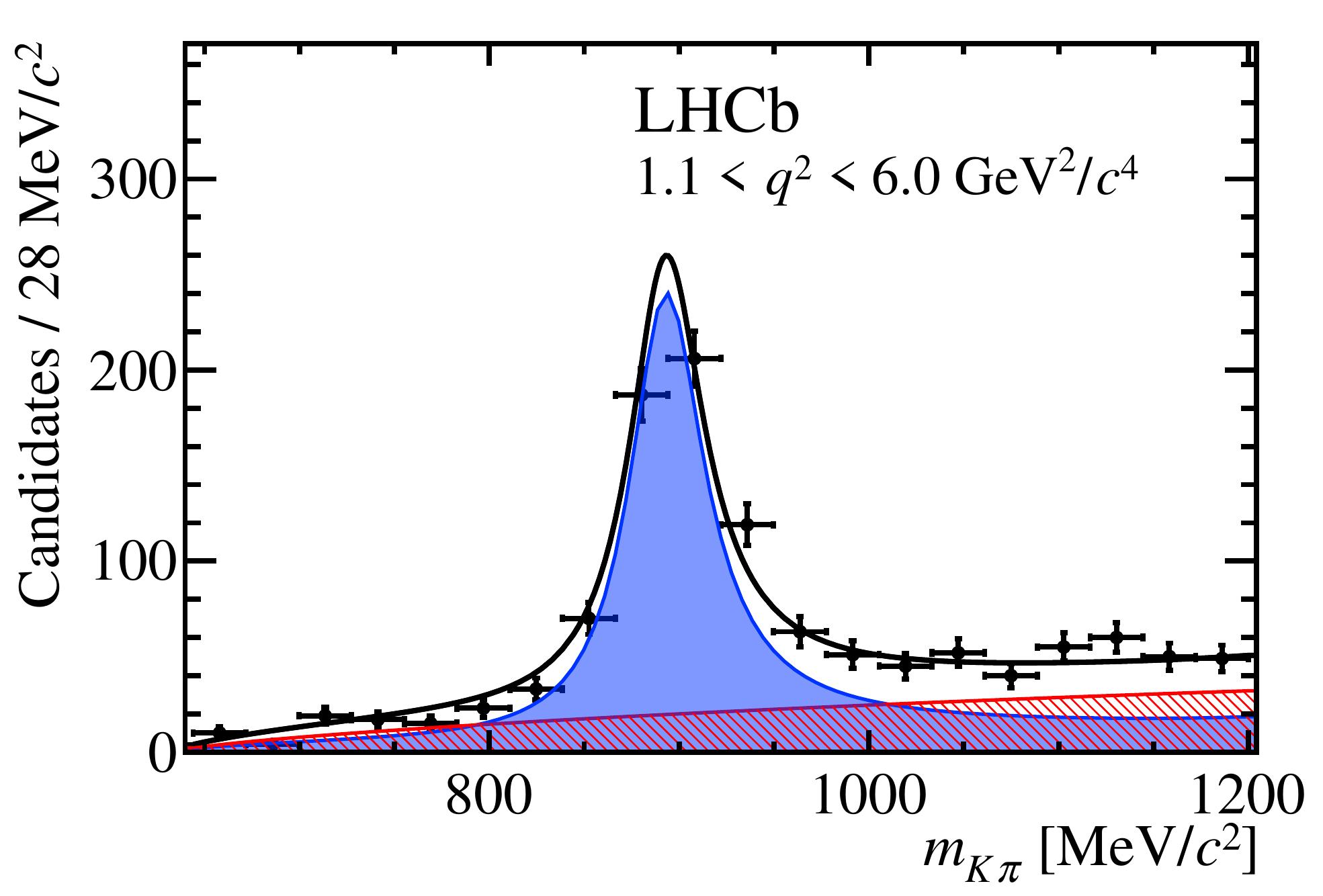}\\
  \includegraphics[width=0.495\textwidth]{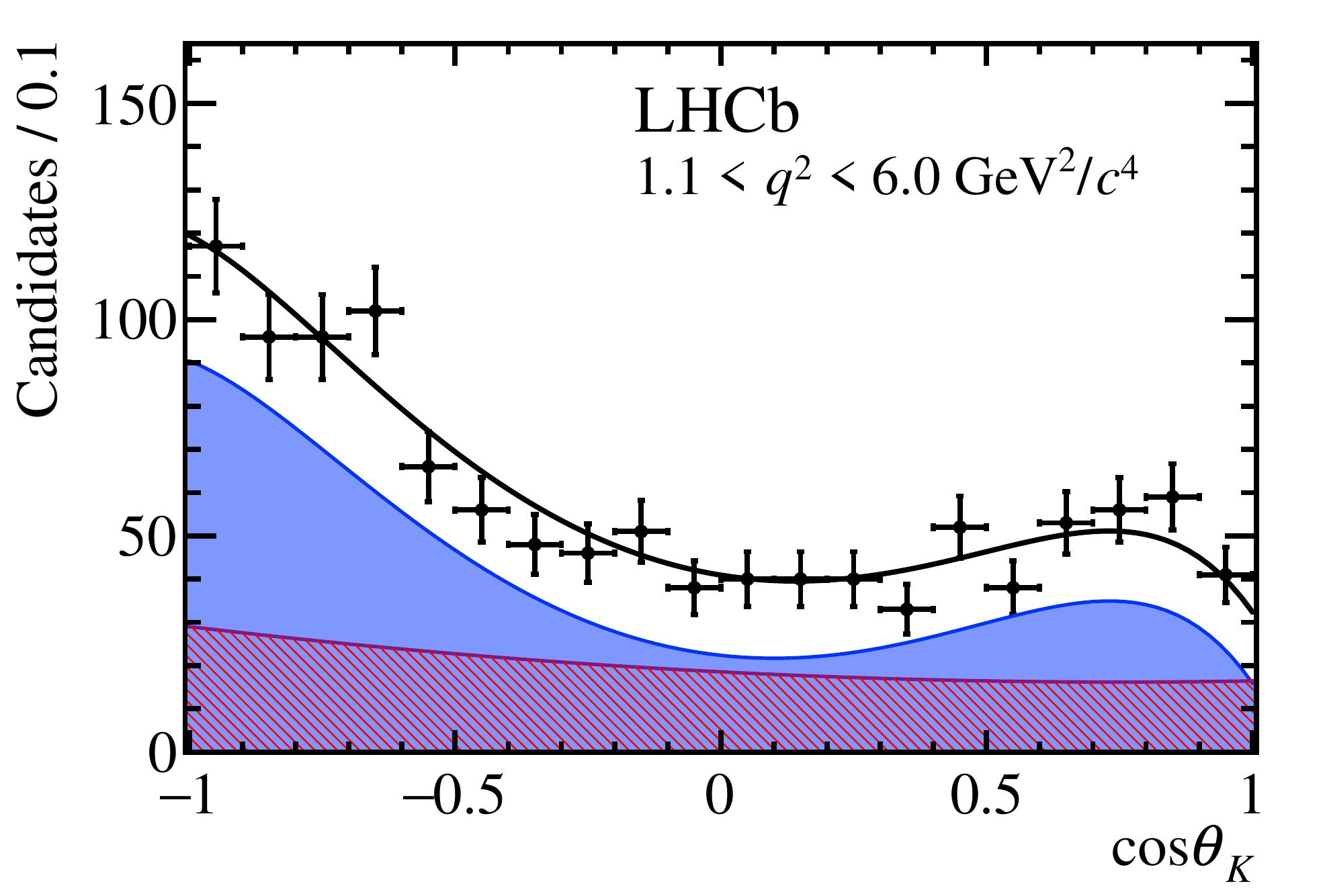}
  \caption{Angular and mass distributions for the $\qsq$ bin $1.1<\qsq<6.0\gevgevcccc$.
    The distributions of $\ctk$ and $\mkpi$ are shown for candidates
    in the signal $\mkpimm$ window of $\pm50\mevcc$ around the known
    $\Bz$ mass. The solid line denotes the total fitted
    distribution. The individual components, signal (blue shaded area)
    and background (red hatched area), are also shown. \label{fig:mkpifit8}}
\end{figure}

The $\ctk$ distribution of the combinatorial background is modelled
with a second-order polynomial where all parameters are allowed to vary in
the fit. The $\mkpi$, $\mkpimm$ and $\ctk$ distributions of the
combinatorial background are assumed to factorise. This assumption has
been validated on data from the upper $\mkpimm$ sideband.  
Figure~\ref{fig:mkpifit8} shows the projections of the probability distribution function on the angular
and mass distributions for the $\qsq$ bin
$1.1<\qsq<6.0\gevgevcccc$. Projections of other $\qsq$ bins are
provided in Appendix~\ref{app:fits}.

\subsection{Result for $F_{\rm S}$}
\label{sec:fsresults}
Using Eq.~\ref{eqn:FSdefinition}, $F_{\rm S}$ is determined in the
full $\mkpi$ region of the fit, $\FS{644}{1200}$, and in the narrow
$\mkpi$ region, $\FS{796}{996}$.
The statistical uncertainty on $F_{\rm S}$ is determined using the
following procedure. Values of the parameters of the fit are generated 
according to a multi-dimensional bifurcated Gaussian distribution.
This distribution is constructed out of the correlation
matrix of the fit and the asymmetric uncertainties obtained from a
profile likelihood. For each generated set of parameters of the fit, a value of
$F_{\rm S}$ is computed. The 68\% confidence interval is defined by taking the
$16^{\rm th}$--$84^{\rm th}$ percentiles of the resulting distribution
of $F_{\rm S}$. The correct coverage of this method is validated
using pseudoexperiments generated with a wide range of $F_{\rm S}$ 
values.

\begin{figure}[t]
  \centering
  \includegraphics[width=0.495\textwidth]{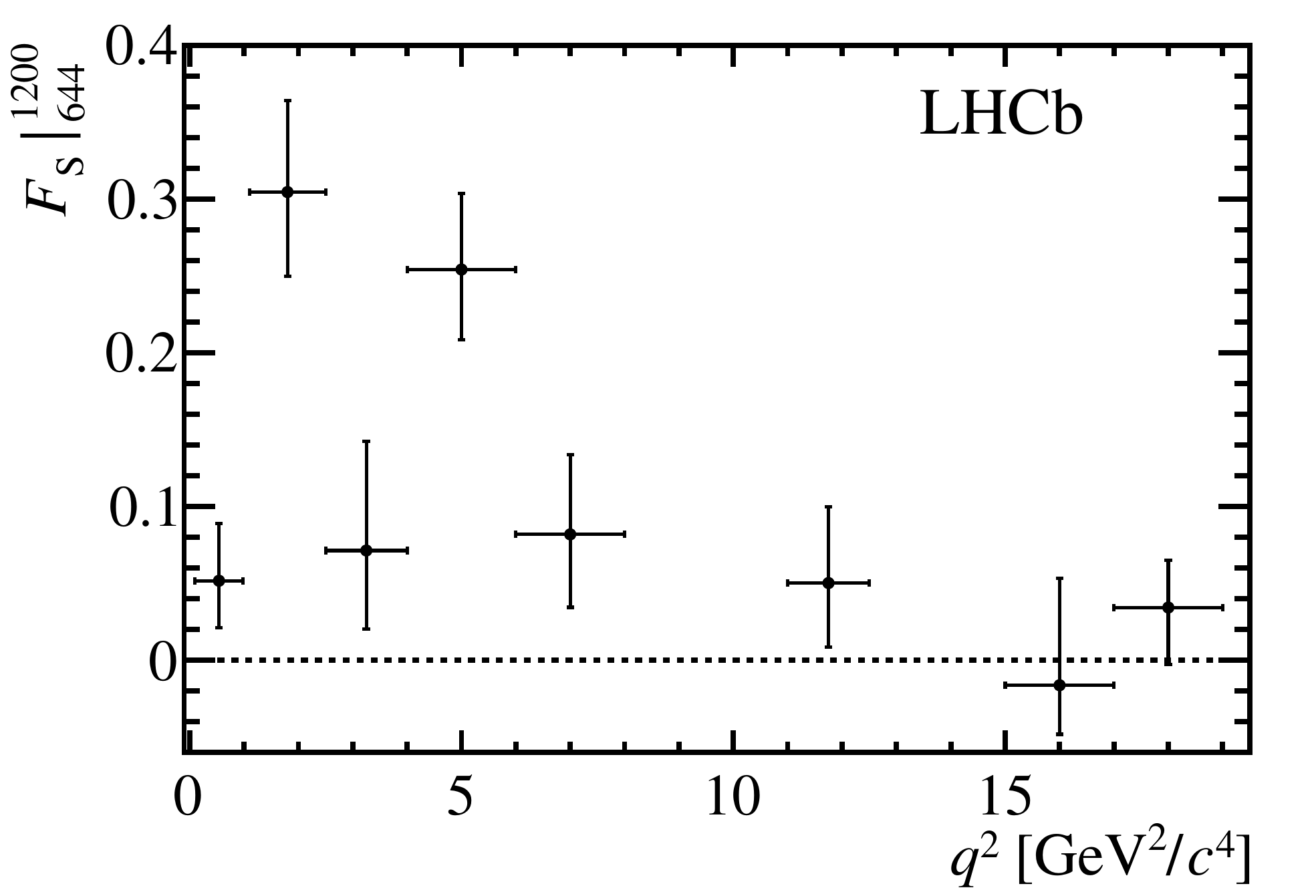}
  \includegraphics[width=0.495\textwidth]{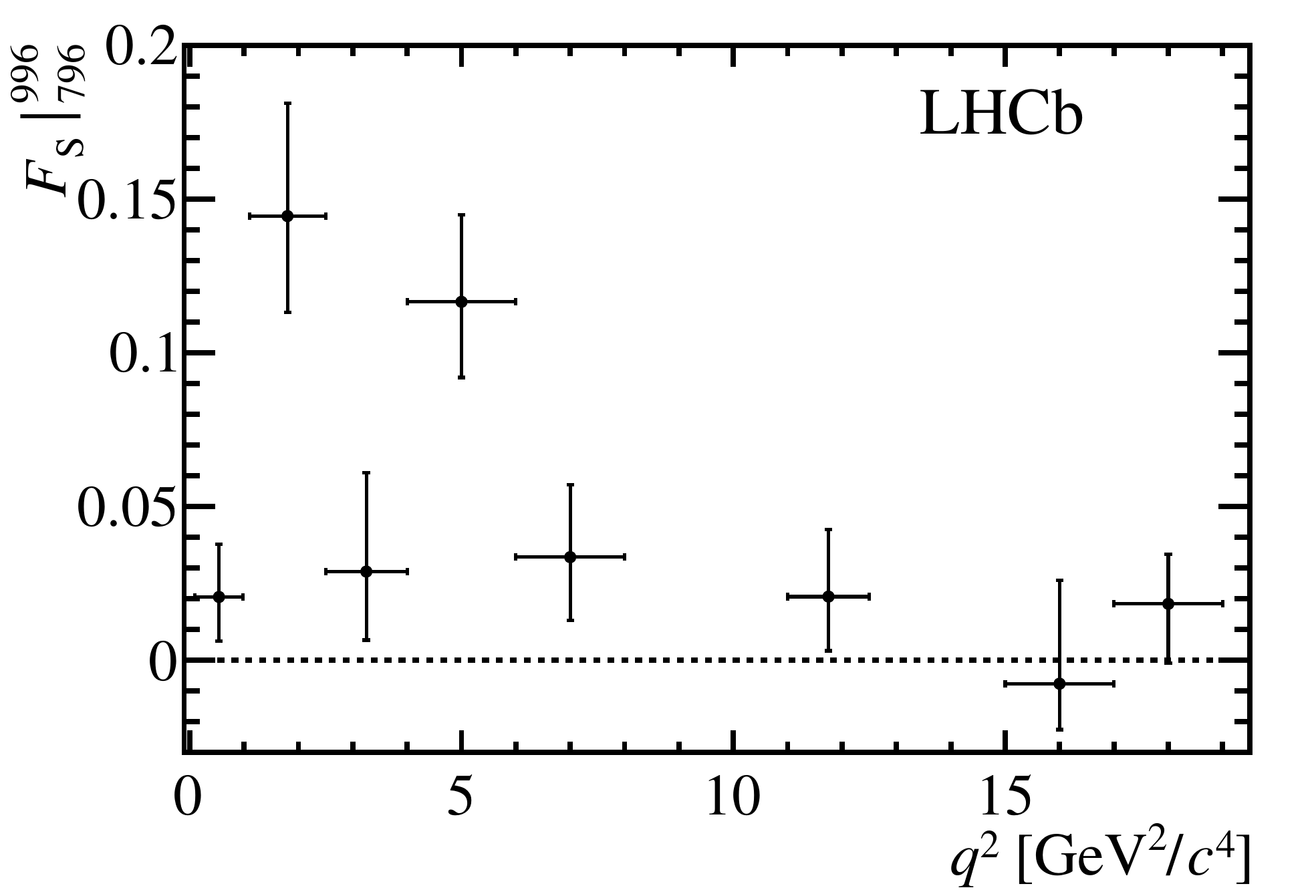}
  \caption{
    Results for the \swave fraction ($F_{\rm S}$) in bins of \qsq in the
    range (left) $644<\mkpi<1200\mevcc$ and (right)
    $796<\mkpi<996\mevcc$. The uncertainties shown are the quadratic sum of the statistical and
    systematic uncertainties. The shape of $F_{\rm S}$ is found to be
    compatible with the smoothly varying distribution of  $F_{\rm L}$, as measured in Ref.~\cite{LHCb-PAPER-2015-051}.
    \label{fig:fsresults}
  }
\end{figure}

Figure~\ref{fig:fsresults} shows the values of $\FS{644}{1200}$ and
$\FS{796}{996}$ in each $\qsq$ bin. The uncertainties given are a
quadratic sum of statistical and systematic uncertainties. The results
are also reported in Table~\ref{tab:fsresults}. The sources of systematic uncertainty
are detailed in Sec.~\ref{sec:systs}. As expected, the shape of the
measured $F_{\rm S}$ distribution is found to be compatible with 
the smoothly varying distribution of $F_{\rm L}$ measured in Ref.~\cite{LHCb-PAPER-2015-051}.

\begin{table}[!th]
  \centering
  \caption{
    \swave fraction ($F_{\rm S}$) in bins of
    $\qsq$ for two $\mkpi$ regions. The
    first uncertainty is statistical and the second systematic. 
    \label{tab:fsresults}
  }
  {
   
    \renewcommand\arraystretch{1.4} \setlength\minrowclearance{2.4pt}
  \begin{tabular}{ c c c }
    \hline
    \qsq bin $(\!\gevgevcccc)$ & \FS{796}{996} & \FS{644}{1200} \\
    \hline\hline
    $0.10<\qsq<0.98$    & $\phantom{+}0.021_{\,-0.011}^{\,+0.015}   \pm{0.009}$   & $\phantom{+}0.052_{\,-0.027}^{\,+0.035}   \pm{0.013}$    \\
    $1.1<\qsq<2.5$        & $\phantom{+}0.144_{\,-0.030}^{\,+0.035}   \pm{0.010}$   & $\phantom{+}0.304_{\,-0.053}^{\,+0.058}   \pm{0.013}$    \\
    $2.5<\qsq<4.0$        & $\phantom{+}0.029_{\,-0.020}^{\,+0.031}   \pm{0.010}$   & $\phantom{+}0.071_{\,-0.049}^{\,+0.069}   \pm{0.015}$    \\
    $4.0<\qsq<6.0$        & $\phantom{+}0.117_{\,-0.023}^{\,+0.027}   \pm{0.008}$   & $\phantom{+}0.254_{\,-0.044}^{\,+0.048}    \pm{0.012}$   \\
    $6.0<\qsq<8.0$        & $\phantom{+}0.033_{\,-0.019}^{\,+0.022}   \pm{0.009}$   & $\phantom{+}0.082_{\,-0.045}^{\,+0.049}    \pm{0.016}$   \\
    $11.0<\qsq<12.5$    & $\phantom{+}0.021_{\,-0.016}^{\,+0.021}   \pm{0.007}$   & $\phantom{+}0.049_{\,-0.039}^{\,+0.048}    \pm{0.014}$  \\
    $15.0<\qsq<17.0$    & $-0.008_{\,-0.014}^{\,+0.033} \pm{0.006}$                       & $-0.016_{\,-0.030}^{\,+0.069} \pm{0.012}$                        \\
    $17.0<\qsq<19.0$    & $\phantom{+}0.018_{\,-0.017}^{\,+0.013}   \pm{0.009}$   & $\phantom{+}0.034_{\,-0.032}^{\,+0.024}   \pm{0.019}$    \\
    \hline
    $1.1<\qsq<6.0$        &  $\phantom{+}0.101_{\,-0.017}^{\,+0.017}   \pm{0.009}$  & $\phantom{+}0.224_{\,-0.033}^{\,+0.032} \pm{0.013}$ \\
    $15.0<\qsq<19.0$    & $\phantom{+}0.010_{\,-0.014}^{\,+0.017} \pm{0.007}$     & $\phantom{+}0.019_{\,-0.025}^{\,+0.030} \pm{0.015}$      \\
  \end{tabular}
}
\end{table}
\clearpage
The presence of a nonresonant P-wave component in the $K^+\pi^-$ system
has been suggested in Refs.~\cite{Das:2014sra,Das:2015pna}. However,
no evidence for such a component was found in the current data sample.
The effect of neglecting a nonresonant P-wave contribution with a
relative phase and magnitude varied within the statistical
uncertainties determined in this analysis, was found to be negligible.

\section{Differential branching fraction of the decay \texorpdfstring{$\BdToKstENTmm$}{B0-->K*(892)0mu+mu-}}
The differential branching fraction of the decay $\BdToKstENTmm$ is
estimated by normalising the signal yield, $n_{\Kstarz\mu^+\mu^-}$, 
obtained from the fit described in Sec.~\ref{sec:fsfit}, to the total
event yield of the decay $\BdToJPsiKstarz$, $n_{\jpsi\Kstarz}$. The
number of $\BdToJPsiKstarz$ events is obtained from a fit to the
$\mkpimm$ spectrum using the same $\qsq$ range as for the fit to
determine the $\mkpimm$ mass shape parameters
(Sec.~\ref{sec:massdistr}), 
but for an $\mkpi$ range $796<\mkpi<996~\mevcc$.  This yield has to be corrected for the
S-wave fraction within the narrow $\mkpi$ window of $\BdToJPsiKstarz$
decays, $F_{\rm S}^{\jpsi\Kstarz}$. The value of  $F_{\rm
  S}^{\jpsi\Kstarz}$ is obtained from
Ref.~\cite{LHCb-PAPER-2013-023} and is adjusted to the $\mkpi$ range 
$796<\mkpi<996~\mevcc$. The ratio of $\BdToKstmm$ and
$\BdToJPsiKstarz$ events is corrected for the relative efficiency
between the two decays,
$R_\epsilon=\epsilon_{J/\psi\Kstarz}/\epsilon_{\Kstarz\mu^+\mu^-}$. This
ratio is determined using simulated samples of $\BdToKstENTmm$ and
$\BdToJPsiKstENT$ decays. The angular distributions of these samples
are corrected to account for the presence of P- and S-wave components
with a relative abundance given by
the measurements of Sec.~\ref{sec:fsresults} and Ref.~\cite{LHCb-PAPER-2013-023}. 
The systematic uncertainty associated with this correction is determined by varying the 
components within the uncertainties of the measured values and
recalculating $R_\epsilon$. The resulting uncertainty on $R_\epsilon$ is negligible. 

\begin{figure}[t]
  \centering
  \includegraphics[width=0.65\textwidth]{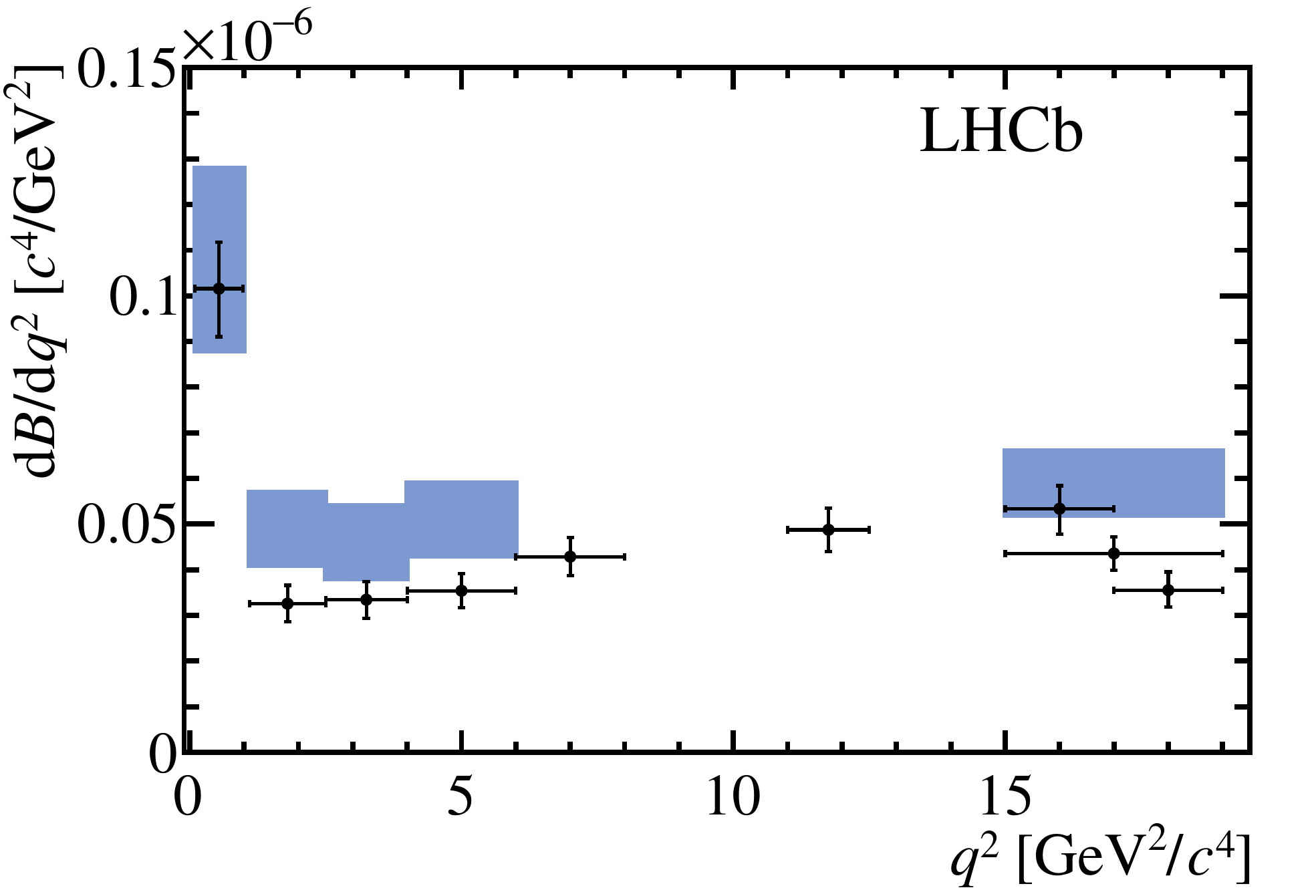}
  \caption{ Differential branching fraction of \BdToKstENTmm decays as
    a function of $\qsq$. The data are overlaid with the SM prediction
    from Refs.~\cite{BSZ,WingateEtAlMay14}. No SM prediction is
    included in the region close to the narrow $c\bar{c}$
    resonances. The result in the wider \qsq bin
    $15.0<\qsq<19.0\gevgevcccc$ is also presented. The uncertainties
    shown are the quadratic sum of the statistical and systematic
    uncertainties, and include the uncertainty on the
      $\BdToJPsiKstarz$ and $\jpsi\to\mu^+\mu^-$ branching fractions.
    \label{fig:bfresult}
  }
\end{figure}

The differential branching fraction of \BdToKstENTmm decays in a $\qsq$ bin
of width $(q^2_{\rm max}-q^2_{\rm min})$ is given by
\begin{equation}\label{eqn:diffbr}
  \begin{split}
    \frac{\diff\mathcal{B}}{\diff\qsq}=&\frac{R_\epsilon}{(q^2_{\rm
        max}-q^2_{\rm min})}
    \frac{(1-\FS{644}{1200})n_{\Kstarz\mu^+\mu^-}}{(1-F_{\rm S}^{\jpsi\Kstarz})n_{\jpsi\Kstarz}}
    \mathcal{B}(\BdToJPsiKstarz)\mathcal{B}(\jpsi\to\mu^+\mu^-),
  \end{split}
\end{equation}
\noindent where $\FS{644}{1200}$, $R_\epsilon$ and $n_{\Kstarz\mu^+\mu^-}$
correspond to quantities measured within the relevant
$\qsq$ bin.  
The branching fraction $\mathcal{B}(\Bz\to\jpsi\KstarENT)$ obtained from
Ref.~\cite{Chilikin:2014bkk} is
\begin{equation*}\label{eqn:bfjpsikst}
  \mathcal{B}(\Bz\to\jpsi\KstarENT)=(1.19\pm0.01\pm0.08)\times10^{-3},
\end{equation*}
where the first uncertainty is statistical and the second systematic.
The branching fraction for $\jpsi\to\mu^+\mu^-$ decays is taken 
from Ref.~\cite{PDG2014}.
The resulting differential branching fraction is shown in
Fig.~\ref{fig:bfresult}. The uncertainties given are a
quadratic sum of statistical and systematic uncertainties and the 
bands shown indicate the SM prediction from
Refs.~\cite{BSZ,WingateEtAlMay14}.
The results are also reported in Table~\ref{tab:bfresults}. The various
sources of systematic uncertainties are described in
Sec.~\ref{sec:systs}. 

\label{sec:bfmeas}

\begin{table}[!htb]
  \centering
  \caption{
    Differential branching fraction of $\BdToKstENTmm$ decays 
    in bins of $\qsq$. The first uncertainty is statistical, the 
    second systematic and the third due to the uncertainty on the
    $\BdToJPsiKstarz$ and $\jpsi\to\mu^+\mu^-$ branching fractions. 
    \label{tab:bfresults}
  }
  {
    \renewcommand\arraystretch{1.4} \setlength\minrowclearance{2.4pt}
  \begin{tabular}{ c c c c }
    \hline
    \qsq bin $(\!\gevgevcccc)$ & $\diff\mathcal B/\dqsq\times 10^{-7}$~$(c^{4}/\gev^{2})$ \\
    \hline\hline
    $0.10<\qsq<0.98$    & $1.016_{\,-0.073}^{\,+0.067} \pm {0.029}  \pm0.069$\\
    $1.1<\qsq<2.5$        & $0.326_{\,-0.031}^{\,+0.032} \pm {0.010} \pm0.022$\\
    $2.5<\qsq<4.0$        & $0.334_{\,-0.033}^{\,+0.031} \pm {0.009}  \pm0.023$\\
    $4.0<\qsq<6.0$        & $0.354_{\,-0.026}^{\,+0.027} \pm {0.009} \pm0.024$\\
    $6.0<\qsq<8.0$        & $0.429_{\,-0.027}^{\,+0.028} \pm {0.010} \pm0.029$\\
    $11.0<\qsq<12.5$    & $0.487_{\,-0.032}^{\,+0.031} \pm {0.012} \pm0.033$\\
    $15.0<\qsq<17.0$    & $0.534_{\,-0.037}^{\,+0.027} \pm {0.020} \pm0.036$\\
    $17.0<\qsq<19.0$    & $0.355_{\,-0.022}^{\,+0.027} \pm  {0.017} \pm0.024$\\
    \hline
    $1.1<\qsq<6.0$        &  $0.342_{\,-0.017}^{\,+0.017}  \pm{0.009} \pm0.023$\\
    $15.0<\qsq<19.0$    &  $0.436_{\,-0.019}^{\,+0.018}  \pm{0.007} \pm0.030$
  \end{tabular}
}
\end{table}

\clearpage
The total branching fraction of the $\BdToKstENTmm$ decay is obtained from
the sum over the eight $\qsq$ bins. To account for the fraction of
signal events in the vetoed $\qsq$ regions, a correction factor of
$1.532\pm0.001({\rm stat})\pm0.010({\rm syst})$ is applied. This factor is determined using the
calculation in Ref.~\cite{Ali:1999mm} and form factors from
Ref.~\cite{BallZwicky}. The systematic uncertainty is determined by
recalculating the extrapolation factor using the form factors from
Ref.~\cite{KMPW2010} and taking the difference to the nominal value.
The resulting total branching fraction is
\begin{equation*}
\mathcal{B}(\BdToKstENTmm)=(0.904^{\,+0.016}_{\,-0.015}\pm0.010\pm0.006\pm0.061)\times10^{-6},
\label{eq:total_bf}
\end{equation*}
where the uncertainties, from left to right, are statistical, systematic, from the
extrapolation to the full $\qsq$ region and due to the
uncertainty of the branching fraction of the normalisation mode.

\section{Systematic uncertainties}
\label{sec:systs}
The sources of systematic uncertainty considered can alter the angular
and mass distributions, as well as the ratio of efficiencies between
the signal and control channels.  In general, the systematic
uncertainties are significantly smaller than the statistical
uncertainties. The various sources of systematic uncertainty are
discussed in detail below and are summarised in
Table~\ref{tab:systematics}. Motivated by Eq.~\ref{eqn:diffbr}, the
systematic uncertainty for $F_{\rm S}$ is presented for the $\mkpi$
region $644<\mkpi<1200\mevcc$.  Typical ranges are quoted in order to
summarise the effect the systematic uncertainties have across the
various $\qsq$ bins.  Sources of systematic uncertainty that can
affect both $F_{\rm S}$ and the differential branching fraction are
treated as 100\% correlated.

\begin{table}[ht]
\caption{Summary of the main sources of systematic uncertainty on $\FS{644}{1200}$ and 
$\diff\mathcal{B}/\diff\qsq$. Typical ranges are quoted in order to summarise
the effect the systematic uncertainties have across the various $\qsq$
bins. 
\label{tab:systematics}}
\begin{center}
\scalebox{1.00}{
    \renewcommand\arraystretch{1.4} \setlength\minrowclearance{2.4pt}
\begin{tabular}{r|cc}
\hline
  Source                                              & $\FS{644}{1200}$     &  $\diff\mathcal{B}/\dqsq$~$\times 10^{-7}(c^4/\rm{GeV}^2)$ \\
\hline\hline
Data-simulation differences              & $0.008$--$0.013$   & $0.004$--$0.021$ \\
Efficiency model                                 & $0.001$--$0.010$   & $0.001$--$0.012$ \\   
S-wave $\mkpi$ model                      & $0.001$--$0.017$   & $0.001$--$0.015$ \\ 
$\Bd\to\KstarENT$ form factors        & --                             & $0.003$--$0.017$ \\
\hline
$\mathcal{B}(\BdToJPsiMuMuKst)$     & --                             & $0.025$--$0.079$
\end{tabular}
}
\end{center}
\end{table}

\subsection{Systematic uncertainties on the S-wave fraction}
\label{sec:fssyst}
The impact of each source of systematic uncertainty on $F_{\rm S}$ is estimated
using pseudoexperiments, where samples are generated varying one or
more parameters. The value of $F_{\rm S}$ is determined using both the
nominal model and the alternative model. For every
pseudoexperiment, the difference between the two values of $F_{\rm S}$
is computed. In general, the systematic uncertainty is then taken as
the average of this difference over a large number of
pseudoexperiments. The exception to this is the statistical
uncertainty of the efficiency correction. In order to account for this
statistical variation, the standard deviation of the difference
between the two values of $F_{\rm S}$ from each pseudoexperiment is
used instead.
The systematic uncertainty is evaluated in each $\qsq$ bin separately. The pseudodata are generated with
signal and background yields many times larger than those of the data,
rendering statistical effects negligible. The main systematic
uncertainties on $F_{\rm S}$ originate from the efficiency correction
function and the choice of model used to describe the S-wave component
of the $\mkpi$ distribution of the signal.

There are two main systematic uncertainties associated with the
efficiency correction function used for determining $F_{\rm S}$.
Firstly, an uncertainty arises from residual data-simulation
differences. After all corrections to the simulation are applied, a 
difference at the level of 10\% remains in the momentum spectrum of
the pions between simulated and genuine $\BdToJPsiKst$ decays. A new
efficiency correction is derived after weighting the simulated
phase-space sample to account for this difference. The second main
systematic uncertainty associated with the efficiency correction is
due to the order of the polynomials used to describe the efficiency
function. To evaluate this uncertainty, a new efficiency correction is
derived in which the polynomial order in \qsq is increased by
two. This change is motivated by a small residual difference
between the $\qsq$ dependence of the nominal efficiency
correction and the simulated phase-space sample, near the upper
kinematic edge of the $\qsq$ range.  Uncertainties due to the limited
size of the simulation sample used to derive the efficiency correction,
as well as due to the evaluation of the efficiency correction at the centre of the $\qsq$ bin
are also assessed and are found to be negligible.

To assess the modelling of the S-wave component in the $\mkpi$
distribution, pseudoexperiments are produced where the LASS line shape
is exchanged for the sum of resonant $K^{*}_{0}(800)^0$ (also known as
the $\kappa$ resonance) and $K^{*}_{0}(1430)^0$ contributions. An
additional variation is considered where the parameters of the LASS
distribution, determined in $B^0\to \jpsi\Kstarz$ decays using the
model described in Ref.~\cite{LHCb-PAPER-2014-014}, are exchanged
for those measured by the LASS collaboration~\cite{lass}. The largest
of the two variations is taken as the systematic uncertainty on the
S-wave model. Systematic uncertainties associated with the modelling
of the P-wave $\mkpi$ distribution of the signal are found to be
negligible.

Integrating the differential decay rate given in Eq.~\ref{eqn:dGGGGG}
over $\ctl$ and $\phih$ results in the cancellation of terms involving
the angular observables $S_3$, $A_{\rm FB}$ and $S_9$. However the
integral of the product of the differential decay rate with the
efficiency correction, given in Eq.~\ref{eqn:sig_pdf}, results in
a residual dependence of the signal distribution on these angular
observables. By generating pseudoexperiments with observables $S_3$,
$A_{\rm FB}$ and $S_9$ either set to zero or varied within the
uncertainties measured in Ref.~\cite{LHCb-PAPER-2015-051}, the
systematic uncertainty on $F_{\rm S}$ is assessed. Even considering the 
largest variation observed, the resulting systematic uncertainty 
is negligible.

All other sources of systematic uncertainties described in
Ref.~\cite{LHCb-PAPER-2015-051}, such as the modelling of the
$\mkpimm$ distribution of the signal and background, the choice of the
$\mkpi$ and $\cos\theta_K$ background models and the effect of
residual specific backgrounds, are found to be sub-dominant. The
effect of neglecting a possible D-wave $K^+\pi^-$ component, arising
from the tail of the $K^{*}_{2}(1430)^{0}$, is also assessed and
found to be negligible.

\subsection{Systematic uncertainties on the differential branching fraction}

Systematic uncertainties affecting the differential branching fraction
predominantly arise through: the knowledge of $R_\epsilon$, the ratio
of the reconstruction and selection efficiencies described in
Sec.~\ref{sec:bfmeas}; the uncertainty of the branching fraction of
the decay $\BdToJPsiKst$, which is shown as a separate systematic
uncertainty in Table~\ref{tab:bfresults}; and 
systematic uncertainties related to the determination of $F_{\rm S}$,
which are propagated to the differential branching fraction measurement.

The imperfect knowledge of the $B\to K^*$ form-factor
model used in the generation of the $\BdToKstmm$ simulated sample
affects the determination of the ratio of efficiencies $R_\epsilon$. A systematic
uncertainty is therefore assessed by weighting simulated events to
account for the variations between the models described in Refs.~\cite{BSZ}
and~\cite{KMPW2010}.

As described in Sec.~\ref{sec:fssyst}, after all corrections to the
simulation are applied, a small difference remains in the momentum
spectrum of the pions between simulated and genuine $\BdToJPsiKst$
decays. The ratio $R_\epsilon$, and consequently $\diff\mathcal{B}/\diff\qsq$, is
therefore calculated by weighting the simulated $\BdToKstENTmm$ and
$\BdToJPsiKstENT$ decays to account for the observed differences.

Other sources of systematic uncertainties affecting the determination
of the signal yield, such as the choice of model to describe the
$\mkpimm$ distribution of the signal and the background components, the
choice of the $\mkpi$ and $\cos\theta_K$ models to describe the
background, and the effect of residual specific backgrounds, are found
to be negligible.

\section{Conclusions}
\label{sec:conclusions}
This paper presents the first measurement of the S-wave fraction in
the $K^+\pi^-$ system of $\BdToKstmm$ decays using a data sample
corresponding to an integrated luminosity of $3\invfb$ collected at
the LHCb experiment. Accounting for the measured S-wave fraction in
the wide $\mkpi$ region, the first measurement of the P-wave component
of the differential branching fraction of $\BdToKstENTmm$ decays is
reported in bins of $\qsq$. All previous measurements of the
differential branching fraction have compared the combination of S-
and P-wave components to the theory prediction, which is made purely
for the resonant P-wave part of the $K^+\pi^-$ system. The
measurements of the S-wave fraction presented in this paper are
compatible with theory
predictions~\cite{Doring:2013wka,WeiWang,Becirevic:2012dp} and
support previous estimates~\cite{LHCb-PAPER-2013-019}. In the absence
of any previous measurement, such estimates have
been used to assign a systematic uncertainty for a possible S-wave
component~\cite{LHCb-PAPER-2013-019}. The measurements of the S-wave
fraction presented in this paper allow these estimates to be replaced
with an accurate assessment of the scalar component in $\BdToKstmm$ decays.
The resulting measurements of the differential branching
fraction of $\BdToKstENTmm$ decays are the most precise to date and
are in good agreement with the SM predictions.

\section*{Acknowledgements}

\noindent We would like to thank Gudrun Hiller and Martin Jung for useful discussions
regarding the treatment of the $K^+\pi^-$ system. We express our gratitude to our colleagues in the CERN
accelerator departments for the excellent performance of the LHC. We
thank the technical and administrative staff at the LHCb
institutes. We acknowledge support from CERN and from the national
agencies: CAPES, CNPq, FAPERJ and FINEP (Brazil); NSFC (China);
CNRS/IN2P3 (France); BMBF, DFG and MPG (Germany); INFN (Italy); FOM
and NWO (The Netherlands); MNiSW and NCN (Poland); MEN/IFA (Romania);
MinES and FANO (Russia); MinECo (Spain); SNSF and SER (Switzerland);
NASU (Ukraine); STFC (United Kingdom); NSF (USA).  We acknowledge the
computing resources that are provided by CERN, IN2P3 (France), KIT and
DESY (Germany), INFN (Italy), SURF (The Netherlands), PIC (Spain),
GridPP (United Kingdom), RRCKI and Yandex LLC (Russia), CSCS
(Switzerland), IFIN-HH (Romania), CBPF (Brazil), PL-GRID (Poland) and
OSC (USA). We are indebted to the communities behind the multiple open
source software packages on which we depend.  Individual groups or
members have received support from AvH Foundation (Germany), EPLANET,
Marie Sk\l{}odowska-Curie Actions and ERC (European Union), Conseil
G\'{e}n\'{e}ral de Haute-Savoie, Labex ENIGMASS and OCEVU, R\'{e}gion
Auvergne (France), RFBR and Yandex LLC (Russia), GVA, XuntaGal and
GENCAT (Spain), Herchel Smith Fund, The Royal Society, Royal
Commission for the Exhibition of 1851 and the Leverhulme Trust (United
Kingdom).

\clearpage

{\noindent\normalfont\bfseries\Large Appendices}

\appendix

\section{The \texorpdfstring{$\mkpi$}{m(Kpi)} distribution of the signal}
\label{app:mkpi}
The $K^+\pi^-$ invariant mass distribution of the signal candidates is
modelled by two distributions. For the P-wave component, a relativistic
Breit-Wigner function is used, given by
\begin{equation}\label{eqn:fR}
  \begin{split}
    f_{\rm BW}(\mkpi)=&\sqrt{kp}\left(\frac {k}{k_{892}}\right)\frac{B'_{1}(k,k_{892},d)
    B'_{0}(p,p_{892},d)}{\mkpi^2 - m_{892}^2 -im_{892}\Gamma_{892}(\mkpi)},
  \end{split}
\end{equation}
where $\sqrt{kp}$ is the phase-space factor, $\Gamma_{892}(\mkpi)$ is given by
\begin{equation}\label{eqn:runningwidth}
  \Gamma_{892}(\mkpi) = \Gamma_{892} B^{\prime\, 2}_{1}(k,k_{892},d)\left(\frac{k}{k_{892}}\right)^{3}\left(\frac{m_{892}}{\mkpi}\right),
\end{equation}
\noindent and $B'$ are Blatt--Weisskopf barrier factors as defined in
Ref.~\cite{PDG2014}. The parameter $d$ is the meson radius parameter and is set to
1.6\invgevc~\cite{LHCb-PAPER-2014-014}. The systematic uncertainty
associated with the choice of this value is negligible. 
The parameters $m_{892}$ and $\Gamma_{892}$
are the pole mass and width of the $\KstarENT$ resonance, and $k$ ($p$) is the momentum of
the $K^+$ ($\Kstarz$) in the rest frame of the $\Kstarz$ ($\Bz$)
evaluated at a given $\mkpi$. The parameters $k_{892}$ and $p_{892}$
are the values of $k$ and $p$ evaluated at the pole mass of the $\KstarENT$ resonance.  
In Eq.~\ref{eqn:fR}, the orbital angular momentum between the $\KstarENT$
and the dimuon system is considered to be zero. The inclusion of a
higher orbital angular momentum component has a negligible effect on
the measurements.

The S-wave component of the signal is modelled using the LASS
parameterisation~\cite{lass}, given by
\begin{equation}\label{eqn:lass}
  f_{\rm LASS}(\mkpi)=
  \sqrt{kp}B'_{1}(k,k_{1430},d)\left(\frac k{k_{1430}}\right)\left(
    \frac1{\cot{\delta_B}-i} + e^{2i\delta_B}\frac1{\cot{\delta_R}-i}
  \right),
\end{equation}
\noindent where $k_{1430}$ is the momentum of the
$\Kstarz$ in the $\Bz$ rest frame, evaluated at the pole mass of the
$\KstarFT$ resonance. The terms $ \cot{\delta_B}$ and $\cot{\delta_R}$
are given by
\begin{equation}
  \cot{\delta_B} = \frac1{ak} + \frac{rk}2
\end{equation}
and 
\begin{equation}
  \cot{\delta_R} = \frac{m_{1430}^2 - \mkpi^2}{m_{1430}\Gamma_{1430}(\mkpi)},
\end{equation}
with the running width $\Gamma_{1430}(\mkpi)$ in turn given by
\begin{equation}
  \Gamma_{1430}(\mkpi) = \Gamma_{1430}\frac{k}{k_{1430}}\frac{m_{1430}}{\mkpi}\,.
\end{equation}
\noindent The parameters $m_{1430}$ and $\Gamma_{1430}$ are the pole mass and
width of the $\KstarFT$ resonance, and $k_{1430}$ is the momentum of the
kaon in the $\Kstarz$ rest frame, evaluated at the pole mass of the $\KstarFT$ resonance.
The second term of Eq.~\ref{eqn:lass} is equivalent to a Breit--Wigner
function for the \KstarFT.
The first term of Eq.~\ref{eqn:lass} contains two empirical parameters $\left\lbrace a,r \right\rbrace$.
These parameters are fixed to the values $a =  3.83\,\gevc^{-1}$ and $ r =  2.86\gevc^{-1}$,  determined in $\BdToJPsiKst$
decays using the model described in Ref.~\cite{LHCb-PAPER-2014-014}. 

In order to assess the systematic effect of this choice, these
parameters are also fixed to values from the LASS experiment,
$a =  1.94\gevc^{-1}$ and $ r =  1.76\gevc^{-1}$.
\noindent The resulting systematic uncertainty is found to be negligible.
\newpage

\section{Likelihood fit projections}
\label{app:fits}

\begin{figure}[b]
  \centering
  \includegraphics[width=0.45\textwidth]{mass_nice_8.pdf}
  \includegraphics[width=0.45\textwidth]{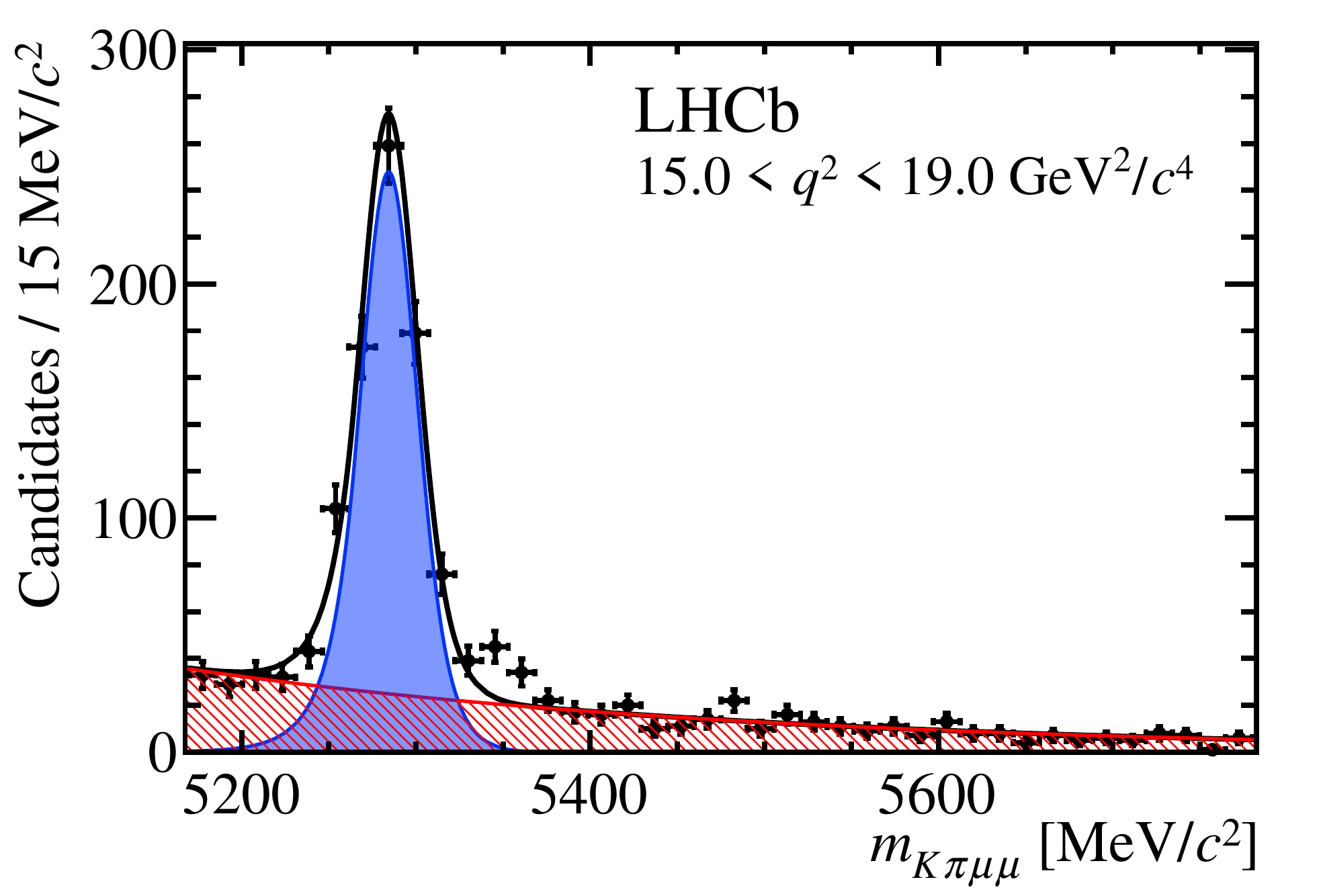}\\
  \includegraphics[width=0.45\textwidth]{mkpi_signal_nice_8.pdf}
  \includegraphics[width=0.45\textwidth]{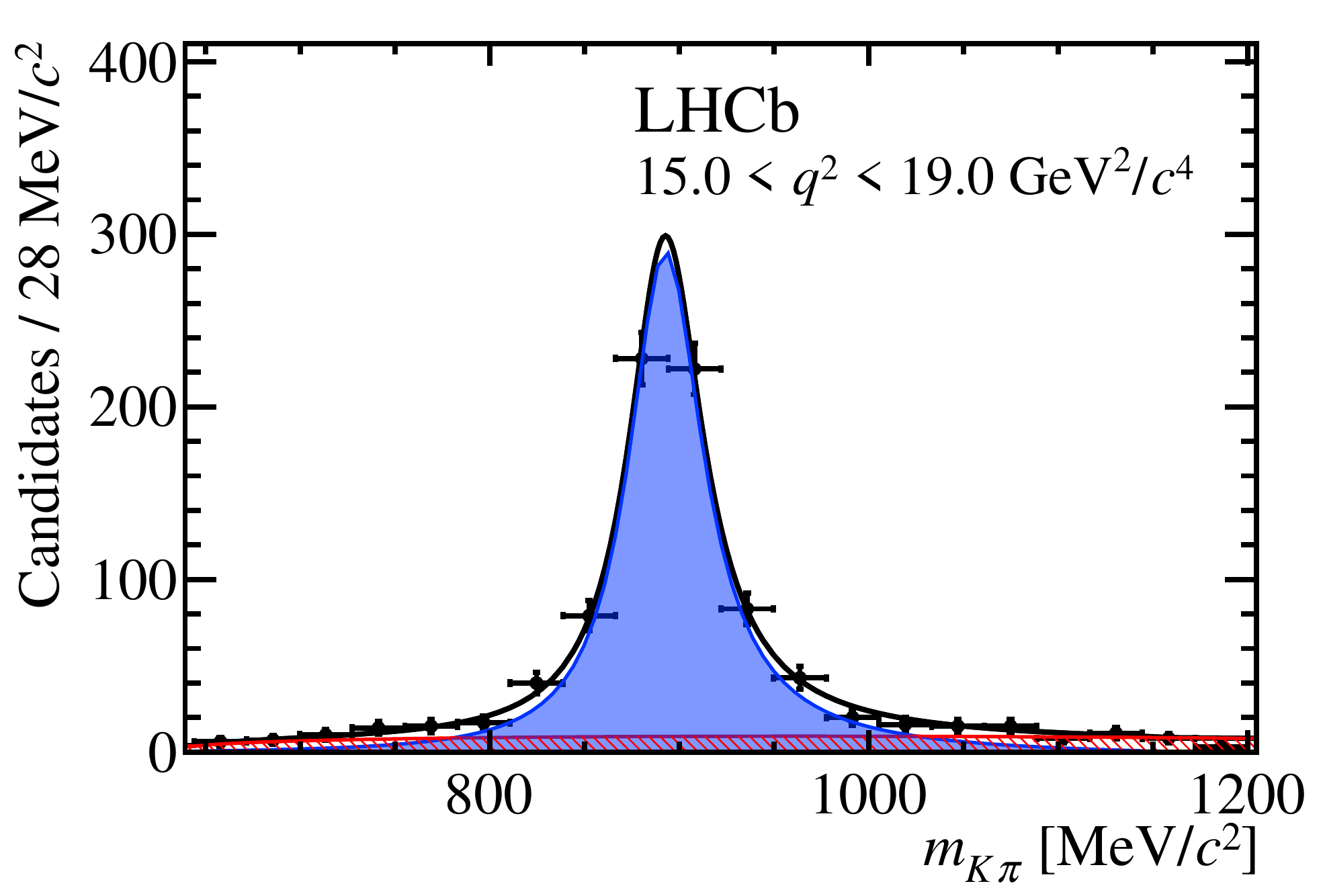}\\
  \includegraphics[width=0.45\textwidth]{ctk_signal_nice_8.pdf}
  \includegraphics[width=0.45\textwidth]{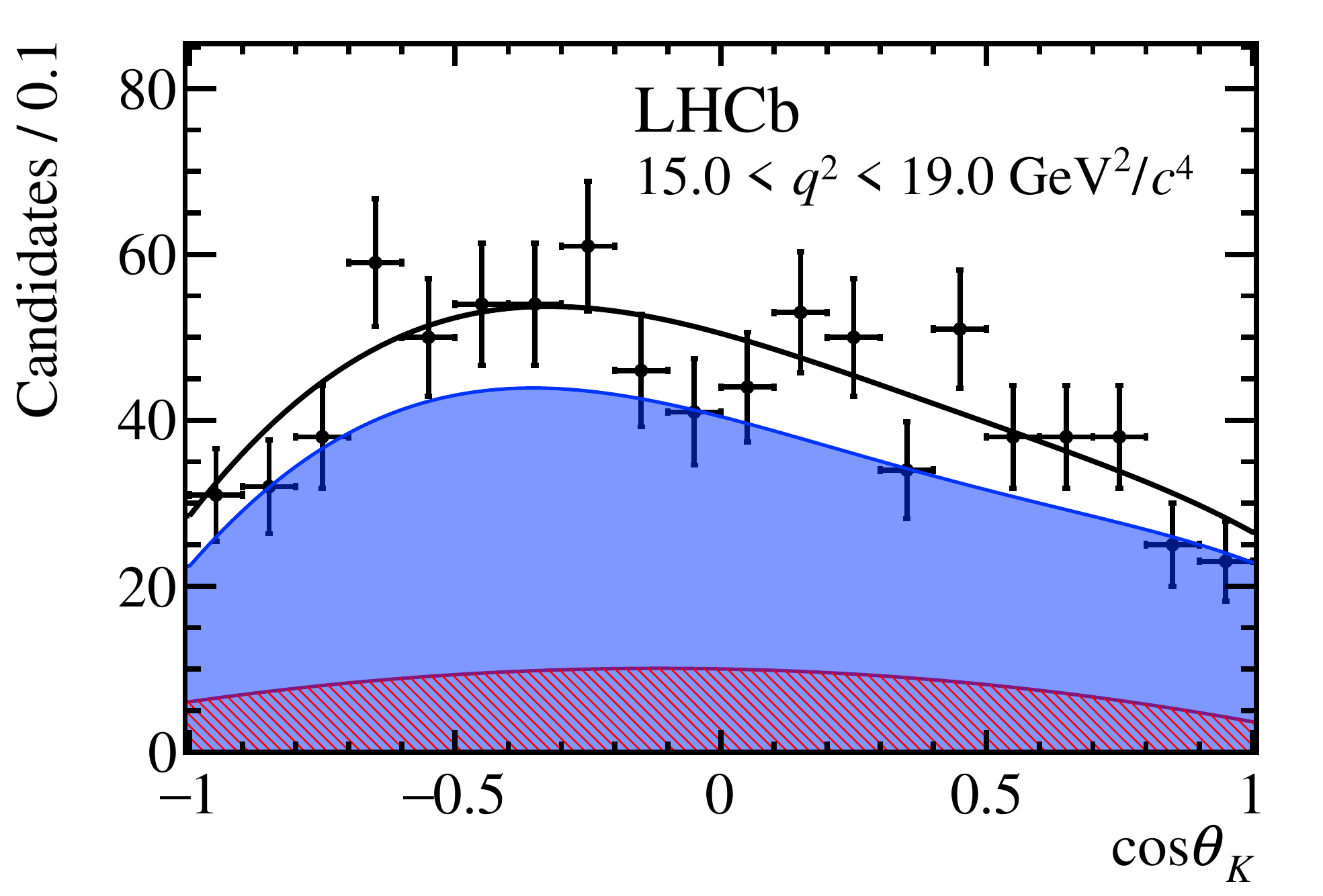}
  \caption{
     Angular and mass distributions for the $\qsq$ bins $1.1<\qsq<6.0\gevgevcccc$
     (left) and $15.0<\qsq<19.0\gevgevcccc$ (right).
     The distributions of $\ctk$ and $\mkpi$ are shown for candidates
     in the signal $\mkpimm$ window of $\pm50\mevcc$ around the known
     $\Bz$ mass.
    \label{fig:appmkpifit0}
  }
\end{figure}

\begin{figure}
  \centering
  \includegraphics[width=0.45\textwidth]{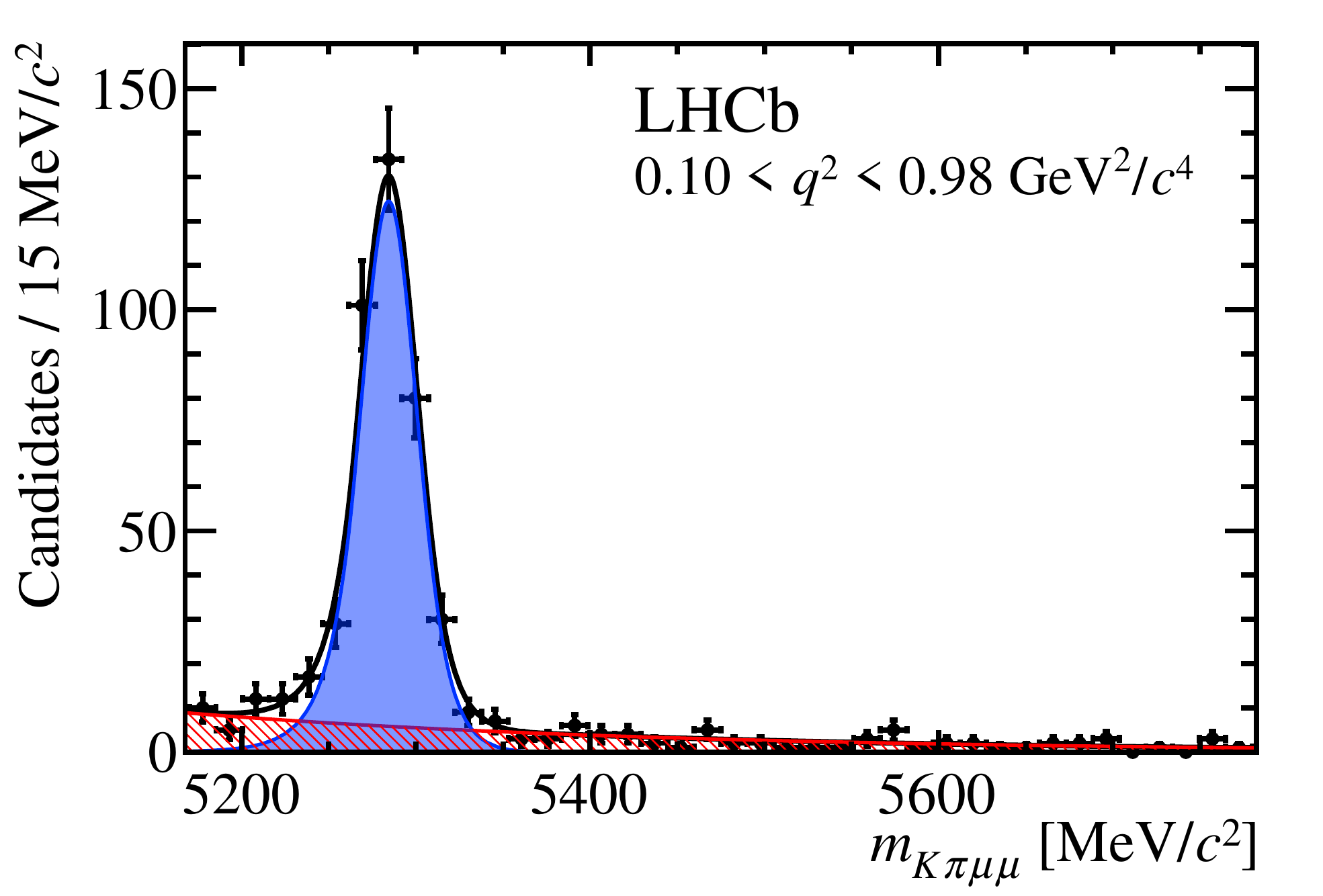}
  \includegraphics[width=0.45\textwidth]{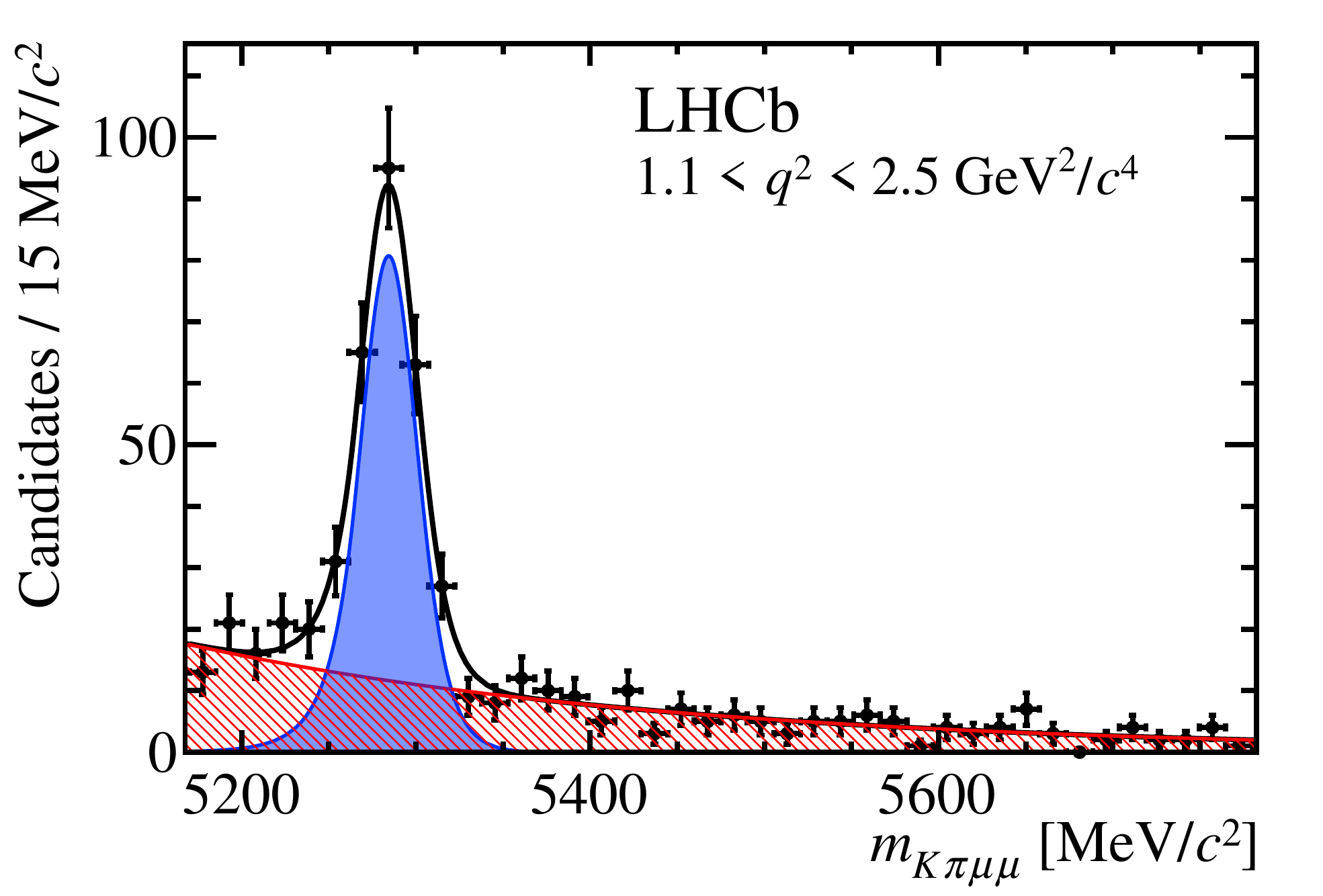}\\
  \includegraphics[width=0.45\textwidth]{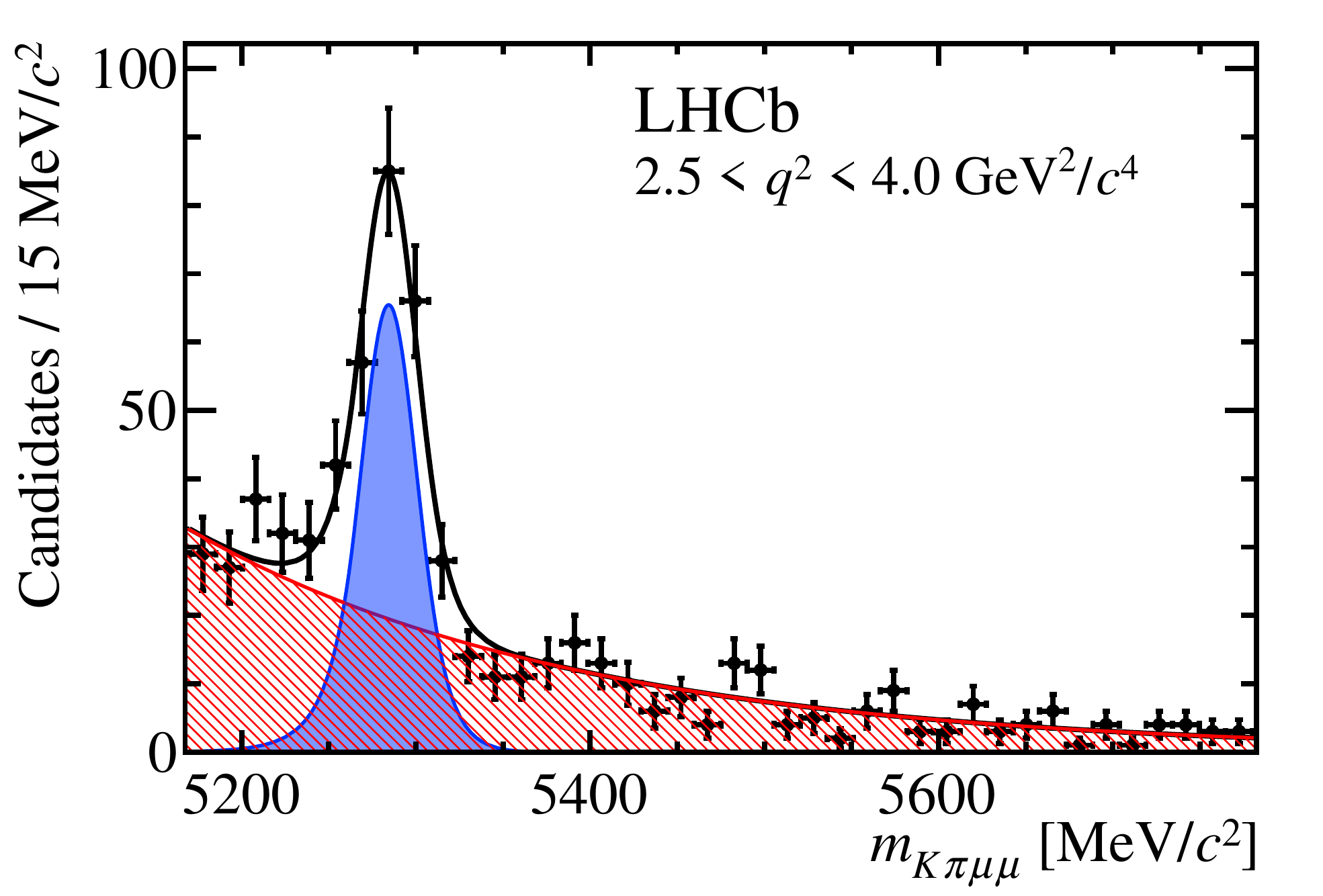}
  \includegraphics[width=0.45\textwidth]{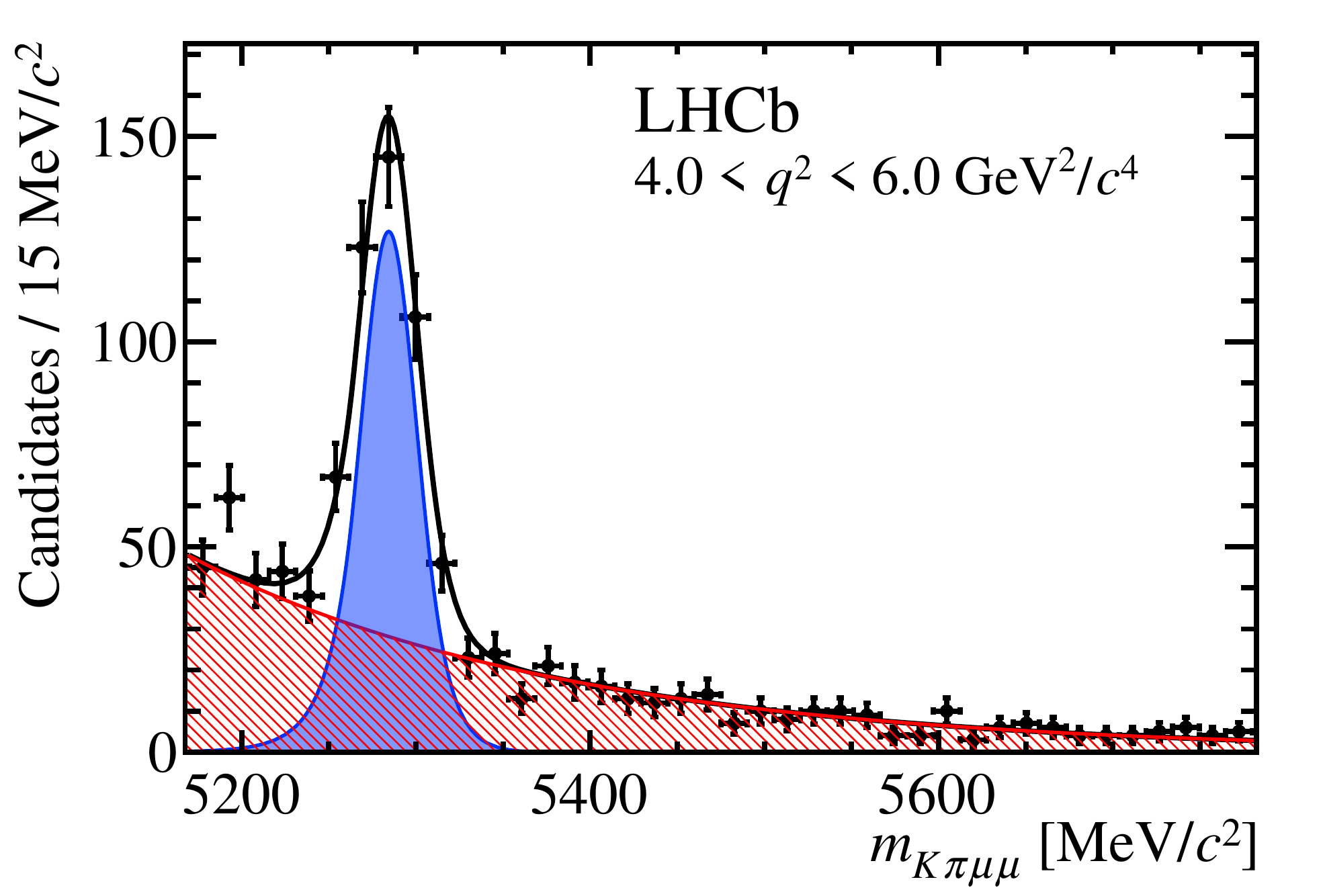}\\
  \includegraphics[width=0.45\textwidth]{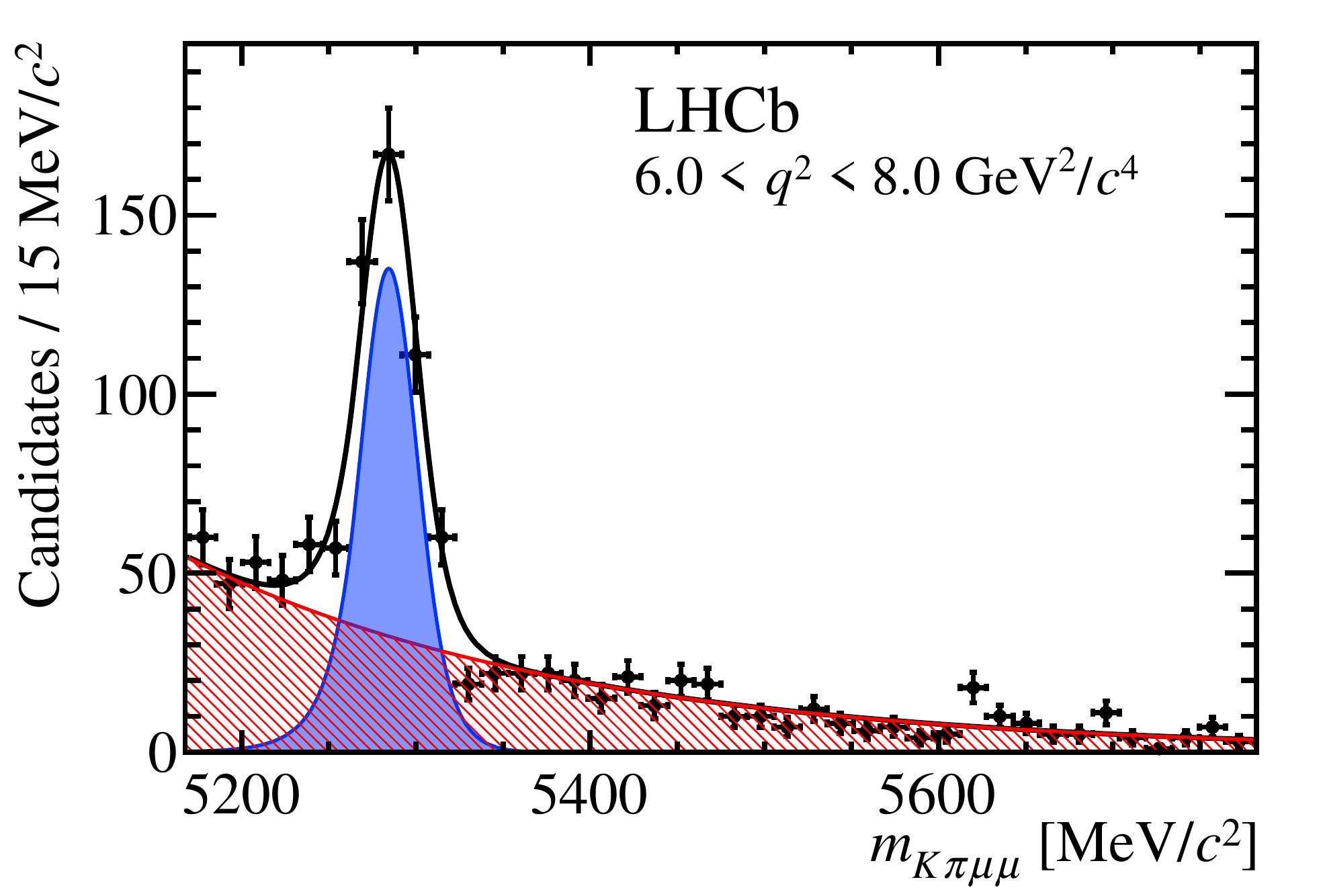}
  \includegraphics[width=0.45\textwidth]{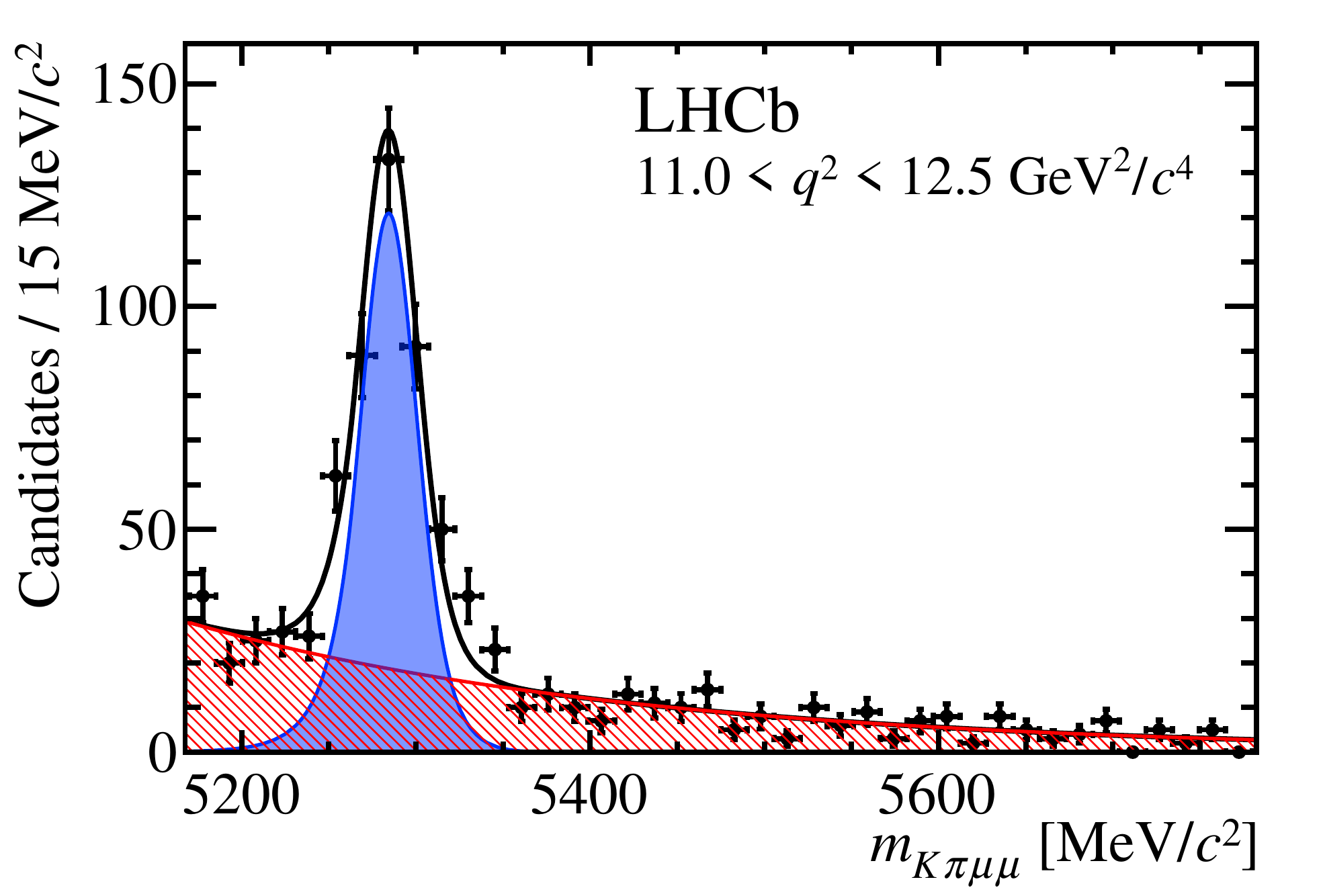}\\
  \includegraphics[width=0.45\textwidth]{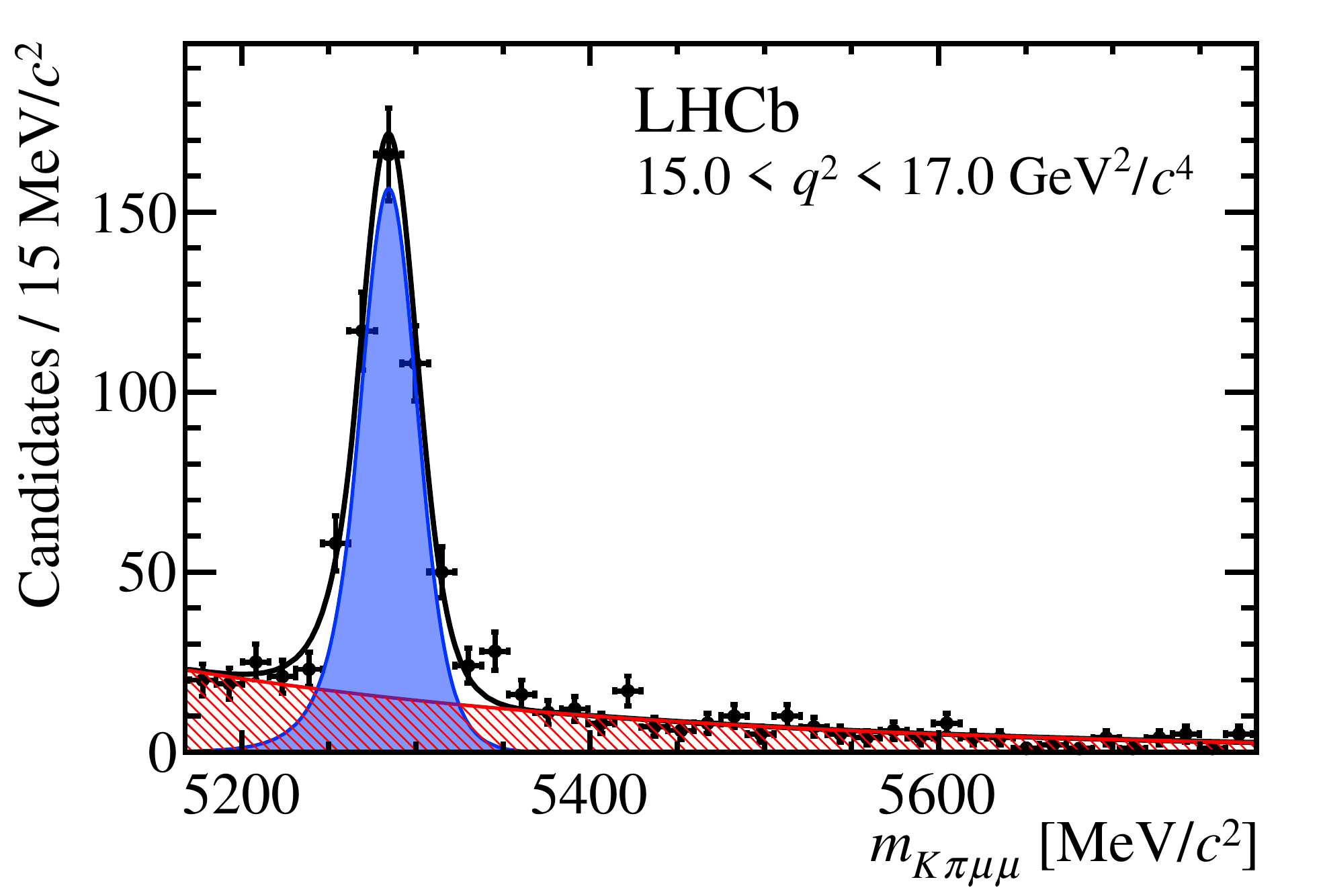}
  \includegraphics[width=0.45\textwidth]{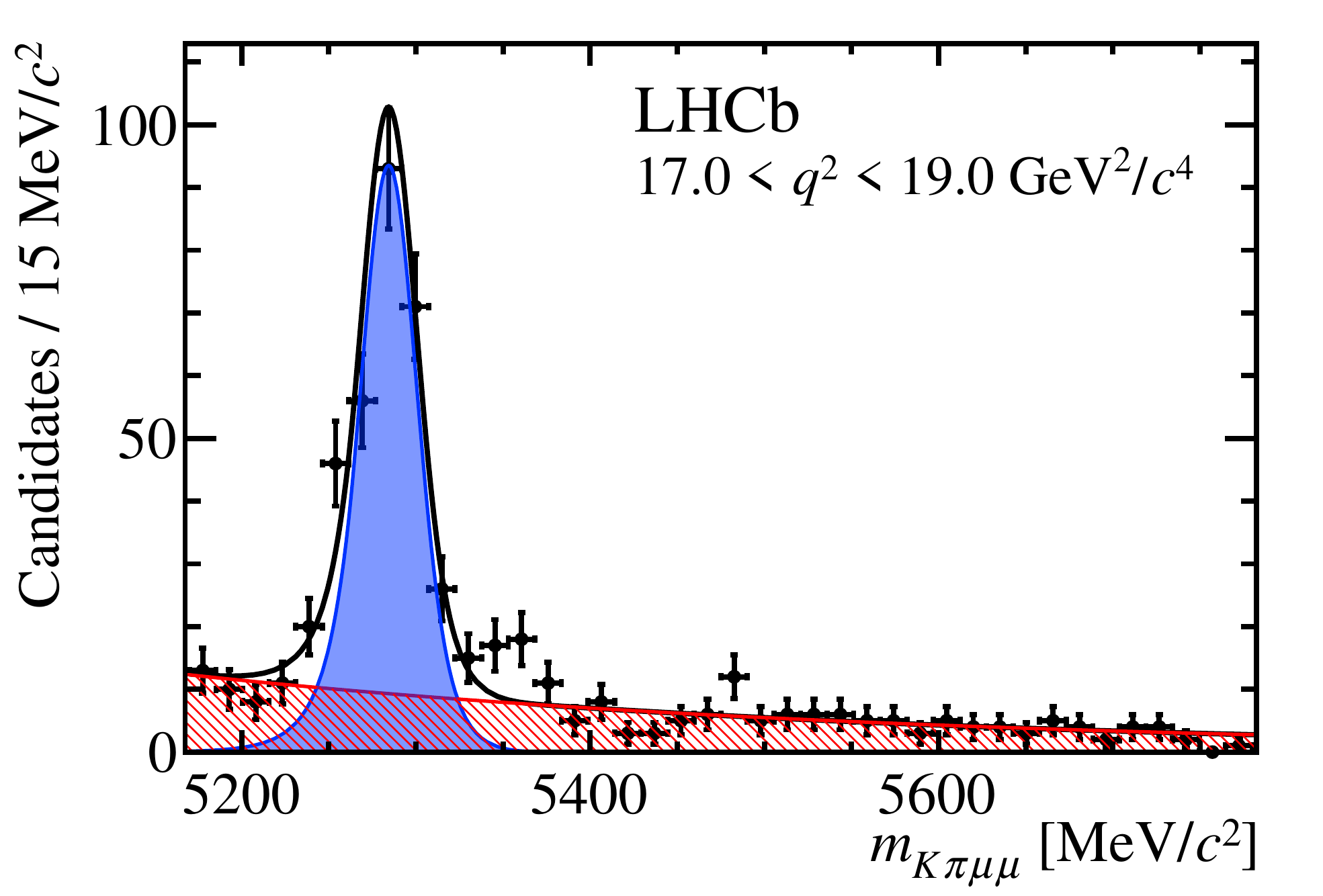}
  \caption{
    The $K^+\pi^-\mu^+\mu^-$ invariant mass distributions for the fine $\qsq$ bins.
    \label{fig:appmkpifit1}
  }
\end{figure}

\begin{figure}
  \centering
  \includegraphics[width=0.45\textwidth]{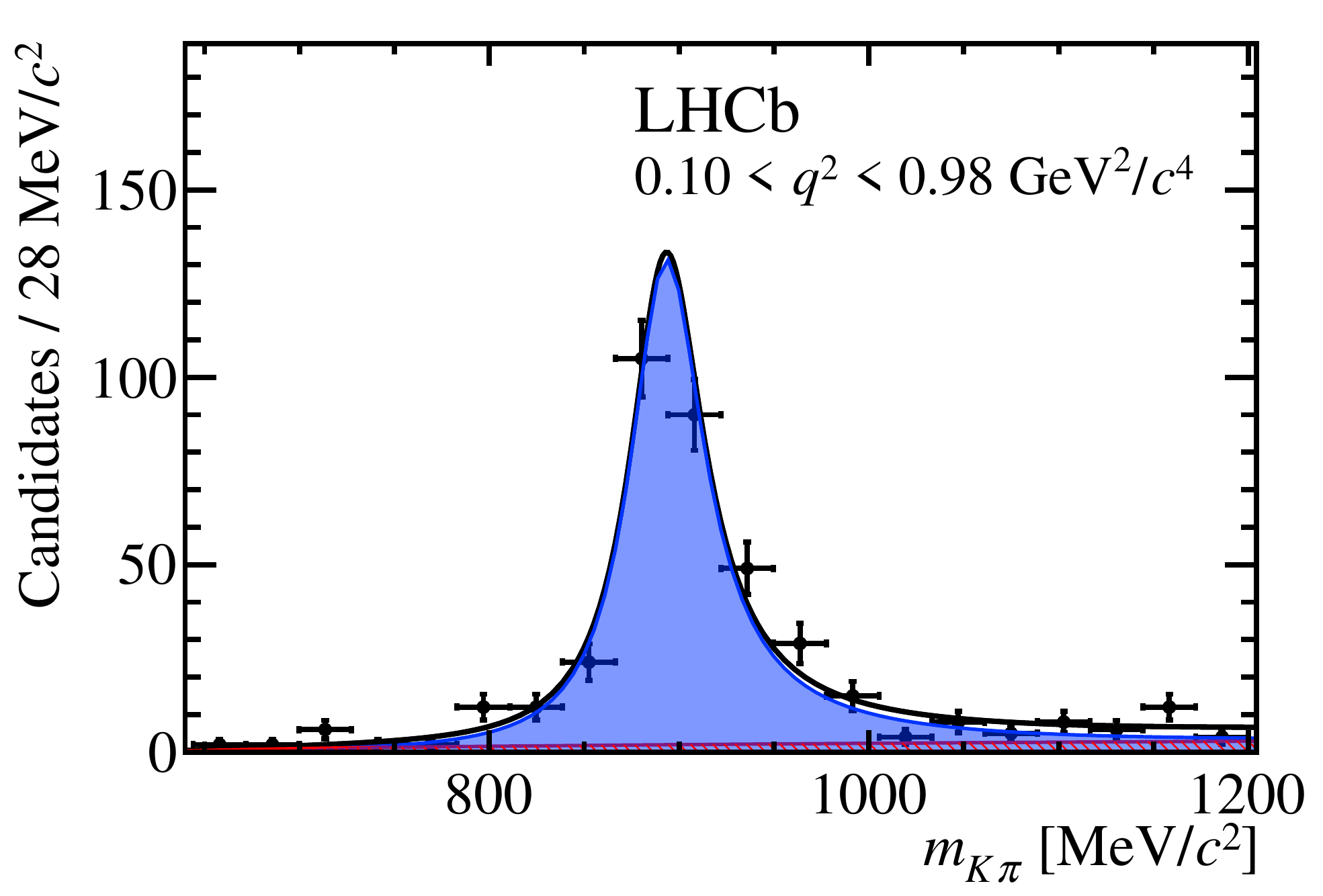}
  \includegraphics[width=0.45\textwidth]{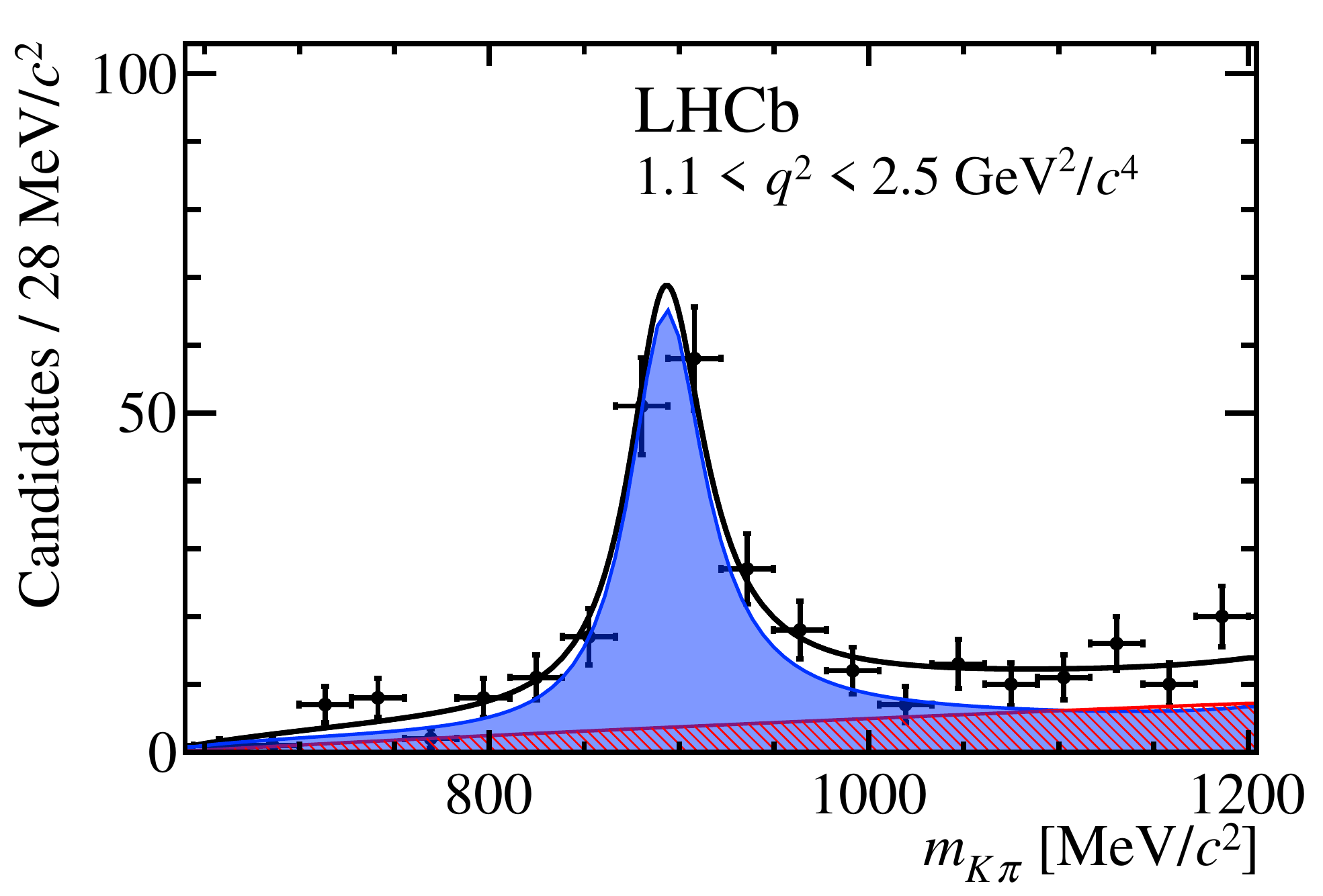}\\
  \includegraphics[width=0.45\textwidth]{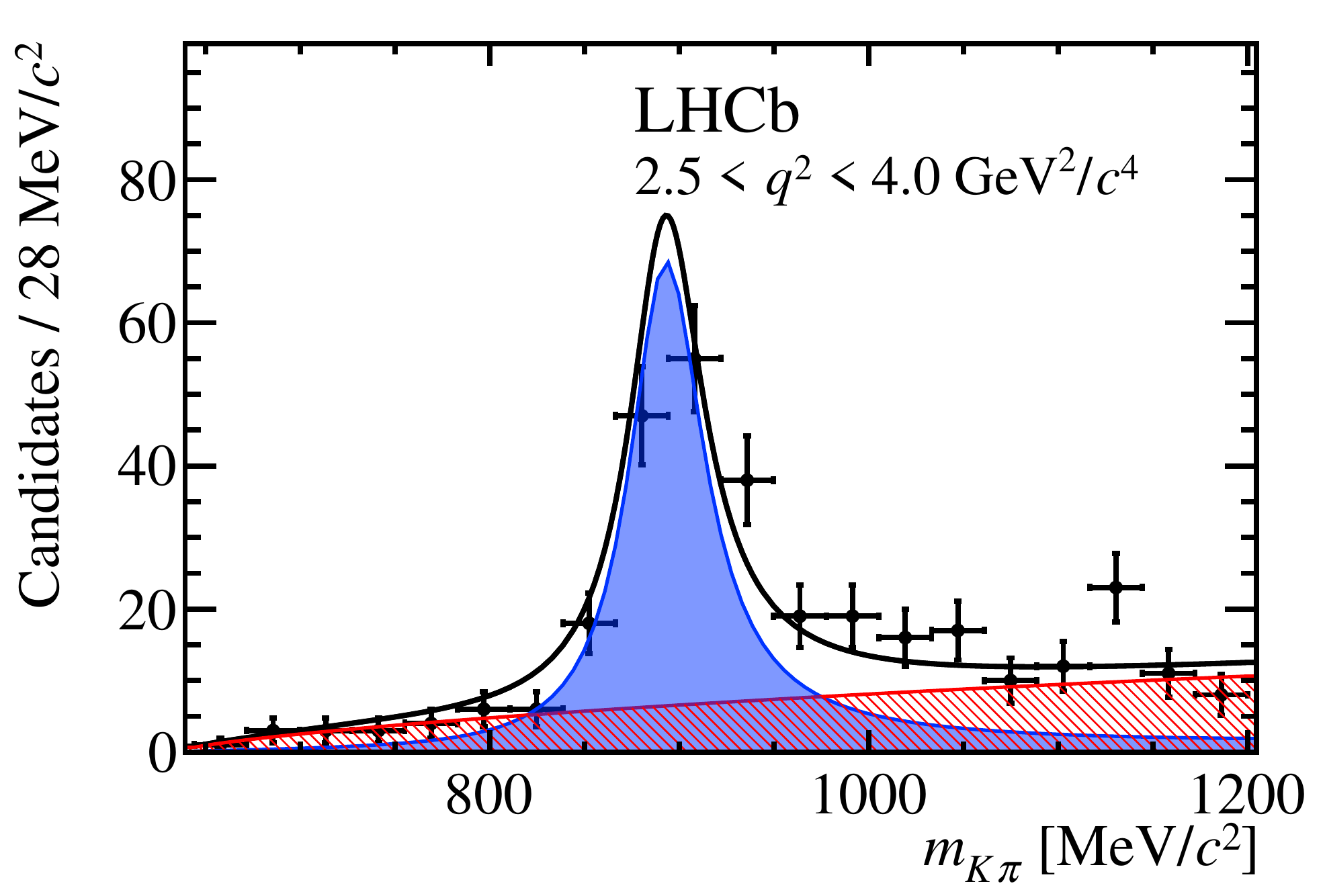}
  \includegraphics[width=0.45\textwidth]{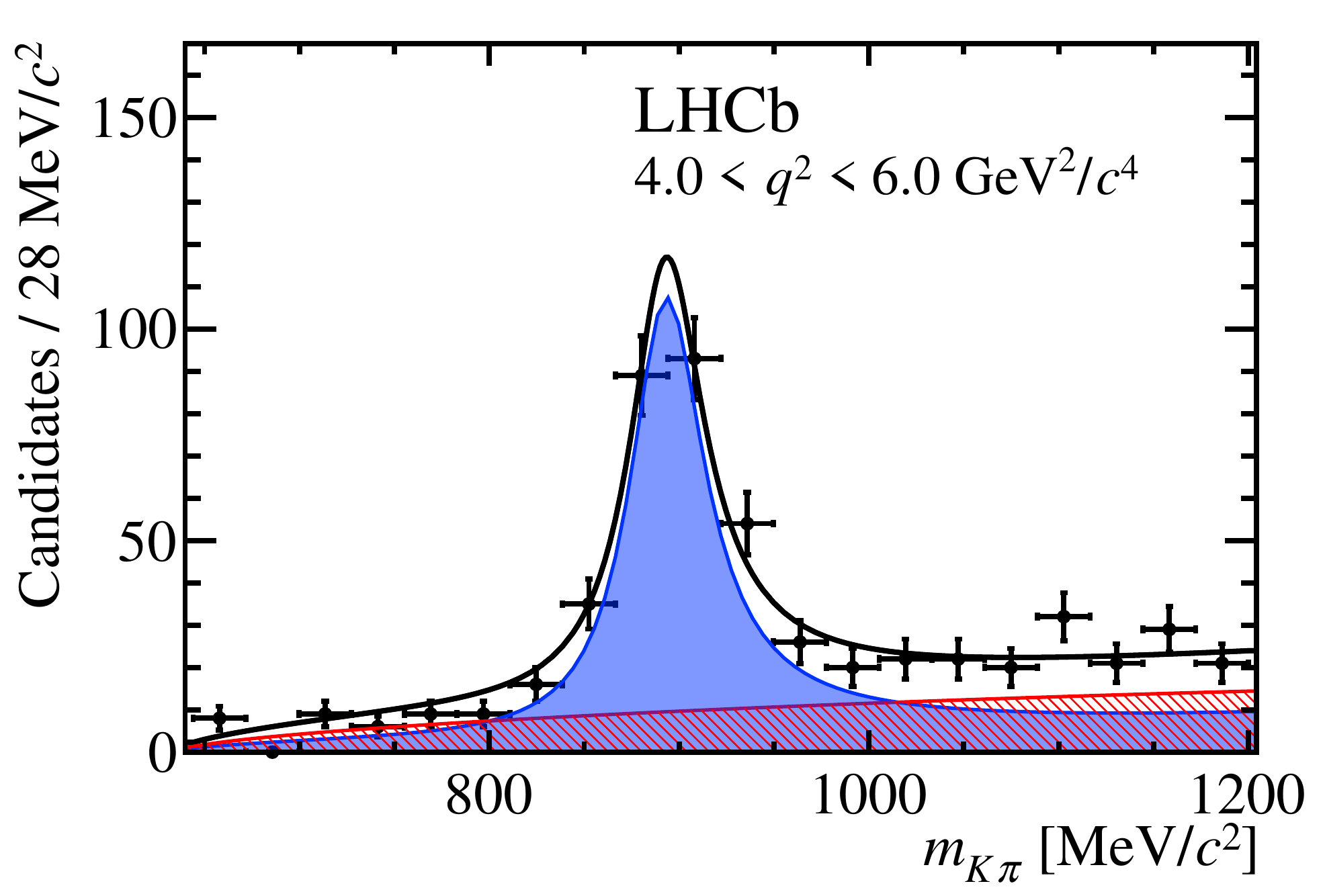}\\
  \includegraphics[width=0.45\textwidth]{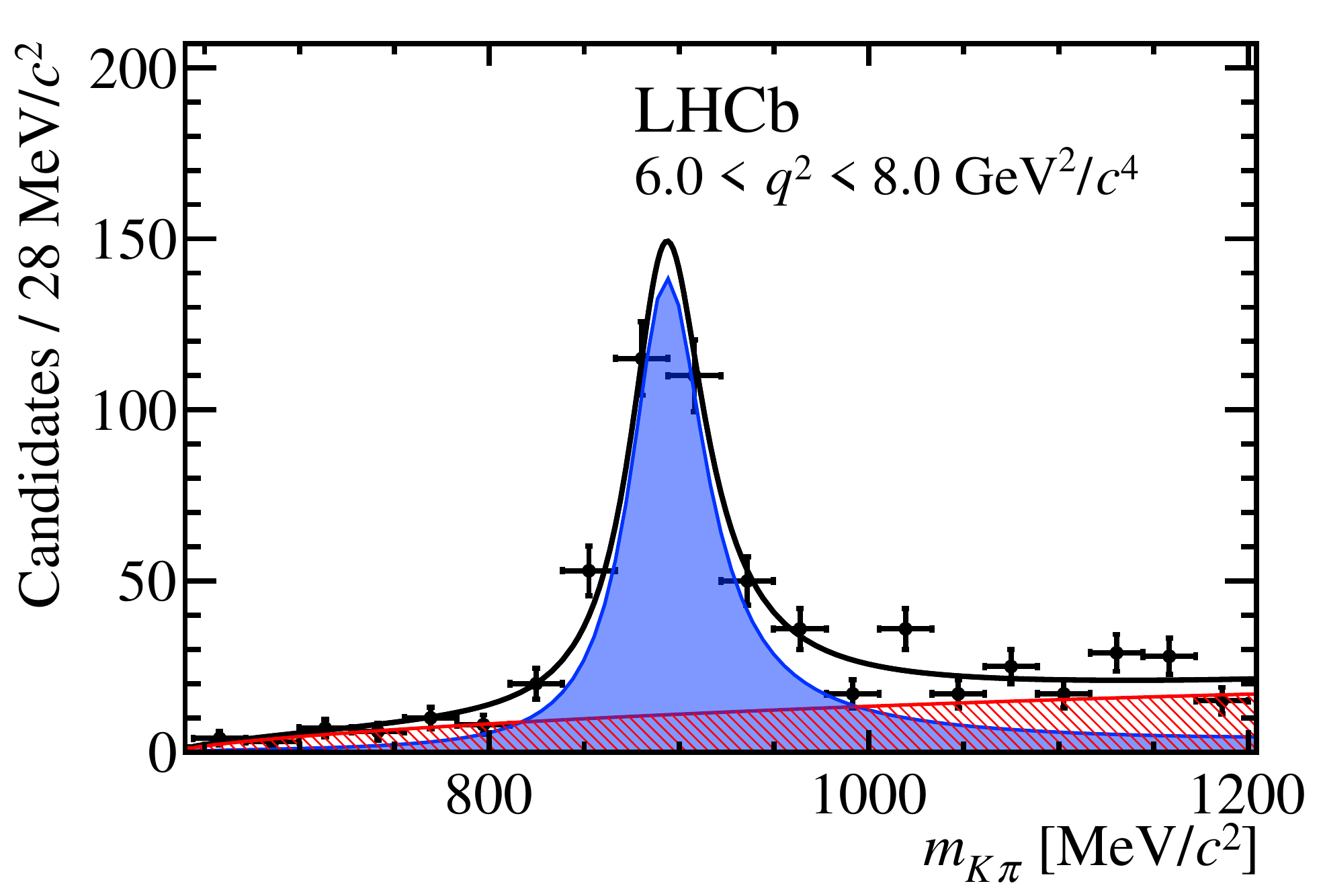}
  \includegraphics[width=0.45\textwidth]{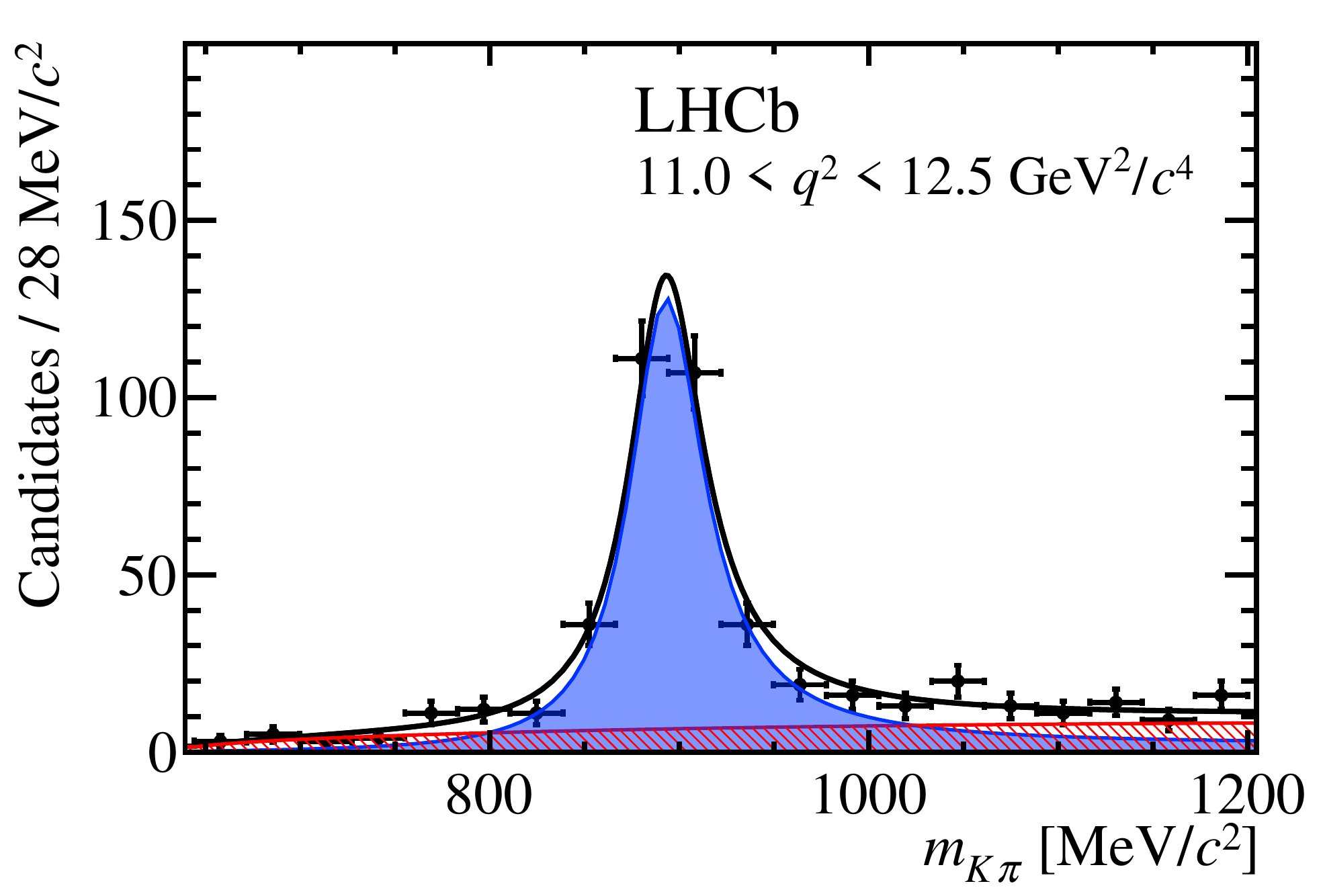}\\
  \includegraphics[width=0.45\textwidth]{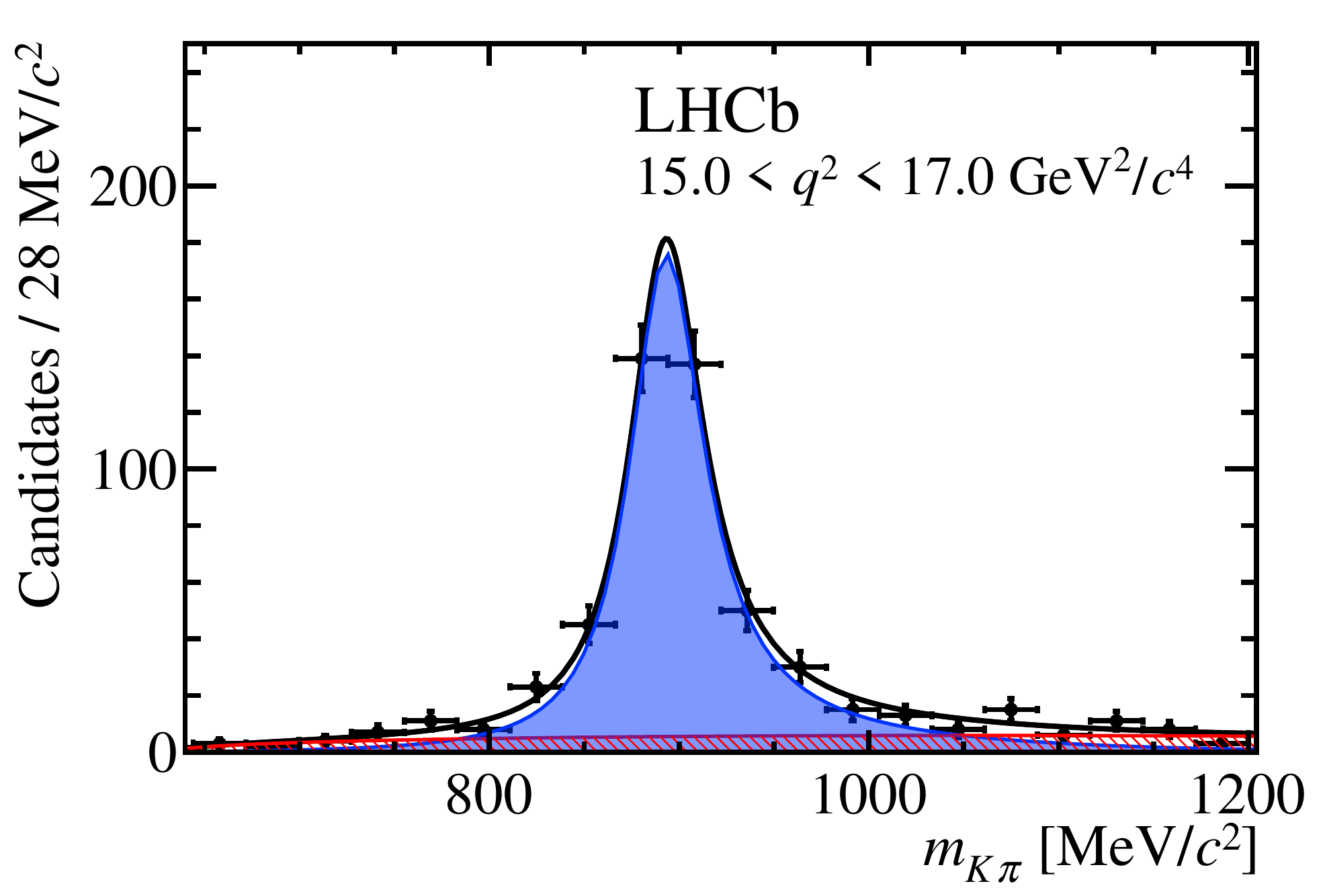}
  \includegraphics[width=0.45\textwidth]{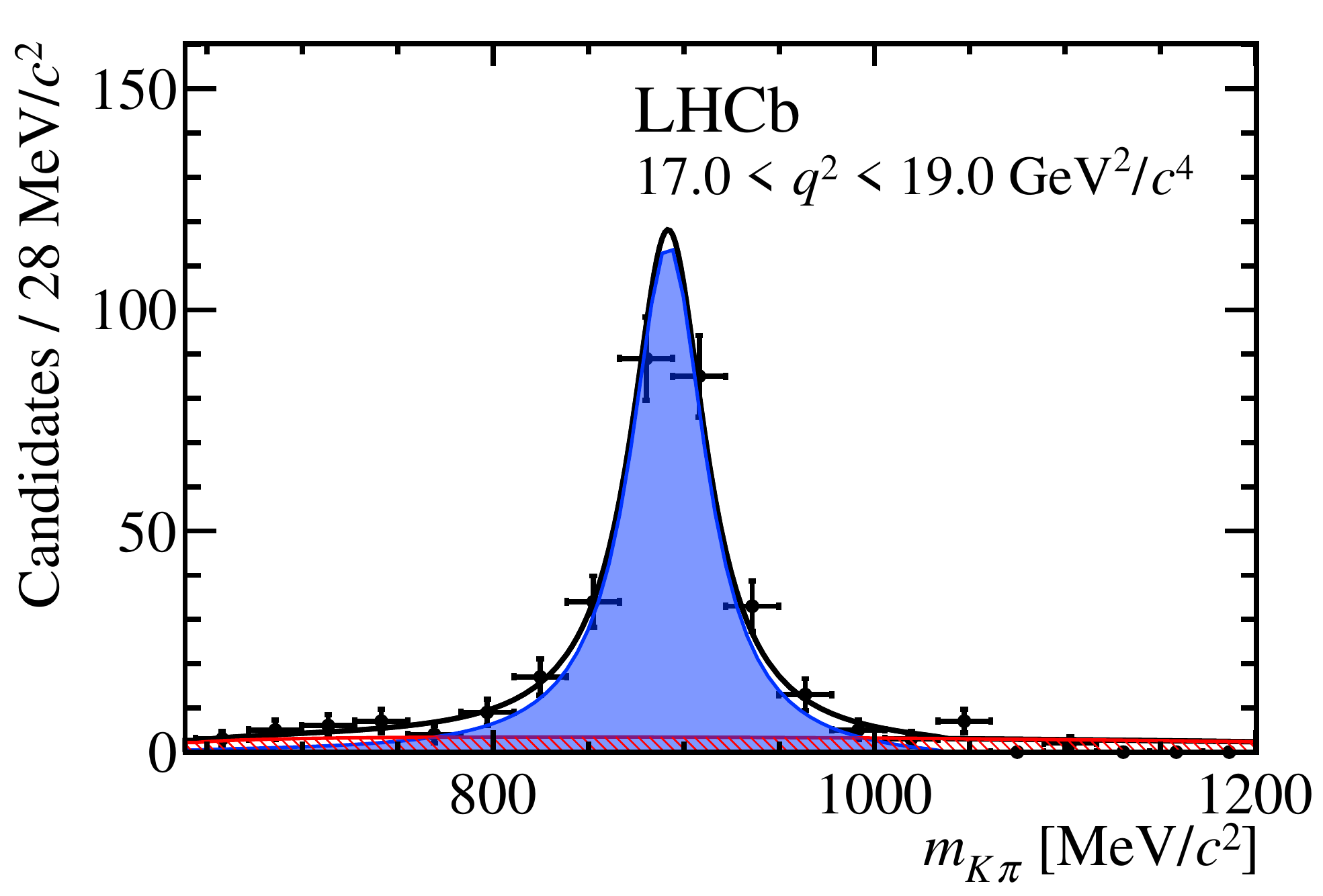}
  \caption{
    The $\Kp\pim$ invariant mass distributions for the fine $\qsq$ bins for candidates
    in the signal $\mkpimm$ window of $\pm50\mevcc$ around the known $\Bz$ mass.
    \label{fig:appmkpifit2}
  }
\end{figure}

\begin{figure}
  \centering
  \includegraphics[width=0.45\textwidth]{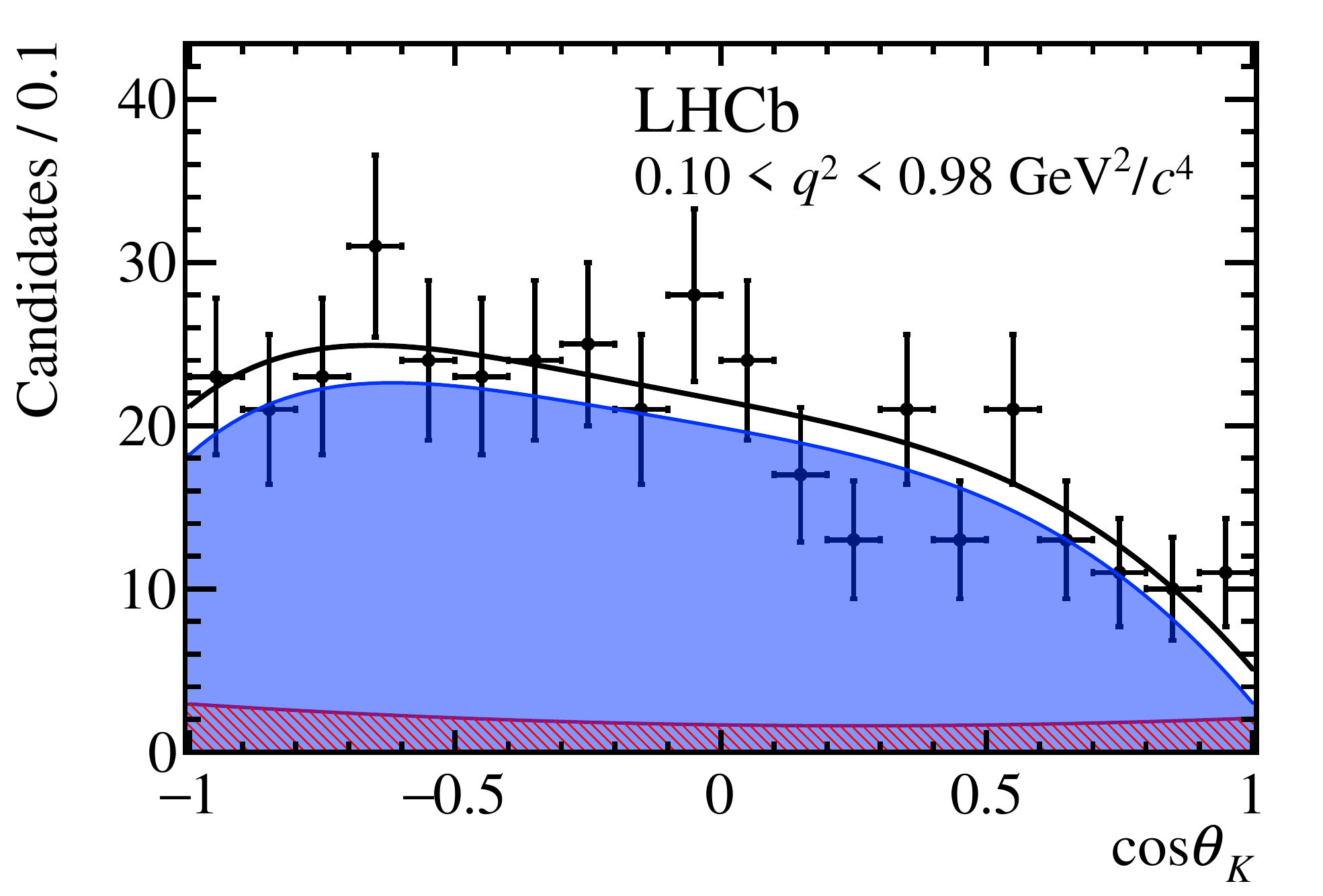}
  \includegraphics[width=0.45\textwidth]{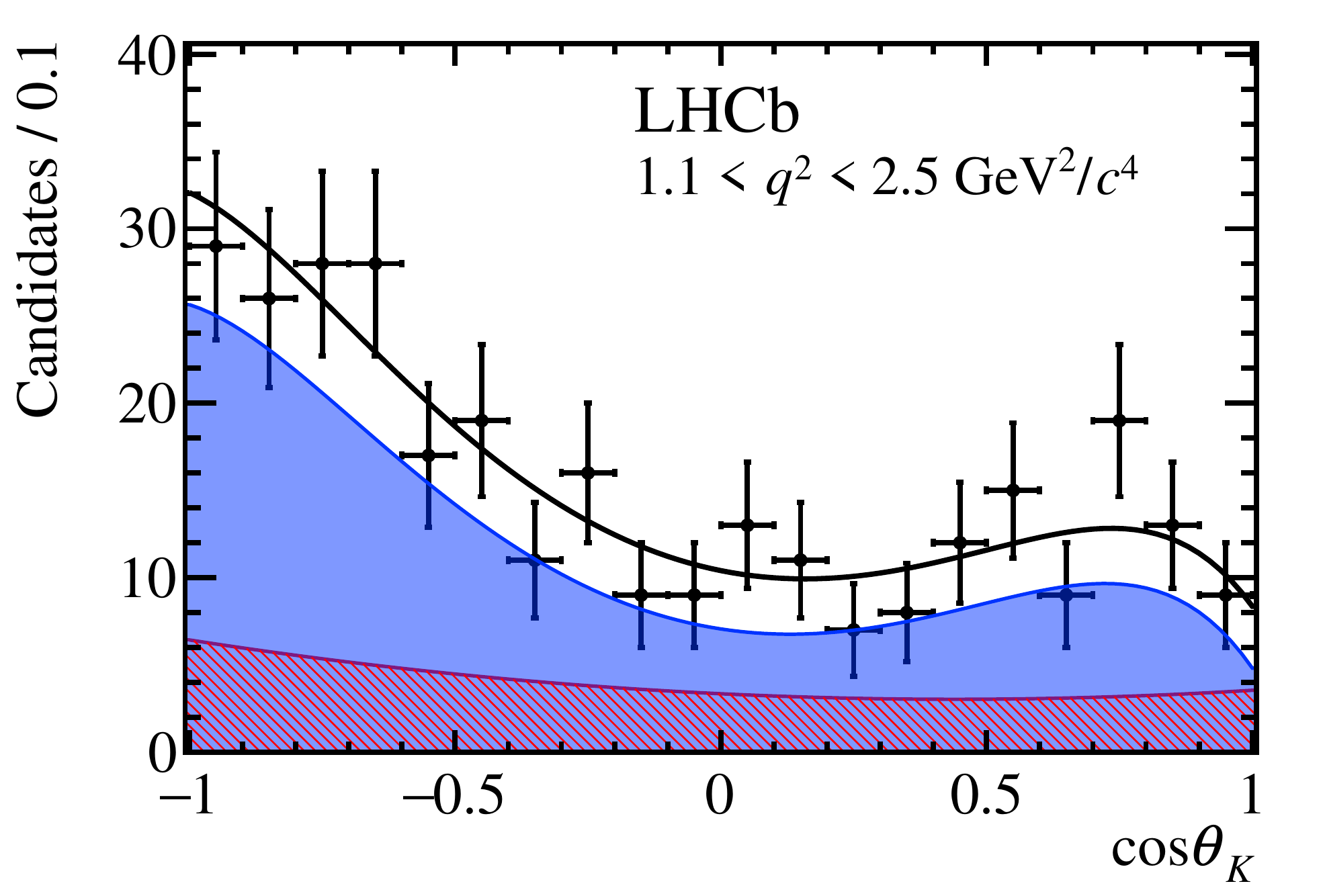}\\
  \includegraphics[width=0.45\textwidth]{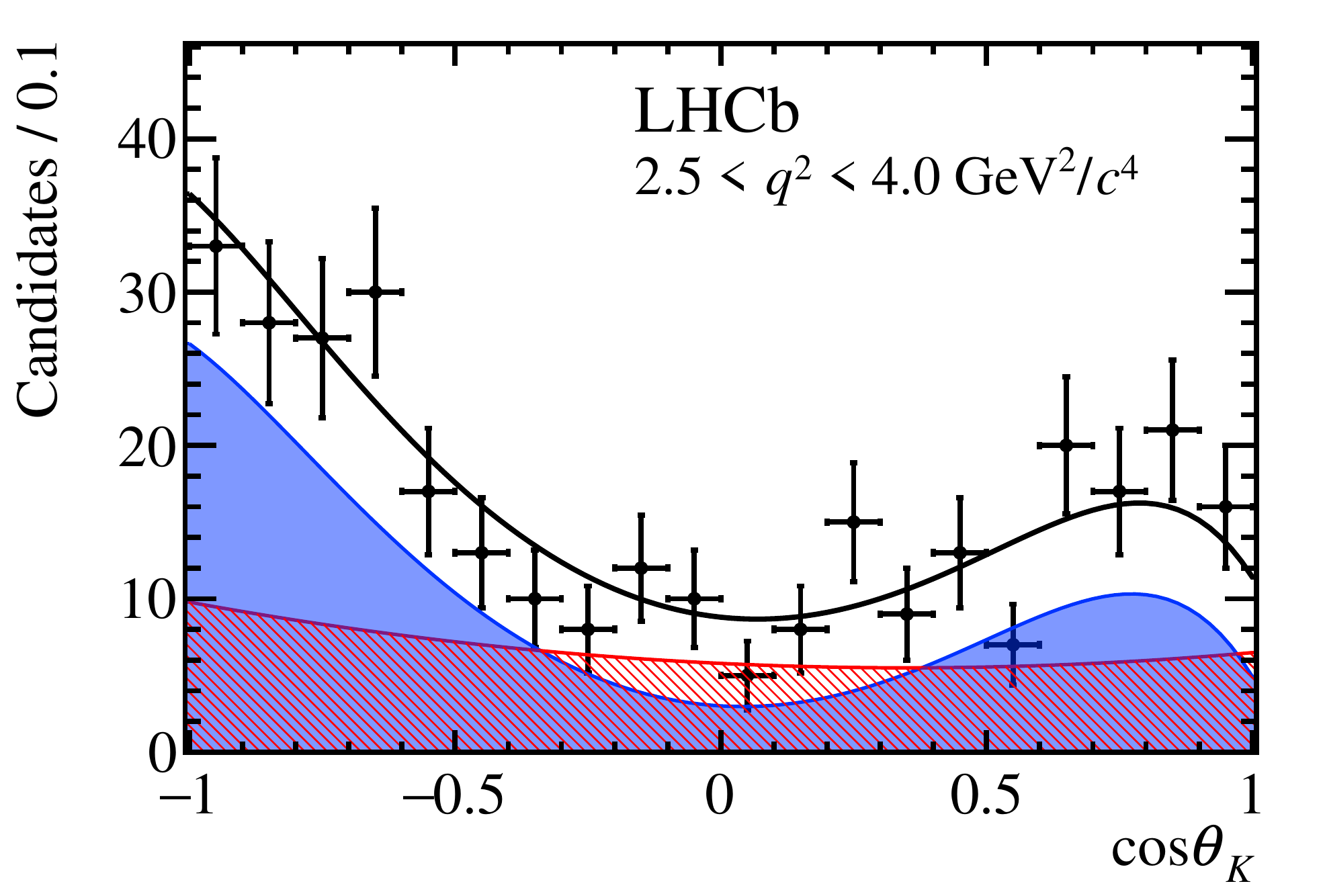}
  \includegraphics[width=0.45\textwidth]{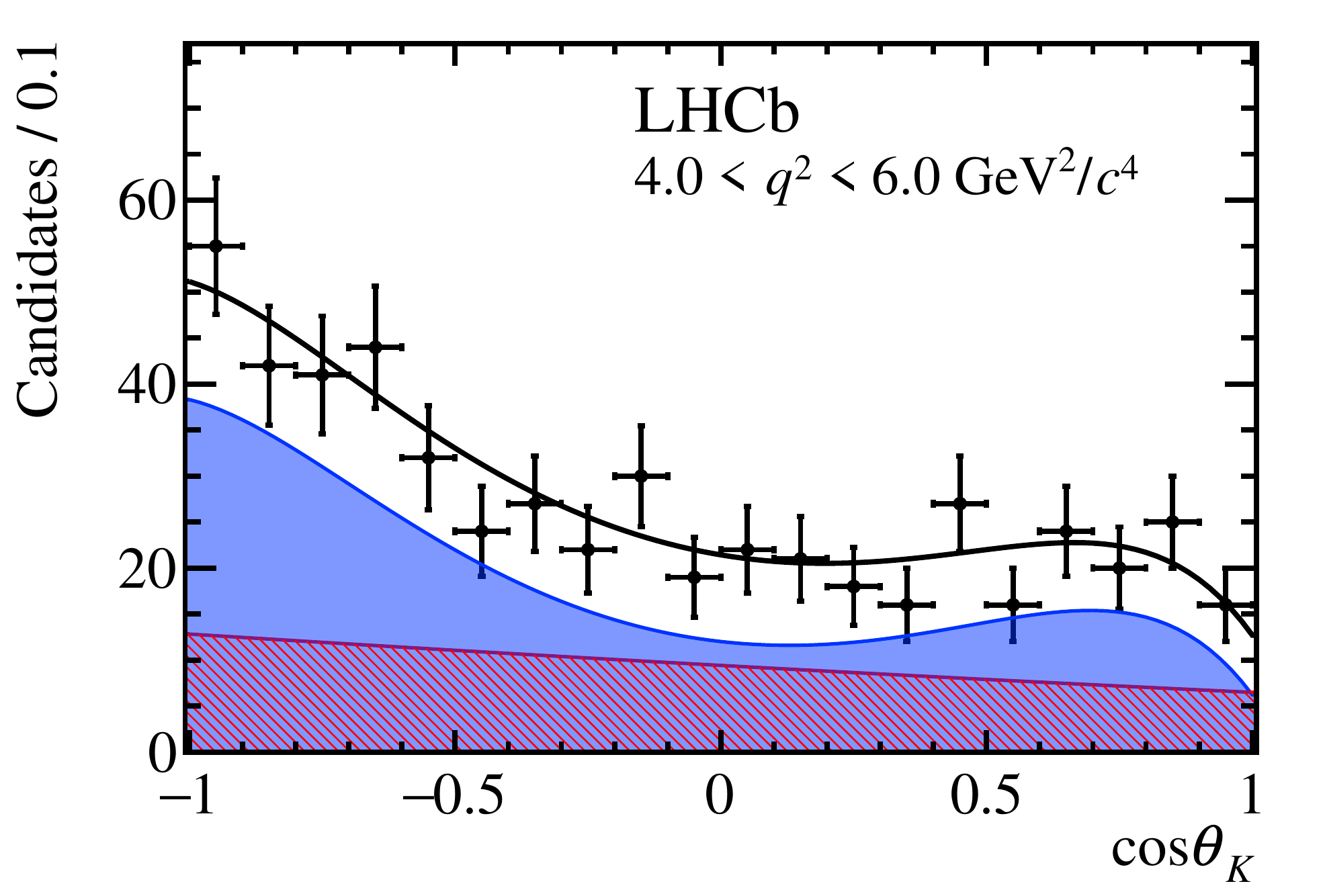}\\
  \includegraphics[width=0.45\textwidth]{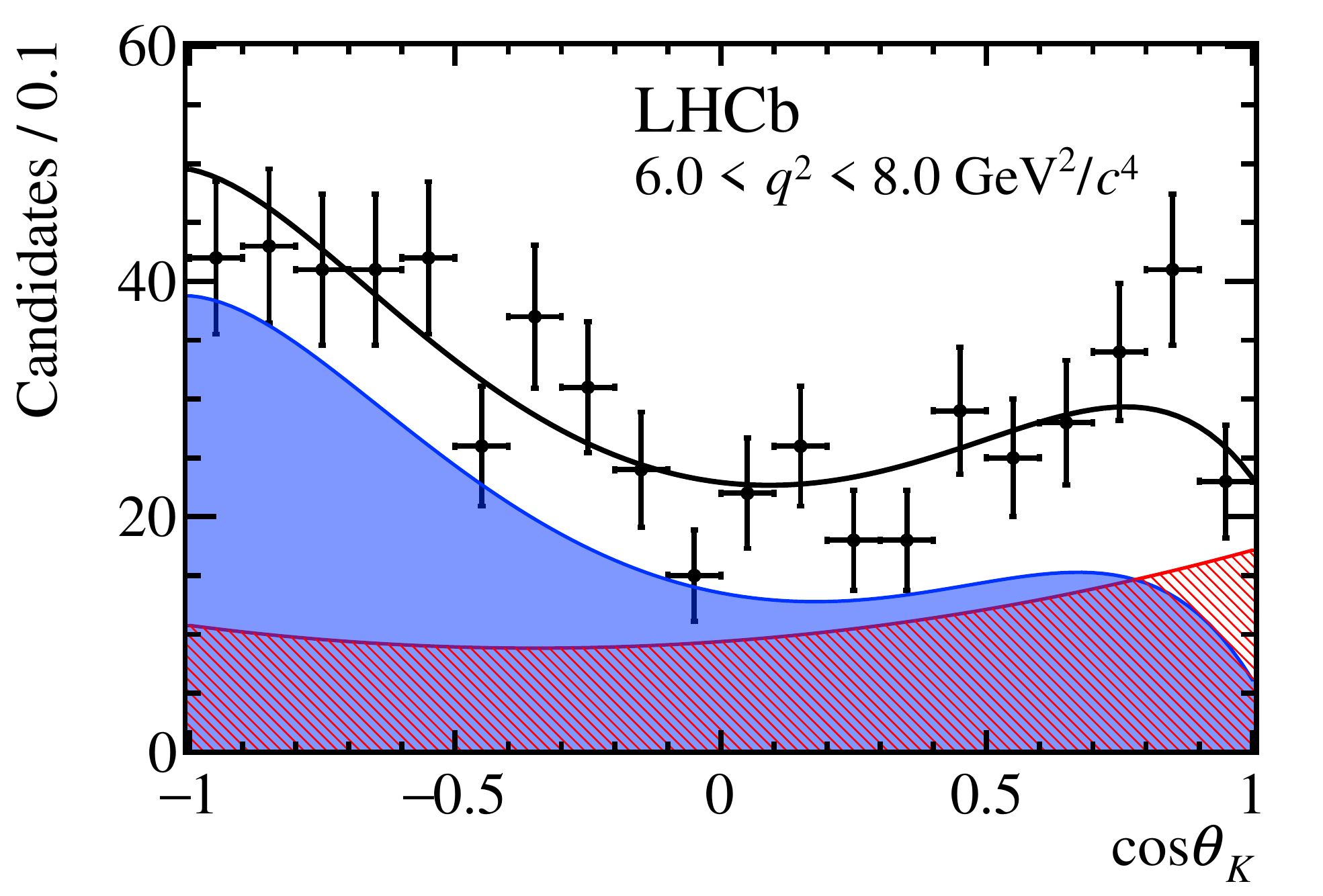}
  \includegraphics[width=0.45\textwidth]{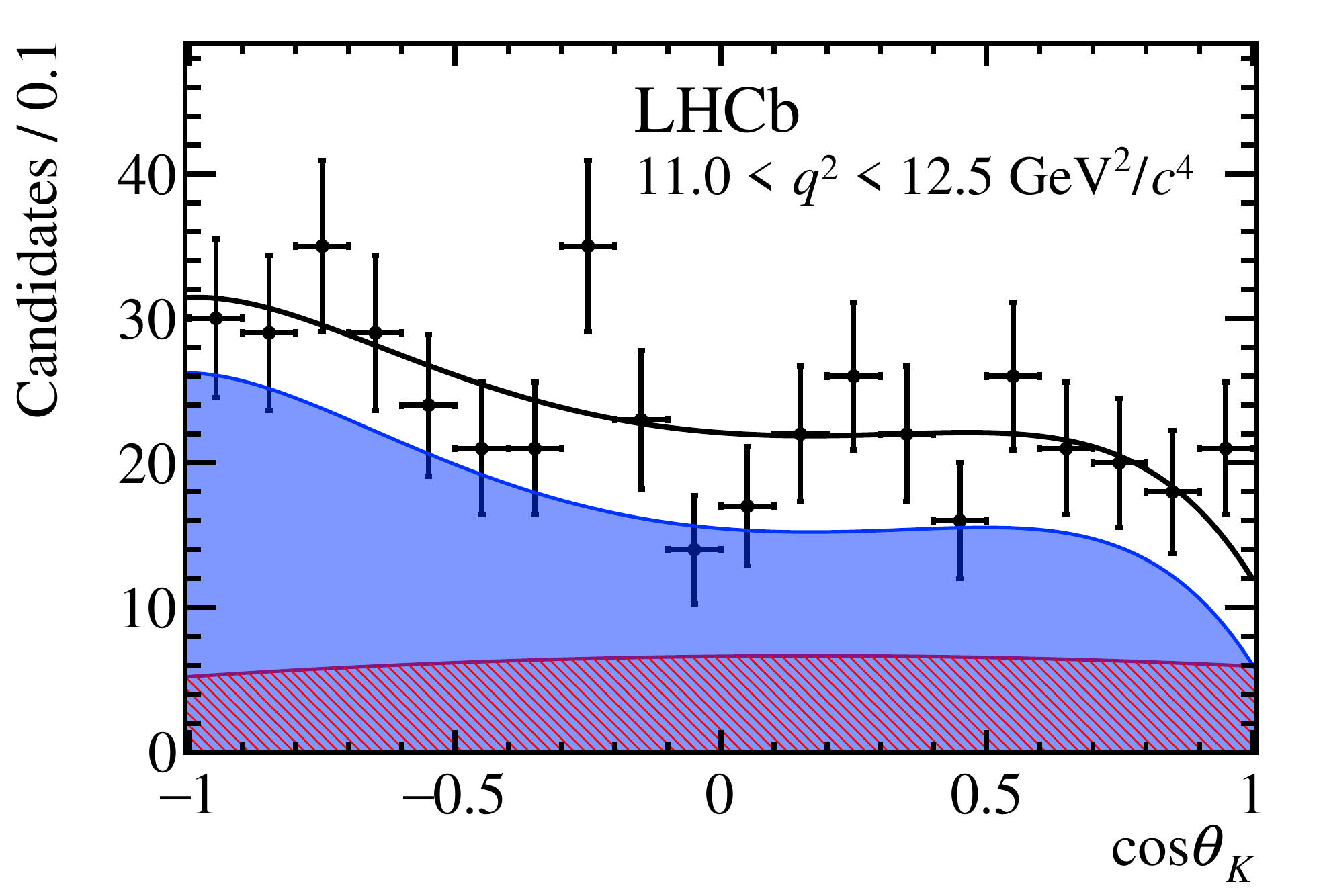}\\
  \includegraphics[width=0.45\textwidth]{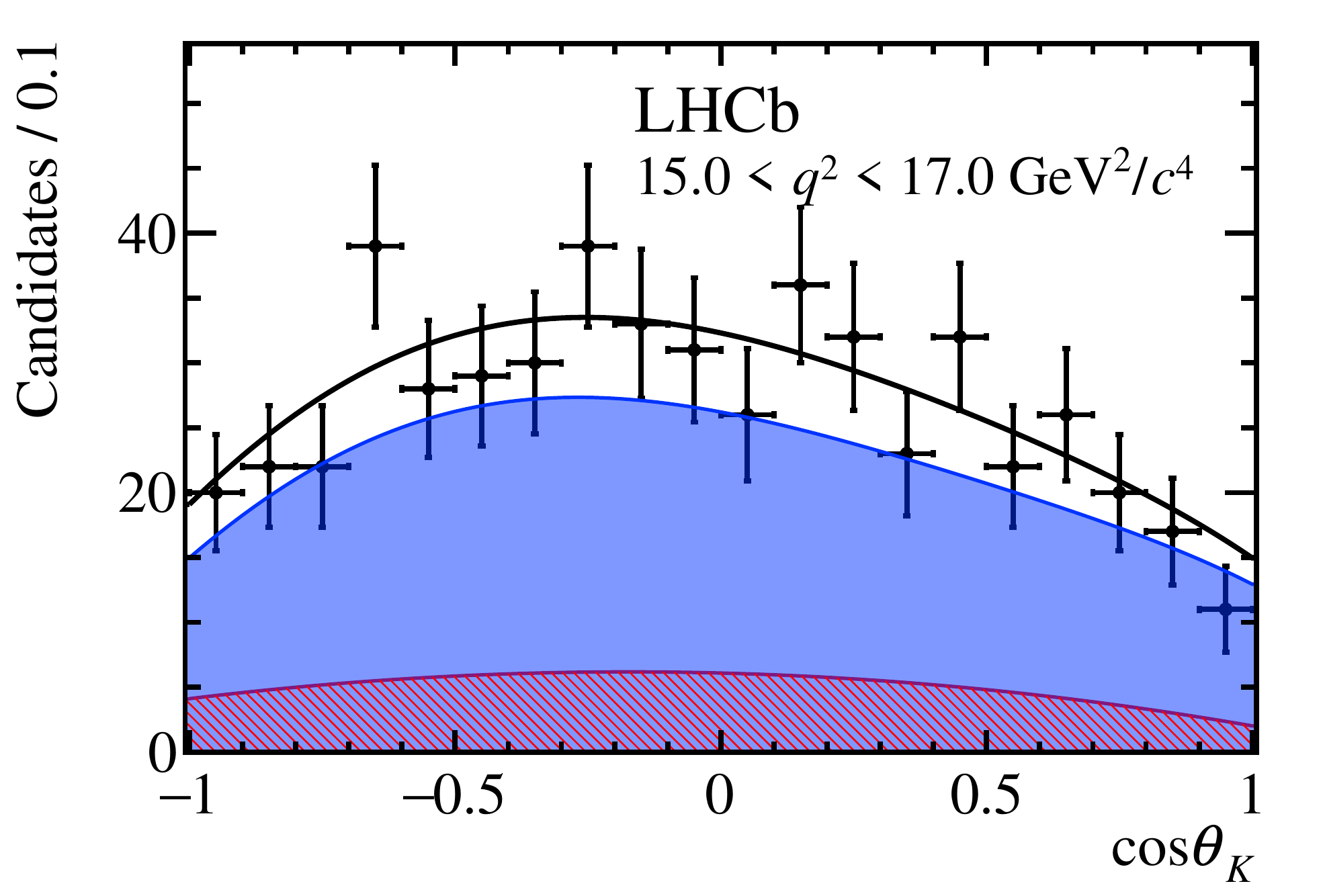}
  \includegraphics[width=0.45\textwidth]{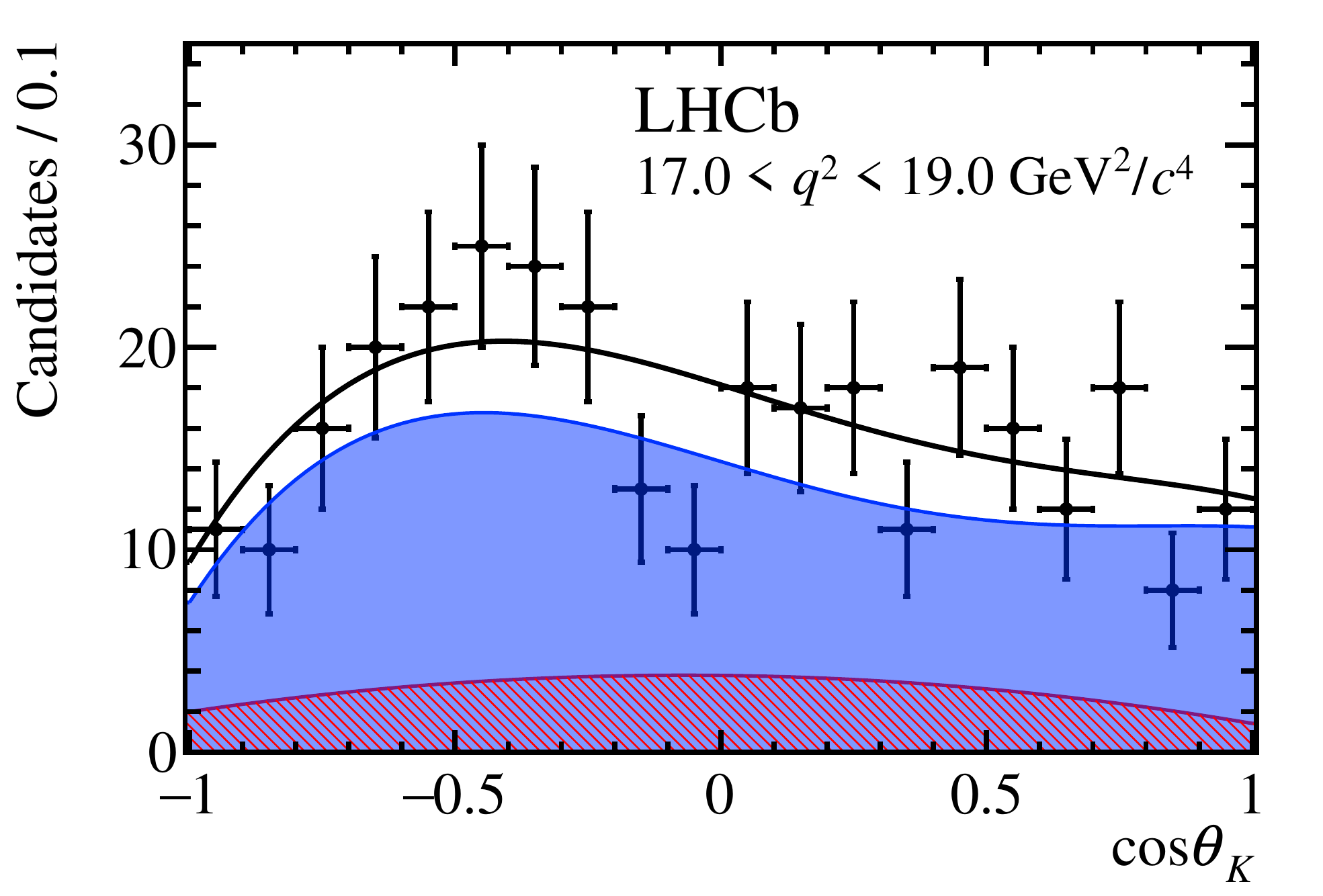}
  \caption{ 
    The \ctk angular distributions for the fine $\qsq$ bins for candidates
    in the signal $\mkpimm$ window of $\pm50\mevcc$ around the known $\Bz$ mass.
    \label{fig:appmkpifit3}
  }
\end{figure}

Figures~\ref{fig:appmkpifit0}--\ref{fig:appmkpifit3} show the projections of the fitted
probability density function on $\mkpimm$, $\mkpi$ and $\ctk$.
Figure~\ref{fig:appmkpifit0} shows the wider $\qsq$ bins of 
$1.1<\qsq<6.0\gevgevcccc$ and $15.0<\qsq<19.0\gevgevcccc$,
Figs.~\ref{fig:appmkpifit1}--\ref{fig:appmkpifit3} show the 
$\mkpimm$, $\mkpi$ and $\ctk$ projections respectively for the finer $\qsq$ bins.
In all figures, the solid line denotes the total fitted
distribution. The  individual components, signal (blue shaded area) and background
(red hatched area), are also shown.

\clearpage

\addcontentsline{toc}{section}{References}
\setboolean{inbibliography}{true}
\bibliographystyle{LHCb}
\bibliography{main,kstmm,LHCb-PAPER,LHCb-CONF,LHCb-DP,LHCb-TDR}

\newpage

\newpage
\centerline{\large\bf LHCb collaboration}
\begin{flushleft}
\small
R.~Aaij$^{39}$,
B.~Adeva$^{38}$,
M.~Adinolfi$^{47}$,
Z.~Ajaltouni$^{5}$,
S.~Akar$^{6}$,
J.~Albrecht$^{10}$,
F.~Alessio$^{39}$,
M.~Alexander$^{52}$,
S.~Ali$^{42}$,
G.~Alkhazov$^{31}$
% ...

{\footnotesize \it
% ...
$ ^{5}$Clermont Universit{\'e}, Universit{\'e} Blaise Pascal, CNRS/IN2P3, LPC, Clermont-Ferrand, France\\
$ ^{6}$CPPM, Aix-Marseille Universit{\'e}, CNRS/IN2P3, Marseille, France\\
% ...
$ ^{10}$Fakult{\"a}t Physik, Technische Universit{\"a}t Dortmund, Dortmund, Germany\\
% ...
$ ^{31}$Petersburg Nuclear Physics Institute (PNPI), Gatchina, Russia\\
% ...
$ ^{38}$Universidad de Santiago de Compostela, Santiago de Compostela, Spain\\
$ ^{39}$European Organization for Nuclear Research (CERN), Geneva, Switzerland\\
% ...
$ ^{42}$Nikhef National Institute for Subatomic Physics, Amsterdam, The Netherlands\\
% ...
$ ^{47}$H.H. Wills Physics Laboratory, University of Bristol, Bristol, United Kingdom\\
% ...
$ ^{52}$School of Physics and Astronomy, University of Glasgow, Glasgow, United Kingdom\\
% ...
}
\end{flushleft}

\end{document}